\pdfoutput=1
\documentclass[12pt]{article}
\usepackage{caption}
\usepackage{amsmath,amsfonts,amsthm,amssymb}
\usepackage{subfigure}
\usepackage{setspace}
\usepackage{Tabbing}
\usepackage{lastpage}
\usepackage{extramarks}
\usepackage{chngpage}
\usepackage[usenames,dvipsnames]{color}
\usepackage{graphicx,float,wrapfig}

\def\rf#1{(\ref{eq:#1})}
\def\lab#1{\label{eq:#1}}

\def\br{\begin{eqnarray}}
\def\er{\end{eqnarray}}
\def\be{\begin{equation}}
\def\ee{\end{equation}}
\def\({\left(}
\def\){\right)}

\def\rlx{\relax\leavevmode}
\def\IR{\rlx\hbox{\rm I\kern-.18em R}}

\def\ve{\varepsilon}

\newcommand{\sbr}[2]{\left\lbrack\,{#1}\, ,\,{#2}\,\right\rbrack}

\def\IZ{\rlx\hbox{\sf Z\kern-.4em Z}}
\def\IR{\rlx\hbox{\rm I\kern-.18em R}}
\def\IC{\rlx\hbox{\,$\inbar\kern-.3em{\rm C}$}}
\def\one{\hbox{{1}\kern-.25em\hbox{l}}}

\begin{document}

\begin{titlepage}
\vspace*{-1cm}

\vskip 2cm

\vspace{.2in}
\begin{center}
{\large\bf Breather-like structures in modified sine-Gordon models}
\end{center}

\vspace{.5cm}

\begin{center}
L. A. Ferreira~$^{\star}$ and Wojtek J. Zakrzewski~$^{\dagger}$

\vspace{.3 in}
\small

\par \vskip .2in \noindent
$^{(\star)}$Instituto de F\'\i sica de S\~ao Carlos; IFSC/USP;\\
Universidade de S\~ao Paulo  \\ 
Caixa Postal 369, CEP 13560-970, S\~ao Carlos-SP, Brazil\\
email: laf@ifsc.usp.br

\par \vskip .2in \noindent
$^{(\dagger)}$~Department of Mathematical Sciences,\\
 University of Durham, Durham DH1 3LE, U.K.\\
email: W.J.Zakrzewski@durham.ac.uk

\normalsize
\end{center}


\begin{abstract}

We report analytical and numerical results on breather-like field configurations in a theory which is a deformation of the integrable sine-Gordon model in $(1+1)$ dimensions.  The main motivation of our study is to test the ideas behind the recently proposed concept of quasi-integrability, which emerged from the observation that some field theories present an infinite number of  quantities which are asymptotically conserved in the scattering of solitons, and periodic in time in the case of breather-like configurations. Even though the mechanism responsible for such phenomena is not well understood yet, it is clear that special properties of the solutions under a space-time parity transformation play a crucial role. The numerical results of the present paper give support for  the ideas on quasi-integrability, and it is found that  extremely long-lived breather configurations satisfy these parity properties. We also report on a mechanism, particular to the theory studied here, that favours the existence of long lived breathers even in cases of significant deformations of  the  sine-Gordon potential.

\end{abstract} 
\end{titlepage}

\section{Introduction}
\label{sec:intro}
\setcounter{equation}{0}

Solitons play a fundamental role in the study of non-linear phenomena because in many situations they can be considered as the ``normal modes'' of the physical system in the strong coupling regime. In fact, in special examples of gauge theories in $(3+1)$ dimensions and integrable field theories in $(1+1)$ dimensions, there exist duality relations interchanging the roles of the fundamental excitations of the fields at weak coupling, and the solitons at the strong coupling regime \cite{duality}.  In $(1+1)$ dimensions, where soliton theory is much better understood, the solitons are often described as  those classical solutions that propagate without dissipation and dispersion, and when two of such solitons are scattered they do not destroy each other. The only effect of the scattering is a shift in their positions in relation to the values they would have had, had they  not participated in the scattering process. The most acceptable explanation for such behaviour is that, in practically all known soliton theories, there exists an infinite number of conserved quantities that dramatically constrains the dynamics, and leaves no options for the solitons after the scattering but to continue being themselves. This is a remarkable fact but it certainly has an annoying drawback. It forces solitons to exist only in the realm of the so-called exactly integrable field theories in $(1+1)$ dimensions. Such theories are, however, very special as they possess highly non-trivial hidden symmetries and they have been used as convenient laboratories to study non-perturbative phenomena and so have lead to the development of new and important techniques in field theories.  

Recently we have observed that some non-integrable field theories in $(1+1)$ dimensions, present properties similar to those of exactly integrable theories \cite{us,us2,recent}. They have soliton-like field configurations that behave in a scattering process in a very similar way to the true solitons, {\it i.e.} they do not destroy each other. We have also found that such theories possess an infinite number of quantities which are not exactly conserved in time, but are, however, asymptotically conserved. By that we mean that the values of these quantities do change, and change a lot, during their scattering process but they  return, after the scattering, to the values they have had before it. This is a remarkable property since from the point of view of the scattering what matters are the asymptotic states, and so such a theory looks a bit as an effective integrable theory. We have also observed that some of such non-integrable  theories possess breather-like solutions that are extremely long-lived, {\it i.e.} they oscillate without loosing much  energy through radiation for very long periods of time. In addition, each from this infinite number of 'almost conserved' quantities, when evaluated on these breather-like field configurations,  does vary in time but in a steady way  by oscillating between two fixed values. For these reasons we have named this phenomenon {\em quasi-integrability}. The mechanisms  responsible for these remarkable properties are not fully  understood yet, but we believe they will play a very important role in the study of many non-linear phenomena. Since exactly integrable theories are rare and in general do not describe realistic physical phenomena, the  {\em quasi-integrable} theories may be very useful in their description
of more realistic physical processes. 

In this paper we report some results of our numerical study of breather like field configurations in a $(1+1)$-dimensional theory of a real scalar field $\phi$ subjected to a potential which is a deformation of the sine-Gordon potential. This theory has already been considered in one of our previous papers \cite{recent}, and the deformed potential depends upon two free pa\-ra\-me\-ters $\ve$ and $\gamma$. The parameter $\ve$ measures the deformation of the potential away from the sine-Gordon potential. The $\gamma$-parameter, when different from zero, makes the potential not symmetrical under the reflection $\phi\rightarrow -\phi$. In \cite{us,recent} we have argued that the phenomenon of quasi-integrability may be related to some properties of the two-soliton and breather field configurations under a very specific space-time parity transformation. When the field $\phi$, eva\-lua\-ted on such configurations, is odd under this parity transformation, we have an infinite number of quasi-conserved quantities, {\it i.e.} quantities which are asymptotically conserved in the case of two-soliton solutions and oscillate in time for breather-like configurations. In order to have this property the potential has to be even under the change $\phi\rightarrow -\phi$. Thus, we would expect the cases when $\gamma\neq 0$ to be less integrable than the cases whith $\gamma=0$. Our numerical simulations do confirm this expectation, but we also observe an effect which had not been foreseen. Due to the way we build our initial field configuration, for the numerical simulations, from the exact sine-Gordon breather solution, the initial kinetic energy decreases with the increase of the $\ve$-parameter, and so does the amplitude of oscillations of the resulting breather-like configurations in the deformed theory. This makes this quasi-breather field to oscillate in a region where the deformed potential differs little from the sine-Gordon potential. So, we find that we can have very long lived breather-like fields  for theories which are (globally) large deformations of the integrable sine-Gordon model.

The paper is organized as follows. In section \ref{sec:model}, for completeness,  we present our ideas about quasi-integrability based on a anomalous quasi-zero curvature condition (Lax equation), and give the algebraic and dynamical arguments of why the properties of the initial field configurations under  this space-time parity transformations are important for the quasi-integrability concept. In section \ref{sec:numerical} we present the results of our numerical simulations. They have involved using the 4th order Runge-Kutta method to simulate the time dependence of field configurations which would allow us to determine and study various properties of breather-like solutions of the full equations of motion of our models for several choices of values of the parameters $\ve$ and $\gamma$ characterizing the potential. In section \ref{sec:conclusions} we present our conclusions. 

\section{The model and the concept of quasi-integrability}
\label{sec:model}
\setcounter{equation}{0}

In this paper we report some results  of our numerical simulations to study breather-like solutions of one of such quasi-integrable theories which corresponds of a particular deformation of the sine-Gordon model introduced in \cite{recent}. It is a $(1+1)$ dimensional theory of a real scalar field $\phi$ described by the Lagrangian
\be
L= \frac{1}{2}\left[ \(\partial_t \phi\)^2-\(\partial_x \phi\)^2\right] - V\(\phi\),
\lab{model}
\ee
where the potential depends on two real parameters $\varepsilon$ and $\gamma$ and is given by
\be
V(\phi)\,=\,\frac{1}{2} \,\frac{\left[1+\varepsilon\phi(\phi-2\gamma)\right]^3}{c^2\,(1-\varepsilon\gamma\phi)^2}\,\sin^2\left[ \psi(\phi)\right],
\lab{v3def}
\ee
where
\be
\psi\(\phi\)\,=\,\frac{c\,\phi}{\sqrt{1+\varepsilon\,\phi\,(\phi-2\,\gamma)}}.
\lab{three.one}
\ee
and
\be
c\,=\,\sqrt{1+\varepsilon\,\pi\left(\pi-2\,\gamma\right)}.
\lab{three.two}
\ee

Note that for $\varepsilon=0$, one gets $c=1$, $\psi =\phi$, and the potential \rf{v3def} reduces to the sine-Gordon potential 
\be
V_{\rm SG}=\frac{1}{2}\,\sin^2(\psi).
\lab{sgpot}
\ee
Thus, the theory \rf{model}, for $\varepsilon=0$, reduces to the sine-Gordon model defined by the Lagrangian
\be
L_{\rm SG}\,=\,\frac{1}{2}\left[(\partial_t \psi)^2\,-\,(\partial_x\psi)^2\right]\,-\,V_{\rm SG}.
\lab{one.onea}
\ee

\begin{figure}[tbp]
    \centering
    \includegraphics[width=1.02\textwidth]{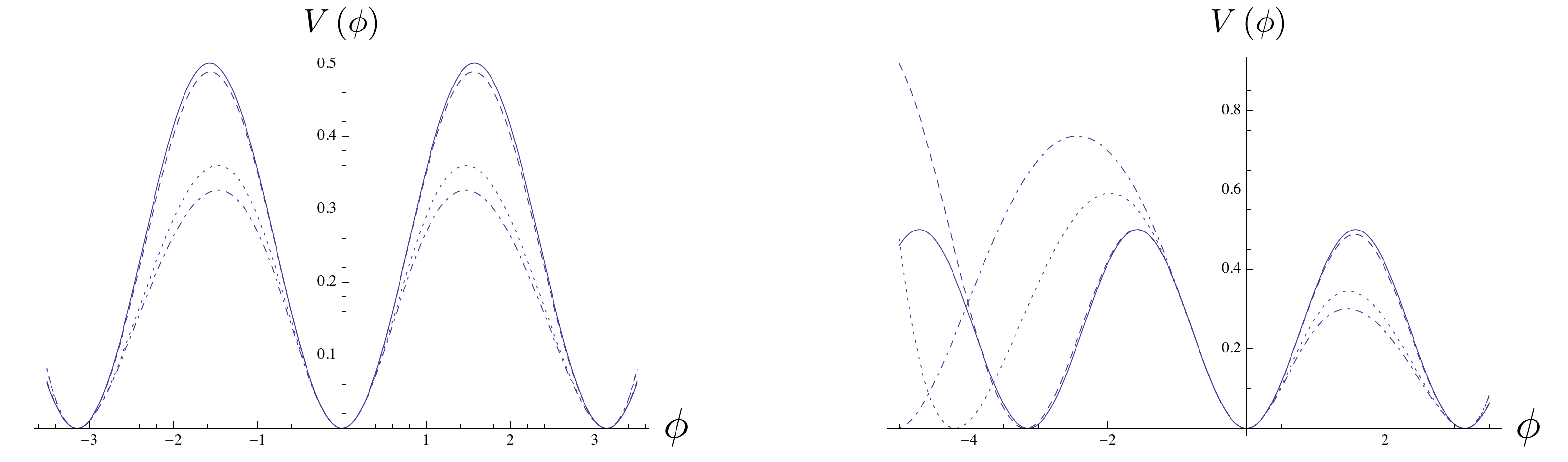}
    \begin{quote}
	\caption[AS]{\small Plot of the potential $V\(\phi\)$, given in \rf{v3def}, as a function of $\phi$ for $\ve=0.0$ (solid line), $\ve=0.01$ (dashed line), $\ve=0.2$ (dotted line) and $\ve=0.3$ (dot-dashed line). The plots on the left correspond to $\gamma=0$, and those on the right to $\gamma=0.2$. Note that  $V\(\phi\)$ is invariant under the change $\phi\rightarrow -\phi$ for the case $\gamma=0$, but not otherwise. In addtion, the vacua $\phi=0$ and $\phi=\pi$ are common to all values of $\ve$ and $\gamma$. For $\ve \neq 0$ the peaks of the potential grow in height, compared to those for $\ve =0$, for $\mid \phi\mid > \pi$, irrespective of the value of $\gamma$.} 
	\label{fig:plotv}
    \end{quote}
\end{figure}

The vacua of the sine-Gordon theory is obviously given by the constant field configurations $\psi = n\, \pi$, with $n$ integer. The theory \rf{model} also has infinitely many degenerate vacua but not equaly spaced like in the sine-Gordon case. However, the parameter $c$ given in \rf{three.two}, was chosen to preserve two of these vacua. Indeed, $\psi\(\phi=0\)=0$ and $\psi\(\phi=\pi\)=\pi$. The parameter $\gamma$ is important in our analysis of the quasi-integrability properties of the theories \rf{model}. Note that the potential \rf{v3def}  is even under the transformation $\phi\rightarrow -\phi$ for the case $\gamma=0$, {\it i.e.} $V_{\gamma=0}\(-\phi\)=V_{\gamma=0}\(\phi\)$ but not otherwise. As we will discuss below the quasi-integrability properties is favoured in the cases when $\gamma=0$. In figure \ref{fig:plotv} we show the potential \rf{v3def} for some values of $\varepsilon$ and $\gamma$. 

The potential \rf{v3def} was introduced in \cite{recent} using the techniques  of \cite{new,bazeia} based on ideas of self-dual or BPS solutions. Indeed, the static one-soliton solutions of the sine-Gordon model \rf{one.onea} given by
\be
\psi=2\,{\rm ArcTan}\( e^{\pm x}\)
\lab{sg1sol}
\ee
satisfy the BPS equation
\be
\partial_{x} \psi=\pm \sqrt{2\, V_{SG}}
\lab{bps-sg}
\ee
In fact, any static solution of the first order BPS equation   \rf{bps-sg} is a solution of the second order Euler-Lagrange (sine-Gordon) equation following from \rf{one.onea}. If one now introduces a field transformation $\psi\(\phi\)$, it follows that the new field $\phi$ satisfies the BPS equation 
\be
\partial_{x} \phi=\pm \sqrt{2\, V}
\lab{bps}
\ee
with the potential being given by
\be
V=\frac{V_{SG}}{\(\frac{d\,\psi}{d\,\phi}\)^2}.
\lab{transfpot}
\ee

The potential \rf{v3def} has been obtained from \rf{transfpot} by the field transformation \rf{three.one}. It then follows that the static solutions of \rf{bps} are solutions of the theory \rf{model}. Indeed, the static one-soliton solutions of \rf{model} are obtained from \rf{sg1sol} by applying the transformation \rf{three.one}. The transformation $\psi\(\phi\)$ maps BPS solutions of the sine-Gordon model \rf{one.onea} into BPS solutions of the theory \rf{model}. Note, however, that  in general, a given solution of the second order equation of motion of the sine-Gordon model is not necessarily mapped into a solution of the second order Euler-Lagrange equation corresponding to \rf{model}. 

The concept of quasi-integrability does not really depend upon the fact that the quasi-integrable theories are obtained from the integrable ones by field transformations of the type described above. However,  many aspects of our analysis get simplified by using such a connection between integrable and non-integrable theories. In particular, the initial configurations used in our numerical simulations for breathers, have been obtained by applying the field transformation  \rf{three.one} to the exact breather solutions of the sine-Gordon model.

As described in \cite{us,recent} our concept of quasi-integrability involves a connection $A_{\mu}$ satisfying an anomalous zero-curvature condition. Indeed, let us consider the connection  or Lax potentials given by 
\br
A_{+}&=& \frac{1}{2}\left[ \(\omega^2 \, V -m\)\, b_{1}
  -i\,\omega\, \frac{d\,V}{d\,\phi}\,F_1\right], 
\nonumber\\
A_{-}&=& \frac{1}{2}\, b_{-1} - \frac{i}{2}\,
\omega\, \partial_{-}\phi\, F_0, 
\lab{potentials}
\er
where we have used light-cone variables 
\be
x_{\pm}=\frac{1}{2}\(t\pm x\) \qquad {\rm with} \qquad  \partial_{\pm}=\partial_t\pm \partial_x \qquad {\rm and} \qquad  \partial_{+}\partial_{-}=\partial_t^2-\partial_x^2\equiv \partial^2.
\lab{lightconedef}
\ee
The quantities $b_n$ and $F_n$ appearing in \rf{potentials} are generators of the $sl(2)$ loop algebra defined as
\be
b_{2m+1}=\lambda^{m}\(T_{+}+\lambda\, T_{-}\),\qquad 
F_{2m+1}=\lambda^{m}\(T_{+}-\lambda\, T_{-}\),\qquad
F_{2m}=2\,\lambda^m\, T_3,
\lab{loopsl2gen}
\ee
where $\lambda$ is the so-called spectral parameter of the loop algebra, and $T_{3,\pm}$ are the generators of the finite $sl(2)$ algebra: 
\be
\sbr{T_3}{T_{\pm}}=\pm\, T_{\pm}, \qquad\qquad \sbr{T_{+}}{T_{-}}=2\, T_3.
\lab{finitesl2}
\ee 

 It is then easy to see that the curvature of the connection \rf{potentials} is given by
\be
F_{+-}\equiv \partial_{+}A_{-}-\partial_{-}A_{+}+\sbr{A_{+}}{A_{-}}= X
\, F_1 -\frac{i\,\omega}{2}\left[\partial^2 \phi + \frac{\partial\,
    V}{\partial\, \phi} \right]\,F_0
\lab{zc}
\ee
with
\be
X = \frac{i\,\omega}{2}\,  \partial_{-}\phi\,
\left[\frac{d^2\,V}{d\,\phi^2}+\omega^2\, V-m\right].
\lab{xdef}
\ee
Note that the Euler-Lagrange equation following from \rf{model} for a general potential $V$ is 
given by
\be
\partial^2 \phi + \frac{\partial\,V}{\partial\, \phi}=0.
\lab{eqofmot}
\ee

Thus, the term proportional to the Lie algebra generator $F_0$ vanishes when the field
configurations satisfy the equation of motion (are 'solutions' of the theory). For the case of  the sine-Gordon model potential
\be
V_{sg}= \frac{m}{\omega^2}\left[1-\cos\(\omega\,\phi\)\right]= \frac{2\,m}{\omega^2}\,\sin^2\(\frac{\omega}{2}\,\phi\)\, ;
\lab{anomalyfreesg}
\ee
the remaining term in \rf{zc}, {\it i.e.}
the anomaly $X$ given in \rf{xdef}, vanishes. In such a case  the curvature \rf{zc} vanishes for sine-Gordon solutions and this is what makes the sine-Gordon model integrable. For the potential \rf{v3def} however, the anomaly $X$ does not vanish irrespective of the choice of values of the parameters $\omega$ and $m$   (except for the trivial case $\omega=0$). 

The infinite number of quantities conserved asymptotically can be constructed using the techniques adapted from those of the integrable field theories. This can be done as follows. We perform the gauge transformation \cite{recent} 
\be
A_{\mu}\rightarrow a_{\mu}=g\, A_{\mu}\,g^{-1}-\partial_{\mu}g\,
g^{-1} \qquad\quad {\rm with} \qquad\quad 
g={\rm exp}\left[\sum_{n=1}^{\infty} \zeta_n\, F_n\right].
\lab{gaugeminus}
\ee
The parameters $\zeta_n$ in $g$ can then be chosen recursively starting from $n=1$ onwards in  such a way that the component $a_{-}$  of the transformed connection has only terms in the direction of the generators $b_{2m+1}$. So these terms generate an infinite dimensional abelian subalgebra of the $sl(2)$ loop algebra. 

Note that due to the non-vanishing anomaly $X$ the component $a_{+}$ has also terms in the direction of the generators $b_{2m+1}$ and $F_m$ as well. For an integrable theory, like the sine-Gordon one, the anomaly $X$ does vanish and the terms proportional to $F_m$ in $a_{+}$ vanish too, and the whole connection can be made to lie in the abelian subalgebra generated by $b_{2m+1}$. In the general case,
{\it i.e.} when the anomaly does not vanish, the transformed curvature becomes  
\be
F_{+-}\rightarrow
g\,F_{+-}\,g^{-1}=\partial_{+}a_{-}-\partial_{-}a_{+}+\sbr{a_{+}}{a_{-}}=
X \, g\, F_1\,g^{-1},
\lab{newcurvature}
\ee
where we have used the equation of motion \rf{eqofmot}. Note that the commutator of any $b_{2m+1}$ with any given $F_n$ produces terms proportional to the $F_m$ generators only. Therefore, for every component of the transformed curvature \rf{newcurvature} in the direction of a given $b_{2m+1}$ we get an equation of the form  
\be
\partial_{+}a_{-}^{(2n+1)}-\partial_{-}a_{+}^{(2n+1)}=
X\,\gamma^{(2n+1)}\qquad\qquad n=0,1,2,\ldots
\lab{quasiconserv}
\ee
with $a_{\pm}^{(2n+1)}$ and $\gamma^{(2n+1)}$ being the coefficients of the generators $b_{2m+1}$ in the expansion of $a_{\pm}$ and $g\, F_1\,g^{-1}$, respectively, in terms of  the elements of the basis of the $sl(2)$ loop algebra.

The relations \rf{quasiconserv} constitute an infinite number of anomalous conservation laws. Indeed, by re-expressing them in the $x$ and $t$ components (see \rf{lightconedef}) one gets the relations   
\be
\frac{d\,Q^{(2n+1)}}{d\,t}=-\frac{1}{2}\,\alpha^{(2n+1)},
\lab{timedercharges}
\ee
where
\be
Q^{(2n+1)}\equiv \int_{-\infty}^{\infty}dx\,a_{x}^{(2n+1)},\qquad\qquad\qquad
\alpha^{(2n+1)}\equiv \int_{-\infty}^{\infty}dx\,X\,\gamma^{(2n+1)}.
\lab{chargeanomalydef}
\ee 
The charges $Q^{(2n+1)}$ are not conserved due to the non-vanishing anomaly $X$. They would, of course, be conserved in an integrable theory like the sine-Gordon one for which $X=0$. 

 Note, however, that traveling solutions {\it i.e.} those which can be set at rest by a $(1+1)$ dimensional Lorentz transformation the charges $Q^{(2n+1)}$ are conserved. To see this we observe that in the rest frame  the charges are obviously $x$-dependent only, and so from \rf{timedercharges} one gets  $\alpha^{(2n+1)}=0$. But from \rf{quasiconserv} one finds that  $X\,\gamma^{(2n+1)}$ is a pseudo-scalar in $(1+1)$ dimensions, and so $\alpha^{(2n+1)}$ vanishes in any Lorentz frame. Therefore, for traveling solutions like the one-soliton solutions, the charges $Q^{(2n+1)}$ are conserved even in non-integrable theories. 

Next we note a striking property that helps us to define what we mean by a {\em quasi-integrable theory}. For some very special subsets of solutions of the theory \rf{model}  the charges $Q^{(2n+1)}$ satisfy what we call a {\em mirror symmetry}. 
For any one of the solutions in such a subset one can find a special point $\(t_{\Delta}\, ,\, x_{\Delta}\)$ in space-time, and define a parity transformation around this point: 
\be
P:\qquad \({\tilde x},{\tilde t}\)\rightarrow \(-{\tilde x},-{\tilde t}\) \qquad\qquad {\rm with} \qquad \quad{\tilde x}= x-x_{\Delta} \qquad \quad{\tilde t}=t-t_{\Delta},
\lab{paritydef}
\ee
The field $\phi$ corresponding to such a solution is odd under such parity, {\it i.e.} 
\be
\phi \rightarrow -\phi+{\rm const.}
\lab{niceparitytransf}
\ee
To find the implications of this observation we combine our parity transformation with the following order two automorphism of the $sl(2)$ loop algebra:
\be
\Sigma\(b_{2n+1}\)=- b_{2n+1},\qquad\qquad \Sigma\(F_{2n}\)=-F_{2n}, \qquad\qquad \Sigma\(F_{2n+1}\)=F_{2n+1}
\lab{automorphism}
\ee 
to build a $\IZ_2$ transformation $\Omega\equiv P\, \Sigma$, as the composition of a space-time and internal  $\IZ_2$ transformations. It turns out that the $A_{-}$ component of the connection \rf{potentials} is odd under such $\IZ_2$ transformation, {\it i.e.} $\Omega\(A_{-}\)=-A_{-}$. This fact can be used to show that the group element $g$ used to perform the gauge transformation \rf{gaugeminus} is even, {\it i.e.}  $\Omega\(g\)= g$. Then, one can use this fact to show that the factor $\gamma^{(2n+1)}$ in the integrand of $\alpha^{(2n+1)}$ is odd under the space-time parity, {\it i.e.} $P\(\gamma^{(2n+1)}\)=-\gamma^{(2n+1)}$. More details of this reasoning can be found in \cite{recent}.

If we now assume that the potential $V\(\phi\)$ in \rf{model} is even under the parity when evaluated on the special solutions satisfying \rf{niceparitytransf}, i.e. $P\(V\)=V$, then it follows that the anomaly $X$, given in \rf{xdef}, is also even, i.e. $P\(X\)=X$. Therefore we get that
\be
\int_{-{\tilde t}_0}^{{\tilde t}_0} dt\,\int_{-{\tilde x}_0}^{{\tilde x}_0} dx\, X\, \gamma^{(2n+1)}=0,
\lab{mm}
\ee
where the integration is performed on any rectangle with center in $\(t_{\Delta}\, ,\, x_{\Delta}\)$, i.e. ${\tilde t}_0$ and ${\tilde x}_0$ are any given fixed values of the shifted time ${\tilde t}$ and space coordinate ${\tilde x}$, respectively, introduced in \rf{paritydef}. Now,  by taking ${\tilde x}_0\rightarrow \infty$, we conclude that the  charges \rf{chargeanomalydef} satisfy the following mirror time-symmetry around the point $t_{\Delta}$: 
\be
Q^{(2n+1)}\(t={\tilde t}_0 +t_{\Delta}\)=Q^{(2n+1)}\(t=-{\tilde t}_0 +t_{\Delta}\) 
\qquad\qquad\qquad n=0,1,2,\ldots
\lab{mirrorcharge}
\ee
That is a remarkable property for the special subsets of solutions satisfying \rf{niceparitytransf} and belonging to a theory of type \rf{model} with an even potential under the parity \rf{paritydef}. Such subset of solutions defines our {\em quasi-integrable} theory. For the case of two-soliton solutions one note that by taking the limit ${\tilde t}_0 \rightarrow \infty$, one gets that the charges are asymptotically conserved, i.e. have the same values before and after the scattering.  

For the case of sine-Gordon theory it is true that for any two-soliton solution or breather solution it is possible to find a point in space-time $\(t_{\Delta}\, ,\, x_{\Delta}\)$,  such that the solution is odd under a parity transformation around such point. Let us now consider the theory \rf{model} with the potential \rf{v3def} and expand its solutions in powers of the parameter $\varepsilon$ around a given solution of the sine-Gordon model $\phi_0^{(-)}$ which is odd under the parity
\be
\phi=\phi_0^{(-)}+\ve\, \phi_1+\ve^2\,\phi_2+\ldots.
\ee
We now split the higher order solutions in even and odd parts as
\be
\phi_n^{(\pm)}\equiv \frac{1}{2}\(1\pm P\)\,\phi_n
\ee
It turns out that for the case where  $\gamma=0$, the equations of motion for the first order solution are of the form
\br
\partial^2 \phi_1^{(+)}+ \frac{\partial^2 V}{\partial \phi^2}\mid_{\ve=0}\, \phi_1^{(+)}&=& 0,
\\
\partial^2 \phi_1^{(-)}+ \frac{\partial^2 V}{\partial \phi^2}\mid_{\ve=0}\, \phi_1^{(-)}&= & f_1\(\phi_0^{(-)}\)
\nonumber
\er 
That means that the odd part of the first order solution $\phi_1^{(-)}$ satisfies a non-homogeneous equation and so it can never vanishes. On the other hand the even part $\phi_1^{(+)}$ satisfies a homogeneous equation and so it can vanish. In fact, if $\phi_1$ is a solution, so is $\phi_1-\phi_1^{(+)}=\phi_1^{(-)}$. Therefore, one can always choose a first order solution which is odd under the parity. If one makes such choice then it turns out that the second order solution has similar properties, i.e. $\phi_2^{(-)}$ satisfies a non-homogeneous equation and $\phi_2^{(+)}$ a homogeneous one. Then one can again choose the second order solution to be odd, and the process repeats in all orders. For a detailed account of that please see \cite{recent}. Such argument works for the potential \rf{v3def} with $\gamma=0$ but not otherwise. Therefore, we can say that the theory \rf{model} with the potential \rf{v3def} with $\gamma=0$ possesses subsets of solutions which constitute a {\em quasi-integrable theory}. Those are the facts that we want to check with our numerical simulations which we now explain. 

\section{The numerical simulations}
\label{sec:numerical}
\setcounter{equation}{0}

Our numerical simulations were performed using the 4th order Runge-Kutta method of simulating time evolution.
As in \cite{recent} we experimented with various grid sizes and numbers of points and most simulations were performed on lattices of 10001 lattice
points with lattice spacing of 0.01  (so they covered the region of (-50.0,\,50.0). Time step $dt$ was 0.0001. 

The breather-like structures were placed at $x\sim 0$ and stretched up $\pm 20.00$ from their positions hence at the
edges of the grid the fields resembled the vacuum configurations which were modified only by waves that were emitted during the scattering. 

At the edges of the grid ({\it i.e.} for $49.50<\vert x\vert <50.00$) we absorbed the waves reaching this region (by decreasing 
the time change of the magnitude of the field there). 

In consequence, the total energy was not conserved but the only energy which was absorbed was the energy of radiation waves.
Hence the total remaining energy was  effectively the energy of the field configuration which we wanted to study.

To start our simulations we took a breather configuration for the sine-Gordon model \rf{breathersol} and then perfomed the change
of variables \rf{three.one} to obtain the corresponding $\phi$ field. We then used  this field  and its derivative at $t=0$
as the initial conditions for the simulations.

The exact breather solution of the sine-Gordon model \rf{one.onea} is given by \cite{massformula} 
\be
\psi=  2\,{\rm ArcTan}\left[\frac{\sqrt{1-\nu^2}}{\nu}\frac{\,\sin
  \Gamma_I}{\cosh \Gamma_R}\right],
  \lab{breathersol}
\ee
where $v$ is the speed of the breather, $\nu$  is its frequency ($-1 <\nu < 1$) and  
\be
\Gamma_R = \sqrt{1-\nu^2}\;\frac{\(x-v\,t\)}{\sqrt{1-v^2}} ,\qquad\qquad 
\Gamma_I = \nu\;\frac{\(t-v\,x\)}{\sqrt{1-v^2}} .
\ee
In all our simulations we looked at the time dependence of breather-like field configurations 
initially at rest, {\it i.e.} with $v=0$. Therefore, the initial configuration of the breather at $t=0$, with $v=0$ is
\be
\psi\vert_{t=0}= 0 \qquad\qquad \qquad\qquad\frac{d\psi}{dt}\mid_{t=0}=\frac{2\,\sqrt{1-\nu^2}}{\cosh\(\sqrt{1-\nu^2}\; x\)}
\lab{initialpsi}
\ee
The input for our program is the initial configuration of the $\phi$-field defined by the transformation \rf{three.one}, and so
\be
\phi\mid_{t=0}= 0 \qquad\qquad \qquad
\frac{d\phi}{dt}\vert_{t=0}=\left[\frac{d\phi}{d\psi}\frac{d\psi}{dt}\right]\vert_{t=0}=
 \frac{1}{\sqrt{1+\varepsilon\,\pi\left(\pi-2\,\gamma\right)}}\;\frac{2\,\sqrt{1-\nu^2}}{\cosh\(\sqrt{1-\nu^2}\; x\)}
 \lab{initialphi}
\ee
From \rf{model} we see that the initial energy of the breather-like configuration of the model was
\be
E= 2\,\int_{-\infty}^{\infty} dx\, \left[\frac{1}{2}\left[ \(\partial_t \phi\)^2+\(\partial_x \phi\)^2\right] + V\(\phi\)
\right]
\lab{energy}
\ee
The factor $2$ in front of the integral was put to match the definition of the energy in the numerical code. 
For the initial configuration \rf{initialphi} we have $\frac{d\phi}{dx}\mid_{t=0}=0$,  and  $V\(\phi=0\)=0$ (see  \rf{v3def}), and so the initial energy was
\be
E = \frac{8\, \sqrt{1-\nu^2}}{\left[1+\varepsilon\,\pi\left(\pi-2\,\gamma\right)\right]}
\lab{niceenergy}
\ee
Note that the energy of the initial configuration has the same  $\nu $-dependence as the sine-Gordon breather, but it is re-scaled by the factor $1/c^2$, with $c$ given by  \rf{three.two}, and so it decreases with the increase of the deformation parameter $\ve$. This rescaling factor and its decrease with $\ve$  has an interesting effect as we will demonstrate in the discussions of the simulations.

We have performed several simulations for different values of the frequency of the breather {\it i.e.} $\nu$ in \rf{breathersol} and for various values of $\varepsilon$ and $\gamma$. For some of these simulations we have also calculated the anomaly of the first non-trivial quasi-conserved charge given in \rf{chargeanomalydef}, namely $\alpha^{(3)}$ and $Q^{(3)}$. We have chosen in the Lax potentials  
\rf{potentials} the parameters as $\omega=2$ and $m=1$. The reason for this choice was that those are the values that make the sine-Gordon potential  \rf{anomalyfreesg}, for which the anomaly \rf{xdef} vanishes, equal to \rf{sgpot} when $\ve$ is set to zero. Then from \rf{quasiconserv} we find  that
\be
\gamma^{(3)}= i\,2\, \partial_{-}^2 \phi 
\ee
 Thus, using \rf{chargeanomalydef} and \rf{xdef}, we see that
\be
\alpha^{(3)}=-2 \, \int_{-\infty}^{\infty} dx\, \partial_{-} \phi\,\partial_{-}^2 \phi\,\left[\frac{d^2\,V}{d\,\phi^2}+4\, V-1\right],
\lab{alpha3}
\ee
 We have also computed the so-called integrated anomaly given by (see \rf{timedercharges}) 
\be
\beta^{(3)}= -\frac{1}{2}\,\int_{t_0}^t dt^{\prime} \, \alpha^{(3)} = Q^{(3)}\(t\)- Q^{(3)}\(t_0\),
\lab{beta3}
\ee
where $t_0$ is the initial time of the simulation, usually taken to be zero. 

In the table below  we summarise the main features of the simulations we had performed:

\vspace{.5cm}

\begin{tabular}{|c|c|c|c|c|c|}
\hline
Figure & $\ve$ & $\nu$ & $\gamma$ & Initial Energy (eq.\rf{niceenergy})& Breather behaviour\\
\hline
\hline
Fig. \ref{fig:700} & $0.01$ & $0.1$ & $0$ & 7.24 & short lived\\
\hline
Fig. \ref{fig:800} & $0.01$ & $0.1$ & $0.3$ & 7.37 & short lived\\
\hline
Fig. \ref{fig:0010500} & $0.01$ & $0.5$ & $0$ & 6.31 & short lived\\
\hline
Fig. \ref{fig:0010502} & $0.01$ & $0.5$ & $0.2$ & 6.38   & short lived\\
\hline
Fig. \ref{fig:0010505} & $0.01$ & $0.5$ & $0.5$ & 6.49  & short lived\\
\hline
Fig. \ref{fig:600} & $0.01$ & $0.95$ & $0$ & 2.27 & long lived\\
\hline
Fig. \ref{fig:900} & $0.01$ & $0.95$ & $0.3$ &  2.31 & long lived\\
\hline
Fig. \ref{fig:150} & $0.01$ & $0.95$ & $0.5$ &  2.34 & long lived\\
\hline
Fig. \ref{fig:350} & $0.01$ & $0.95$ & $0.7$ &  2.37 & long lived\\
\hline
\hline
Fig. \ref{fig:02050} & $0.2$ & $0.5$ & $0$ & 2.33  & long lived\\
\hline
\hline
Fig. \ref{fig:200} & $0.3$ & $0.1$ & $0$ & 2.01  & long lived\\
\hline
Fig. \ref{fig:fig2} & $0.3$ & $0.5$ & $0$ &  1.75 & long lived\\
\hline
Fig. \ref{fig:4} & $0.3$ & $0.5$ & $0.2$ & 1.93  & reasonably long lived\\
\hline
Fig. \ref{fig:fig5} & $0.3$ & $0.5$ & $0.5$ &  2.29 & short lived\\
\hline
Fig. \ref{fig:300} & $0.3$ & $0.9$ & $0$ &  0.88 & long lived\\
\hline
\end{tabular}

\vspace{.5cm}

\begin{figure}
  \centering
  \subfigure[]{\label{fig:energy700a}\includegraphics[trim = 0cm 0cm 1.8cm 1.8cm, width=0.45\textwidth]{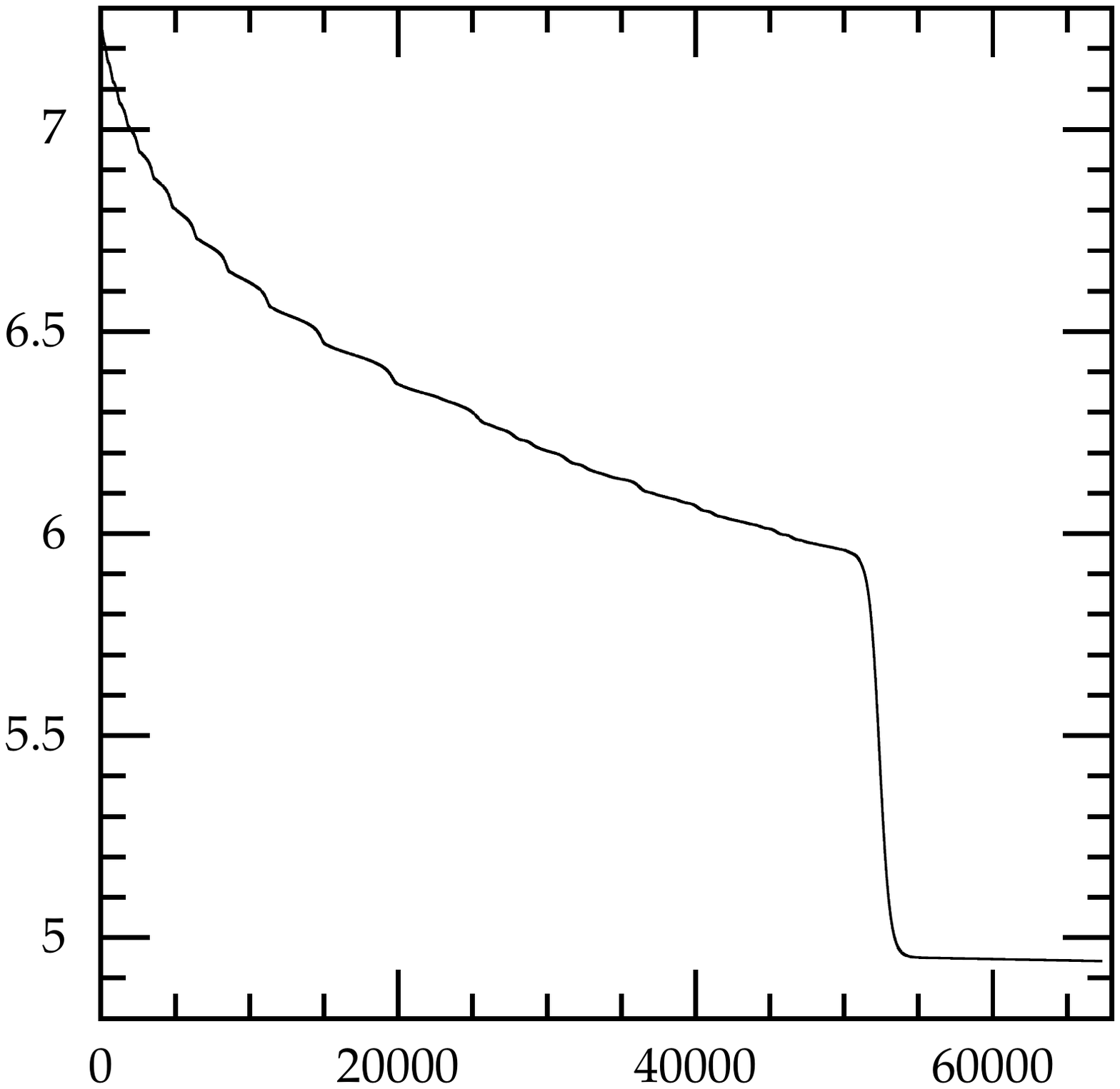}} 
    \subfigure[]{\label{fig:energy700b}\includegraphics[trim = 0cm 0cm 1.8cm 1.8cm, width=0.45\textwidth]{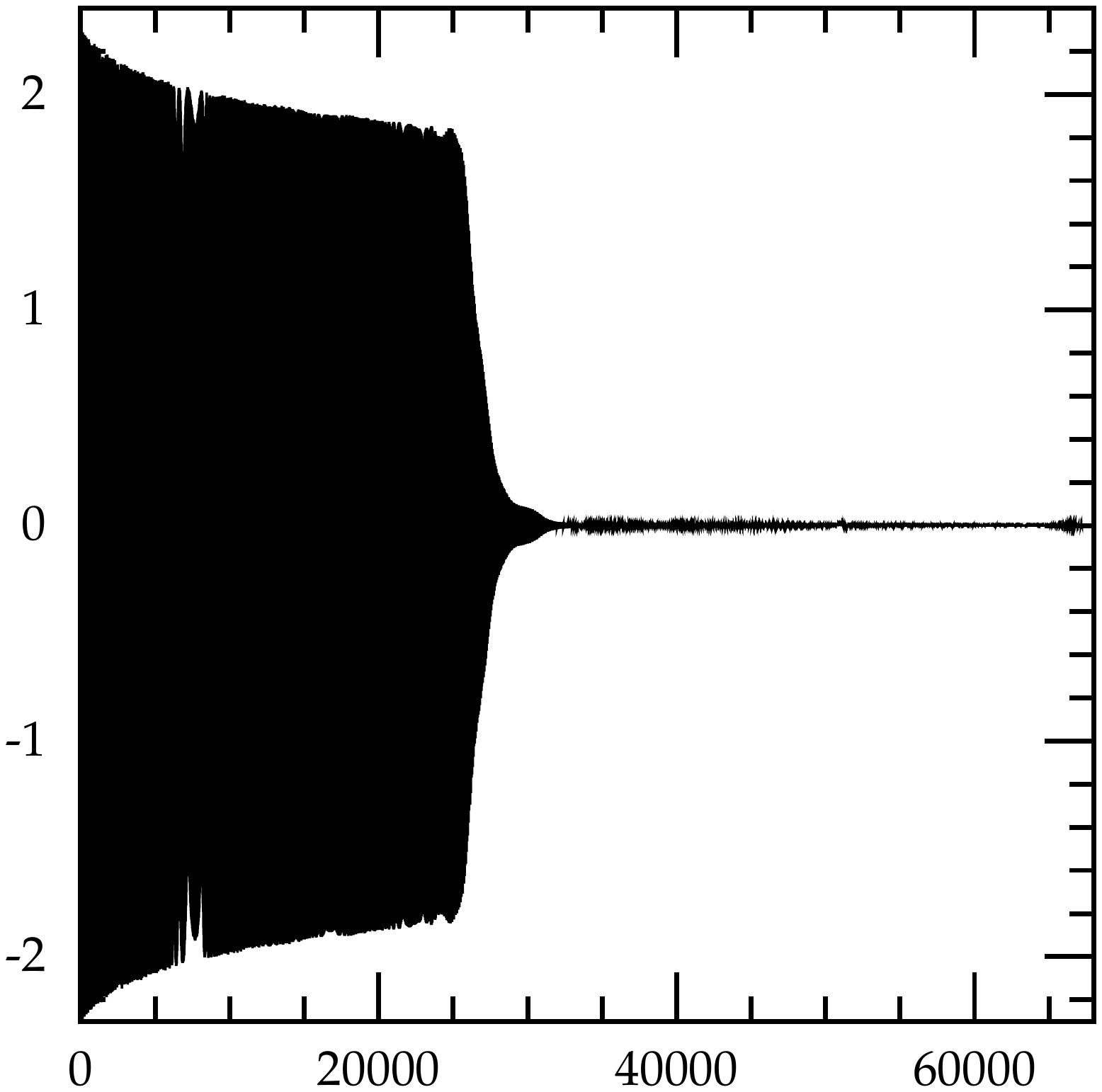}}                               
  \subfigure[]{\label{fig:anomaly700a}\includegraphics[trim = 0cm 0cm 1.8cm 1.8cm,  width=0.45\textwidth]{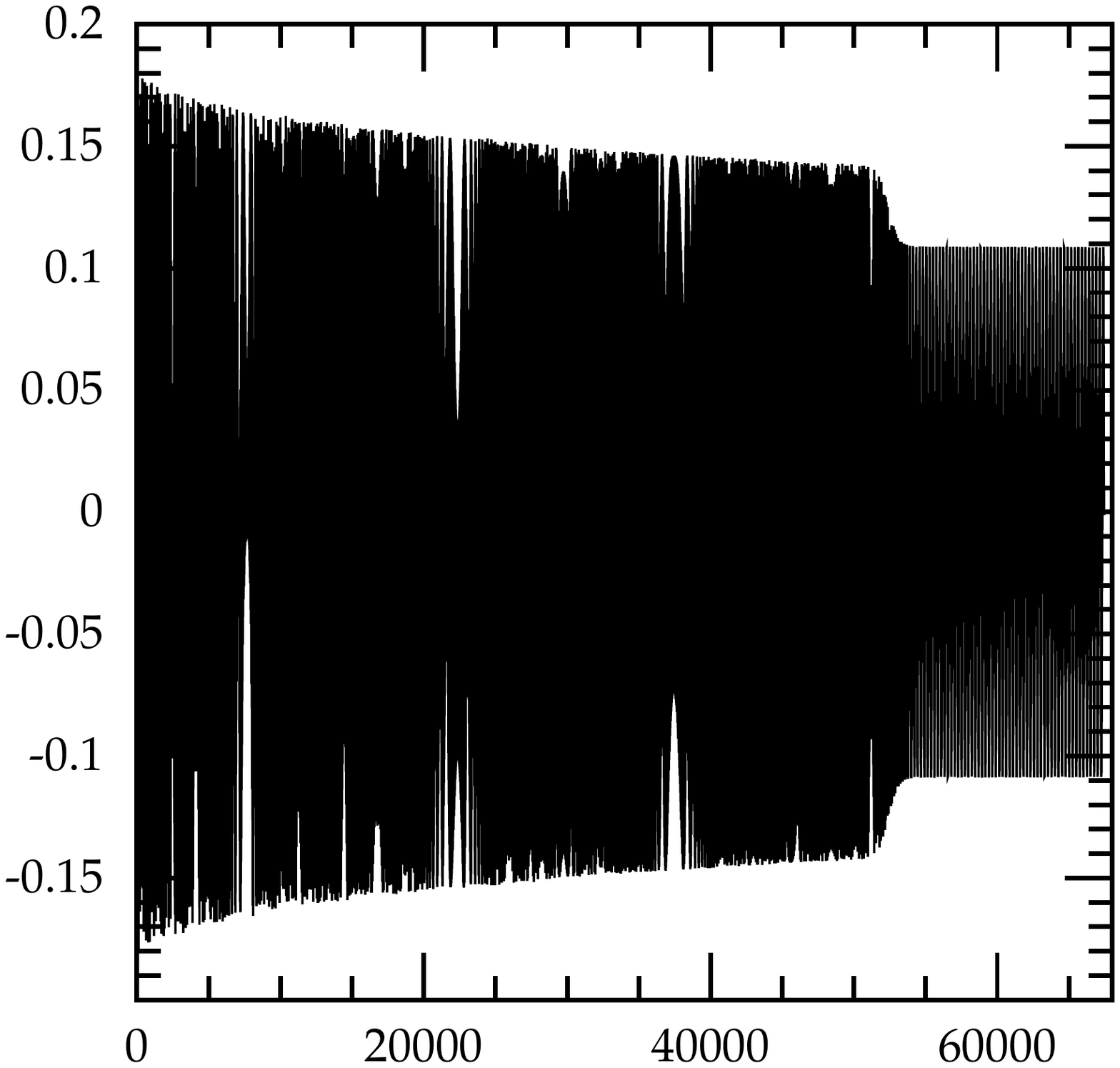}}
  \subfigure[]{\label{fig:anomaly700b}\includegraphics[trim = 0cm 0cm 1.8cm 1.8cm,  width=0.45\textwidth]{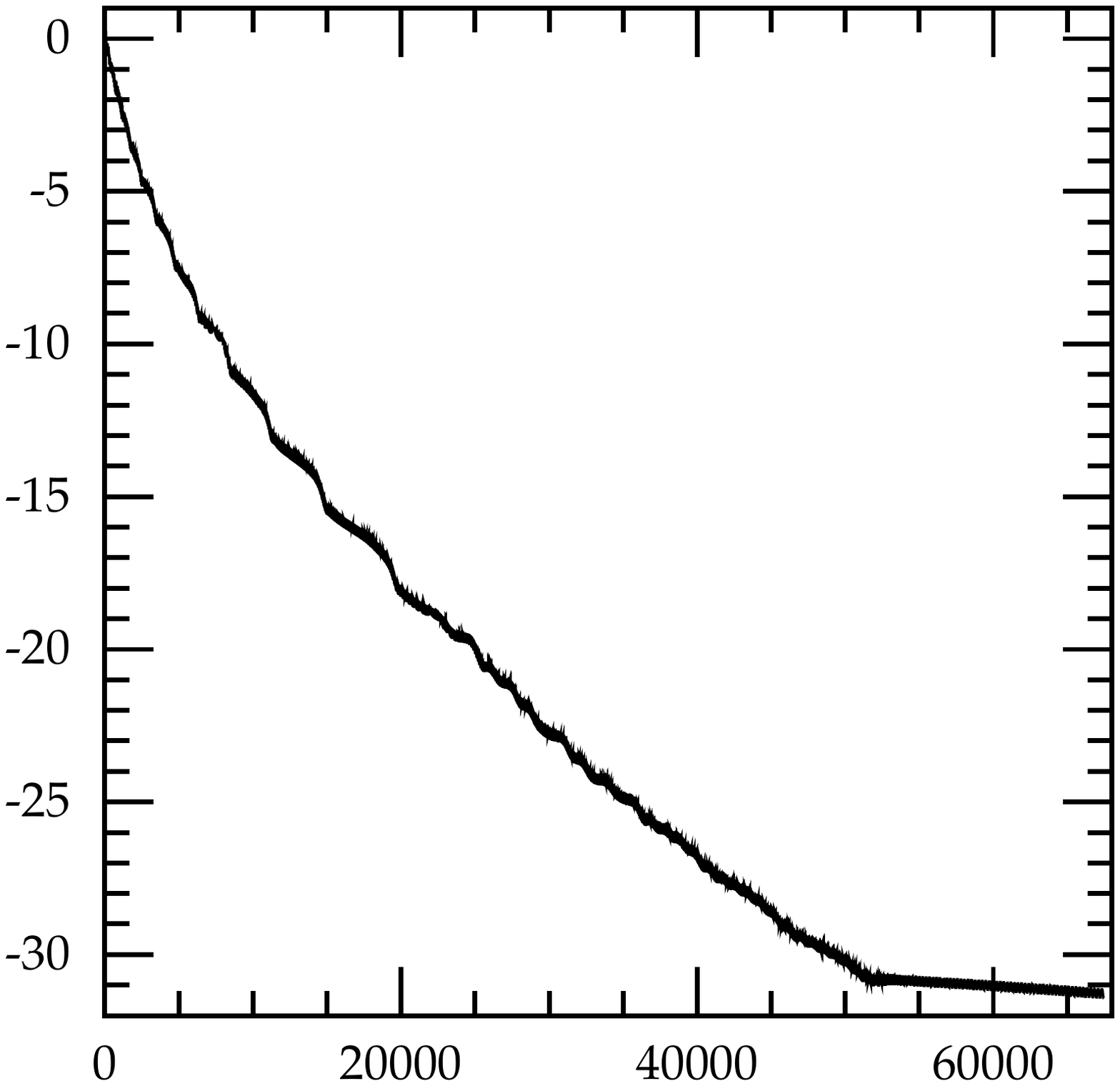}}
  \caption{Breather's simulation in the theory \rf{model} with initial configuration \rf{initialphi},  and with parameters given by $\varepsilon=0.01$, $\nu=0.1$ and $\gamma=0$. The plots show the time dependence of (adimensional units): 
   (a) the energy \rf{energy};  (b) the field $\phi$ at position $x=0$ in the grid; (c) the anomaly \rf{alpha3} and (d) the integrated anomaly \rf{beta3}. }
  \label{fig:700}
\end{figure}

\begin{figure}
  \centering
  \subfigure[]{\label{fig:energy800a}\includegraphics[trim = 0cm 0cm 1.8cm 1.8cm, width=0.45\textwidth]{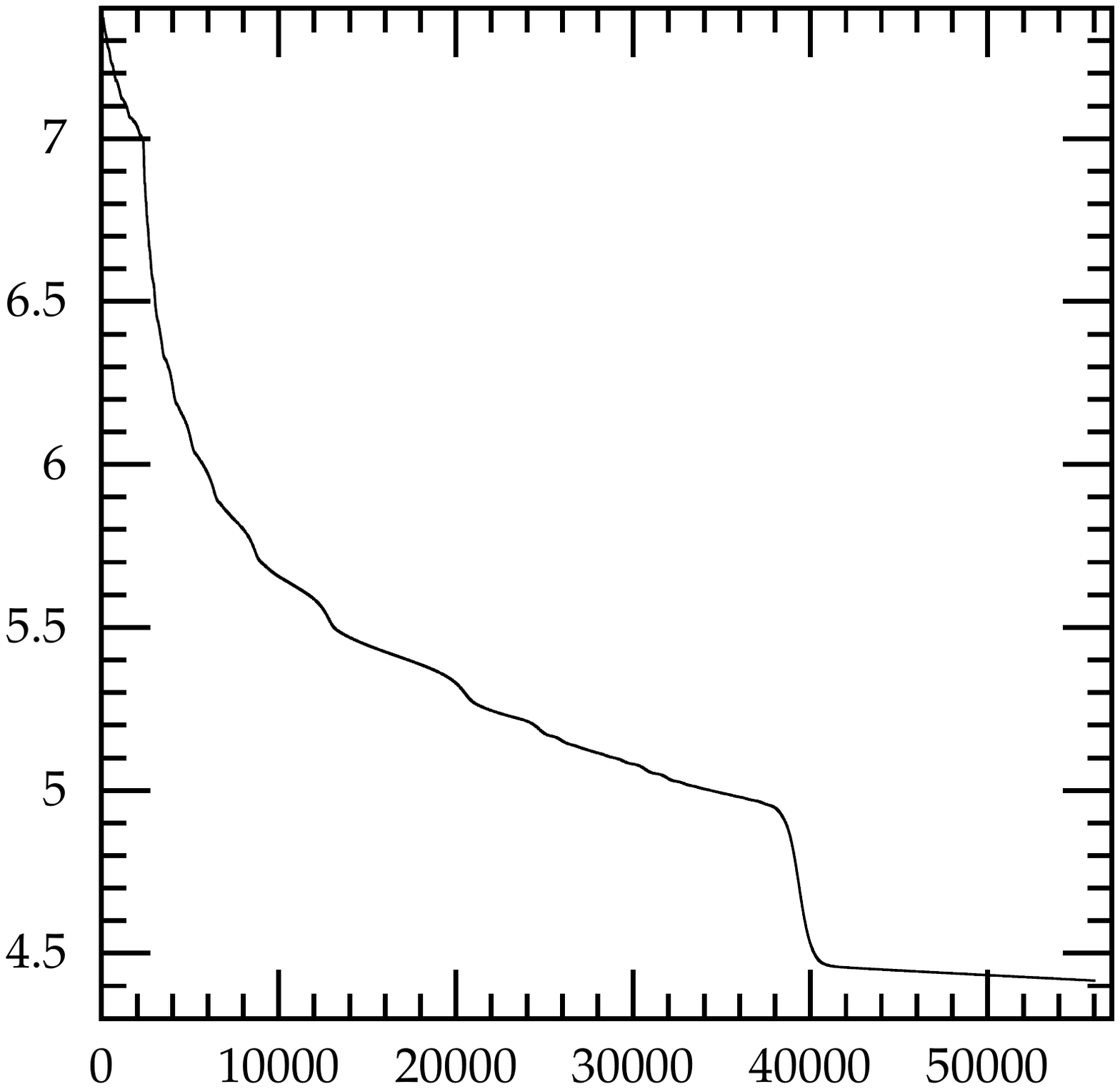}}  
  \subfigure[]{\label{fig:energy800b}\includegraphics[trim = 0cm 0cm 1.8cm 1.8cm, width=0.45\textwidth]{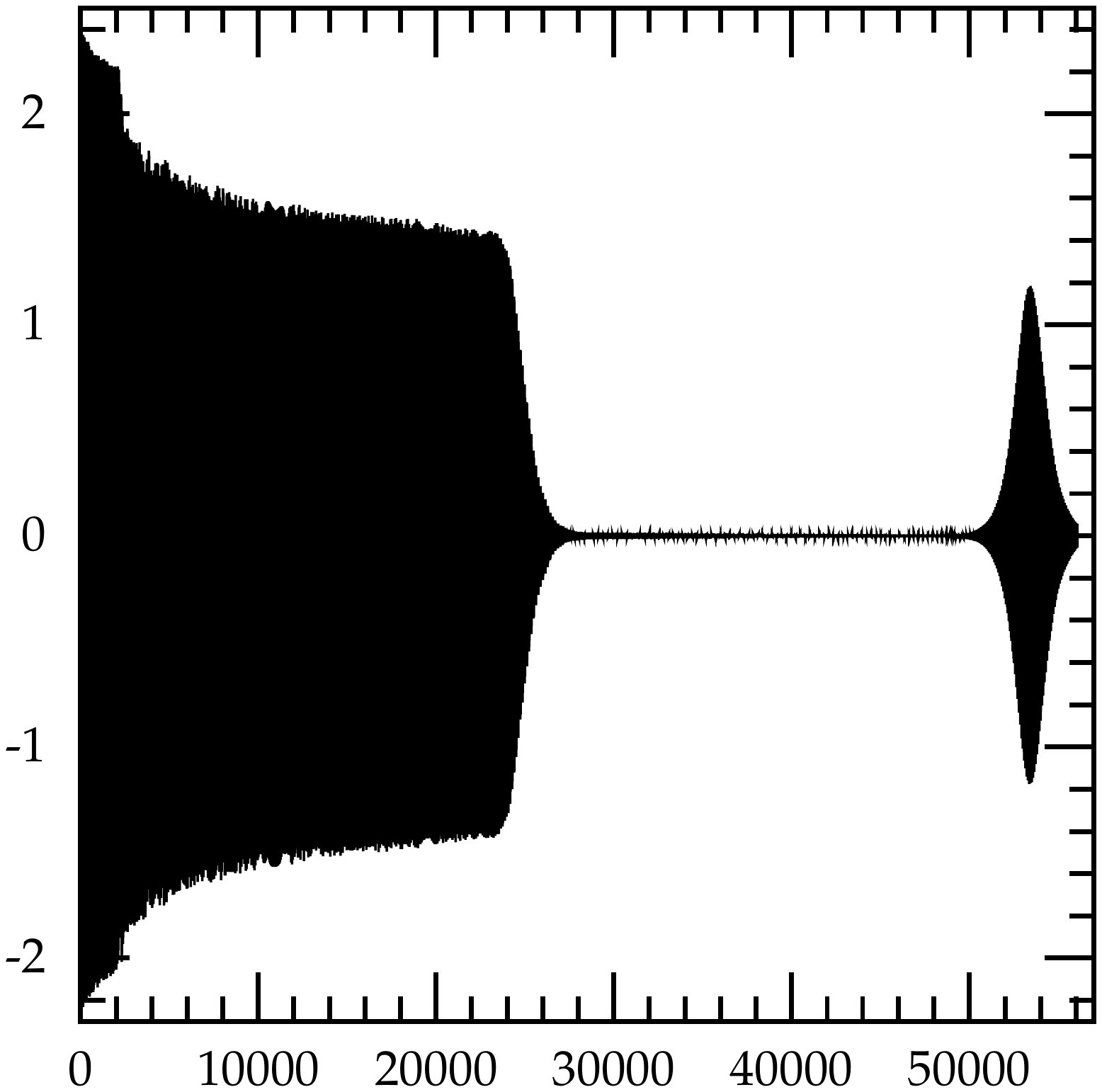}}                
  \subfigure[]{\label{fig:anomaly800a}\includegraphics[trim = 0cm 0cm 1.8cm 1.8cm,  width=0.45\textwidth]{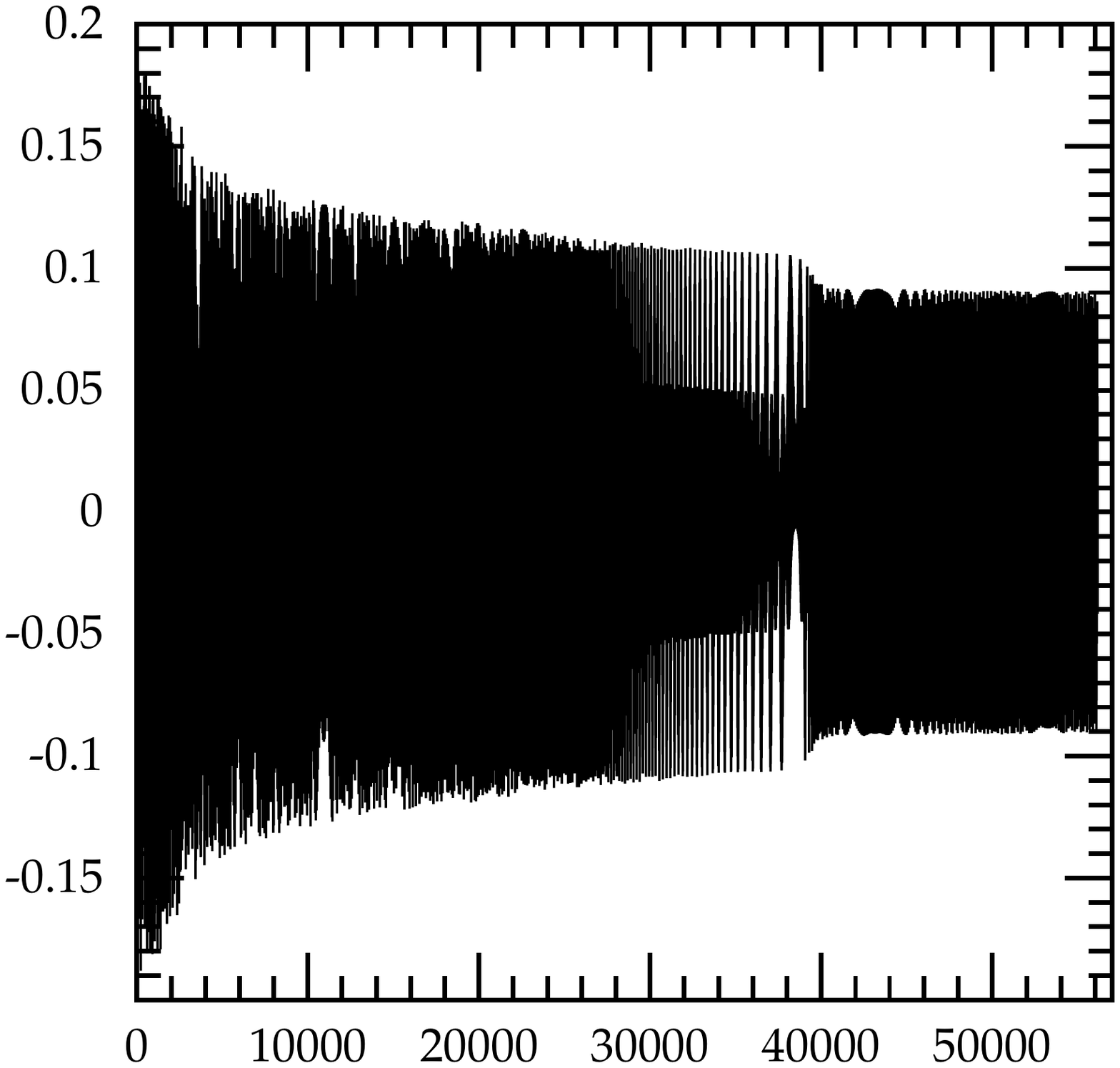}}
   \subfigure[]{\label{fig:anomaly800b}\includegraphics[trim = 0cm 0cm 1.8cm 1.8cm,  width=0.45\textwidth]{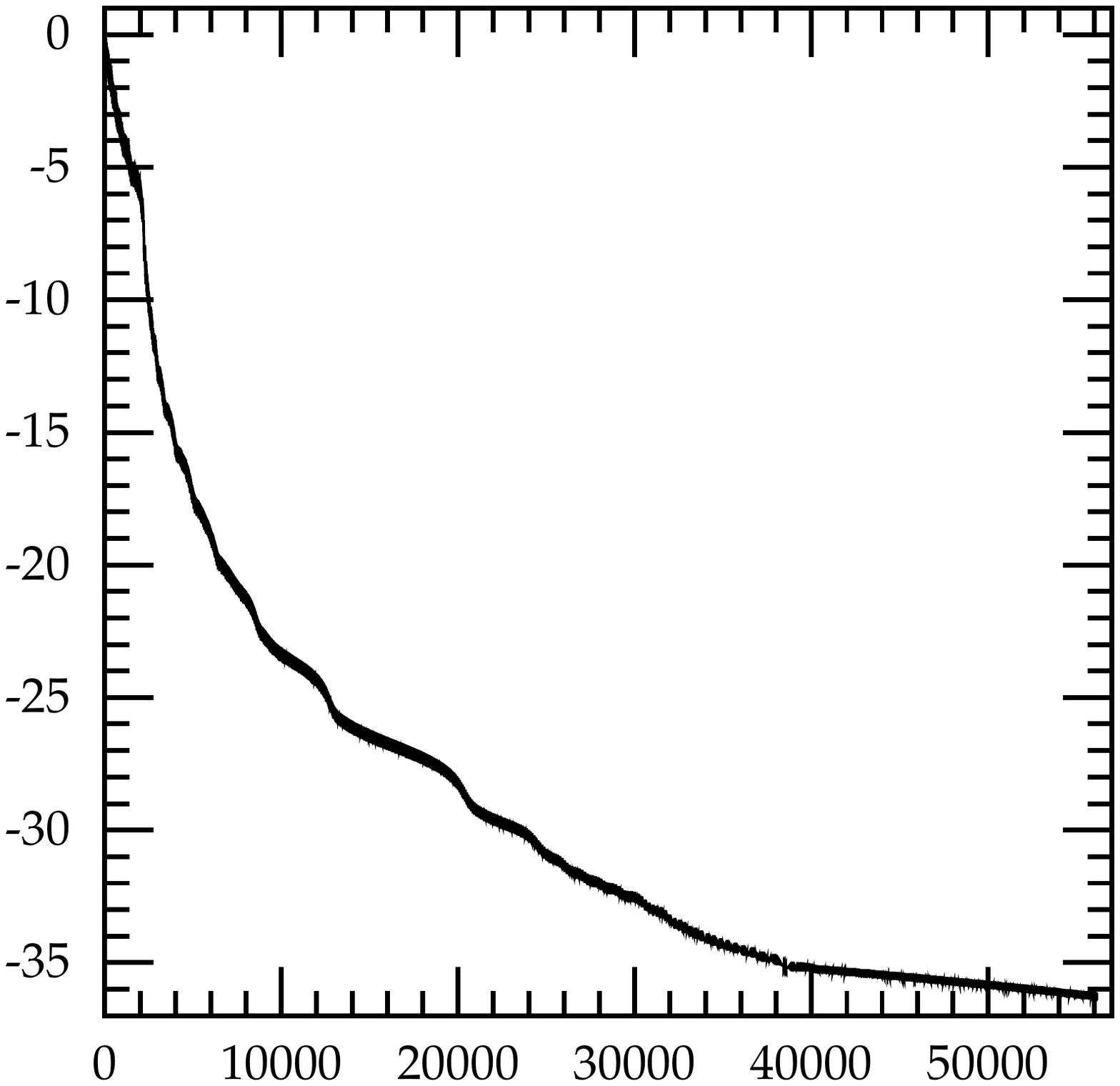}}
    \caption{Breather's simulation in the theory \rf{model} with initial configuration \rf{initialphi},  and  with parameters given by $\varepsilon=0.01$, $\nu=0.1$ and $\gamma=0.3$. The plots show the time dependence of (adimensional units): 
   (a) the energy \rf{energy};  (b) the field $\phi$ at position $x=0$ in the grid; (c) the anomaly \rf{alpha3} and (d) the integrated anomaly \rf{beta3}. }
  \label{fig:800}
\end{figure}

\begin{figure}
  \centering
  \subfigure[]{\label{fig:pot1a}\includegraphics[trim = 0cm 0cm 1.5cm 1.5cm, width=0.30\textwidth]{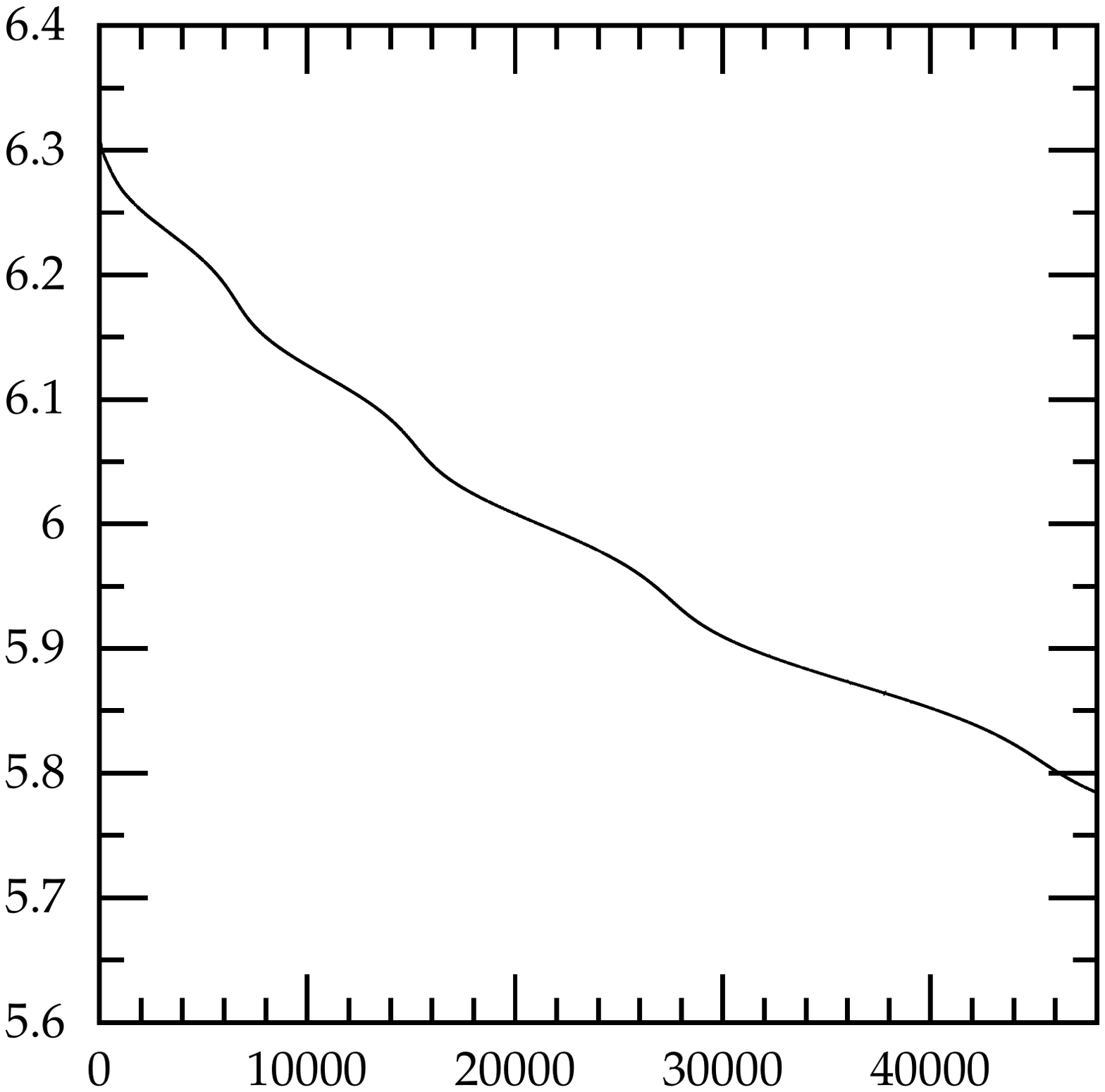}}                
  \subfigure[]{\label{fig:pot2a}\includegraphics[trim = 0cm 0cm 1.5cm 1.5cm,  width=0.30\textwidth]{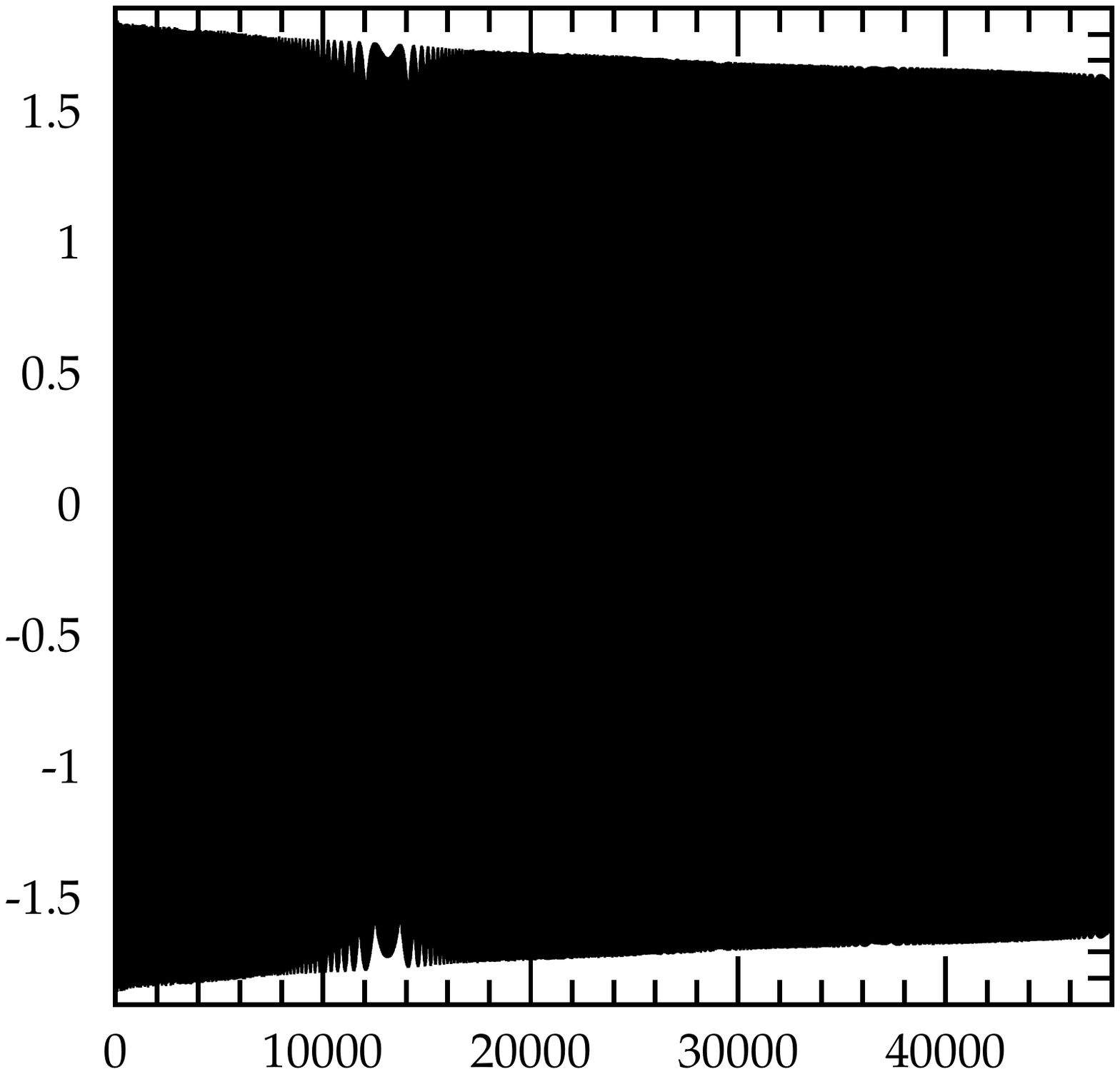}}
  \subfigure[]{\label{fig:pot1b}\includegraphics[trim = 0cm 0cm 1.5cm 1.5cm, width=0.30\textwidth]{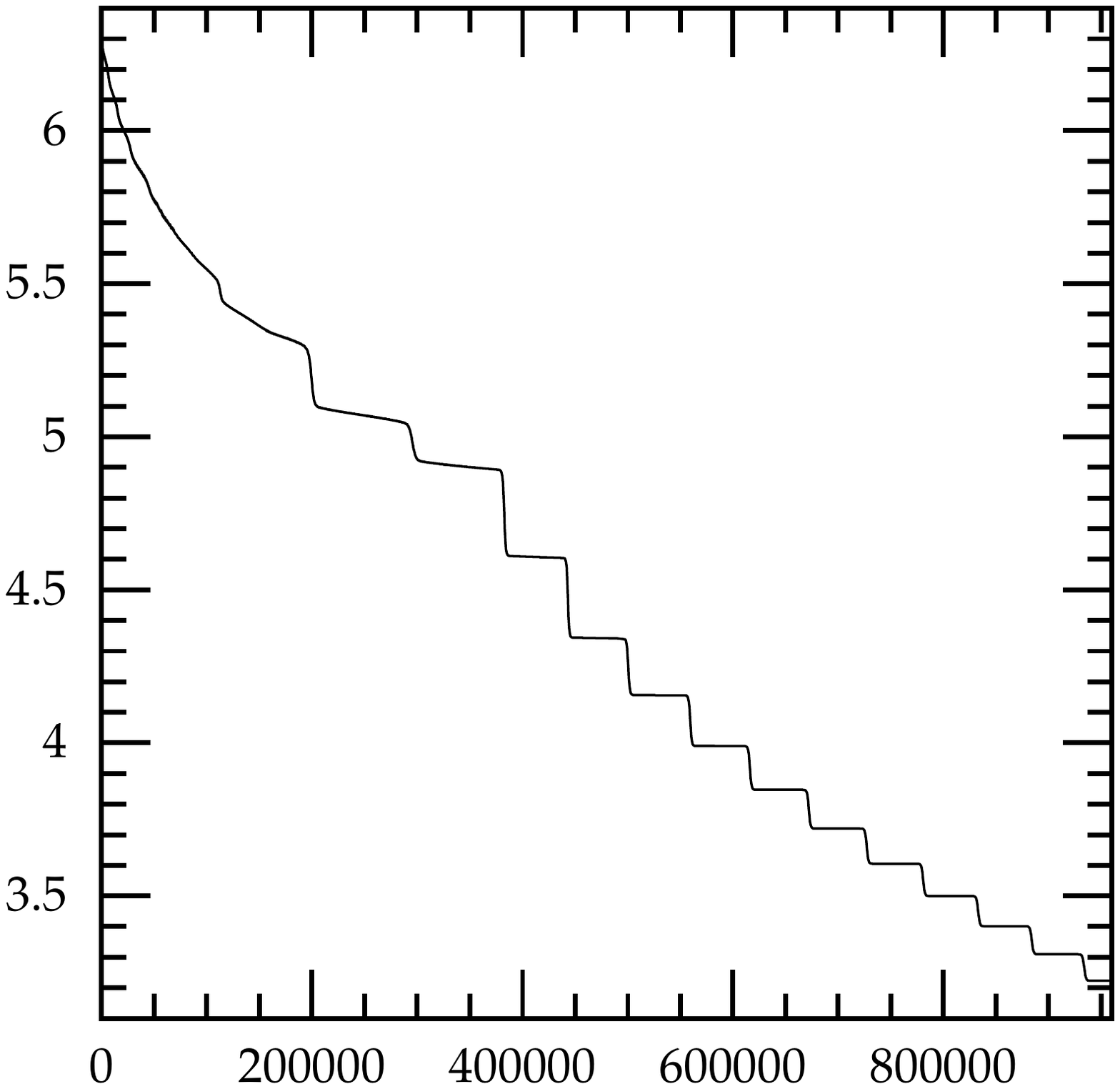}}                
  \subfigure[]{\label{fig:pot2b}\includegraphics[trim = 0cm 0cm 1.5cm 1.5cm,  width=0.30\textwidth]{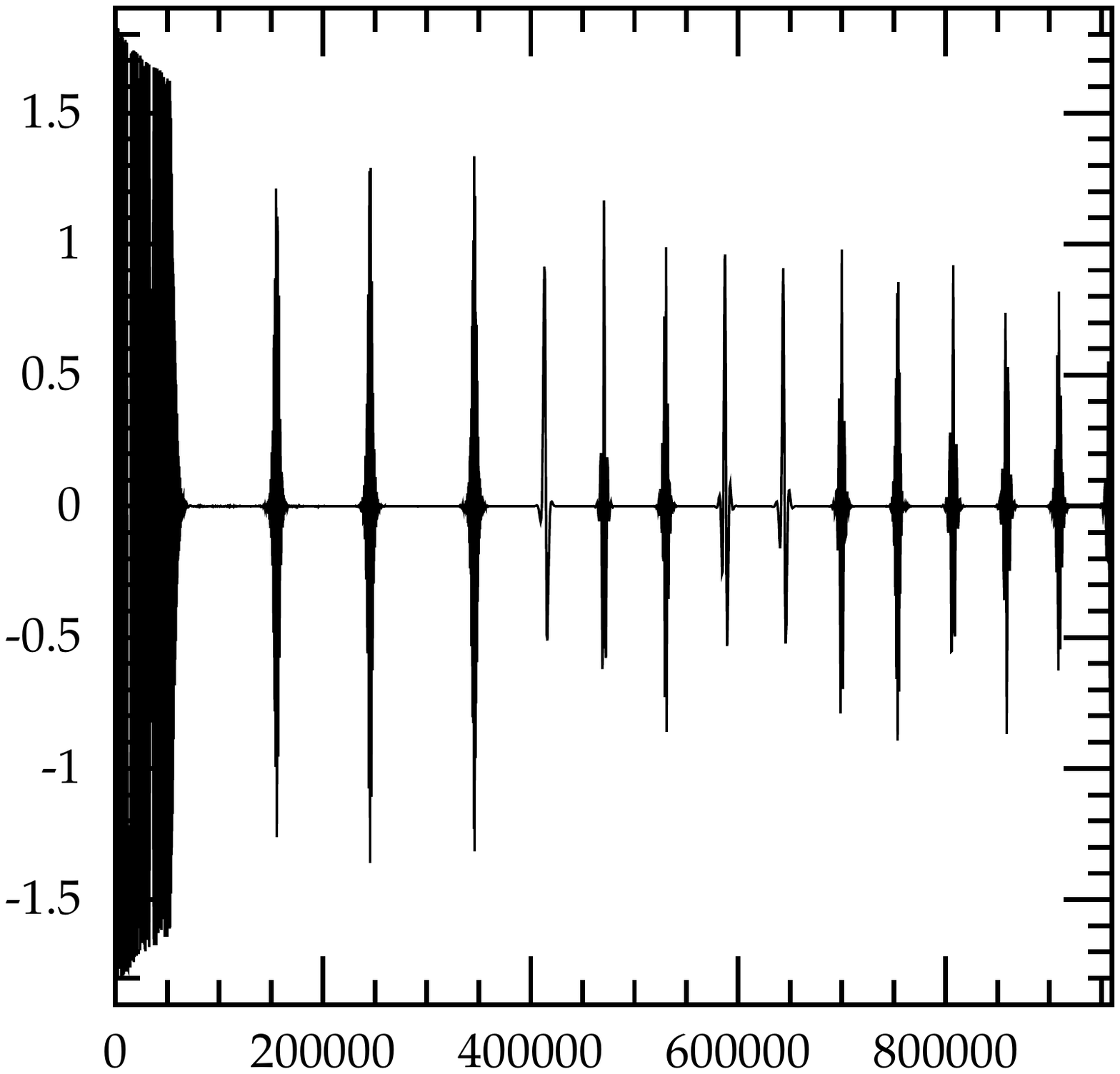}}
 \subfigure[]{\includegraphics[trim = 0cm 0cm 1.5cm 1.5cm,  width=0.30\textwidth]{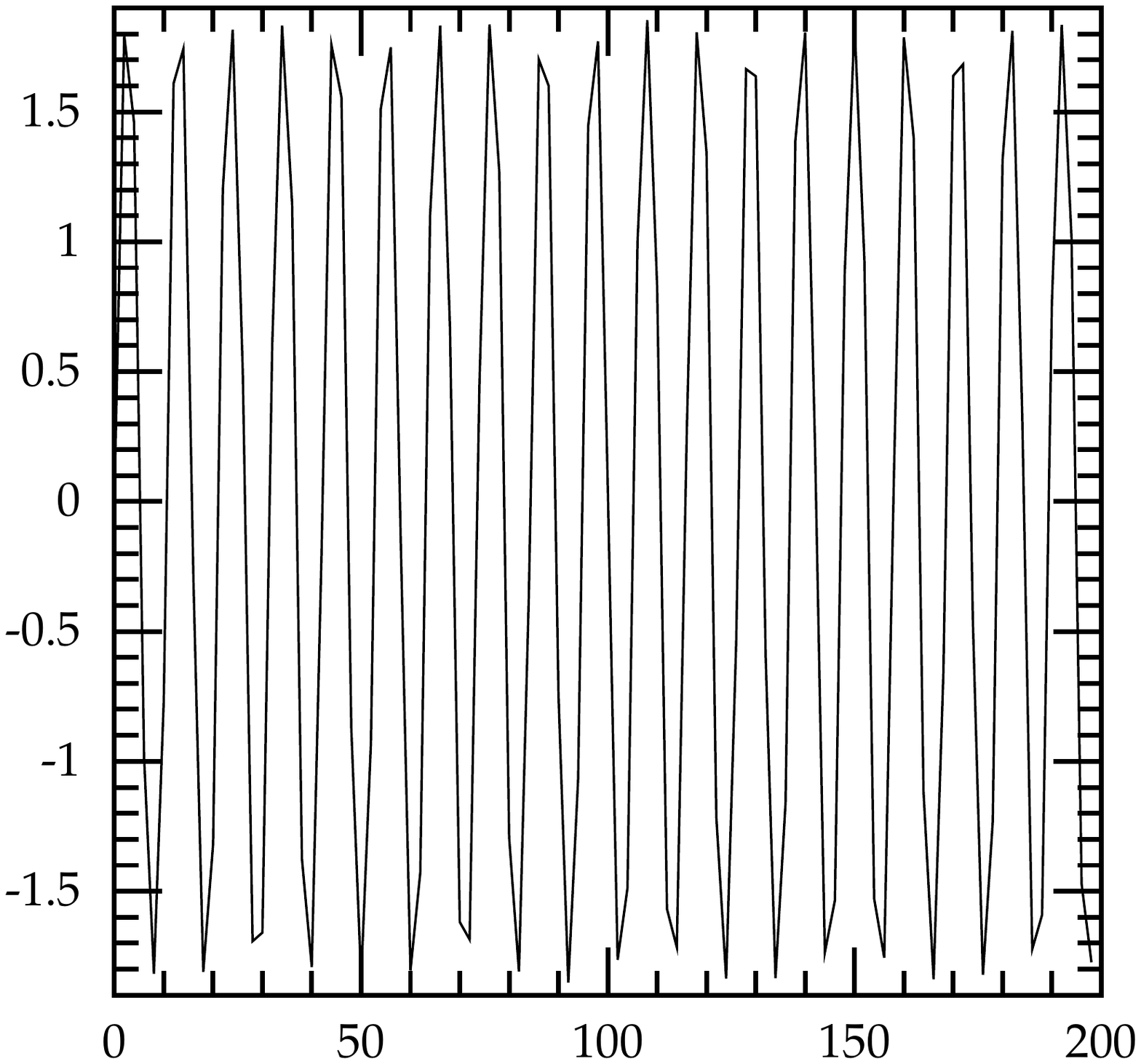}}
 \subfigure[]{\includegraphics[trim = 0cm 0cm 1.5cm 1.5cm,  width=0.30\textwidth]{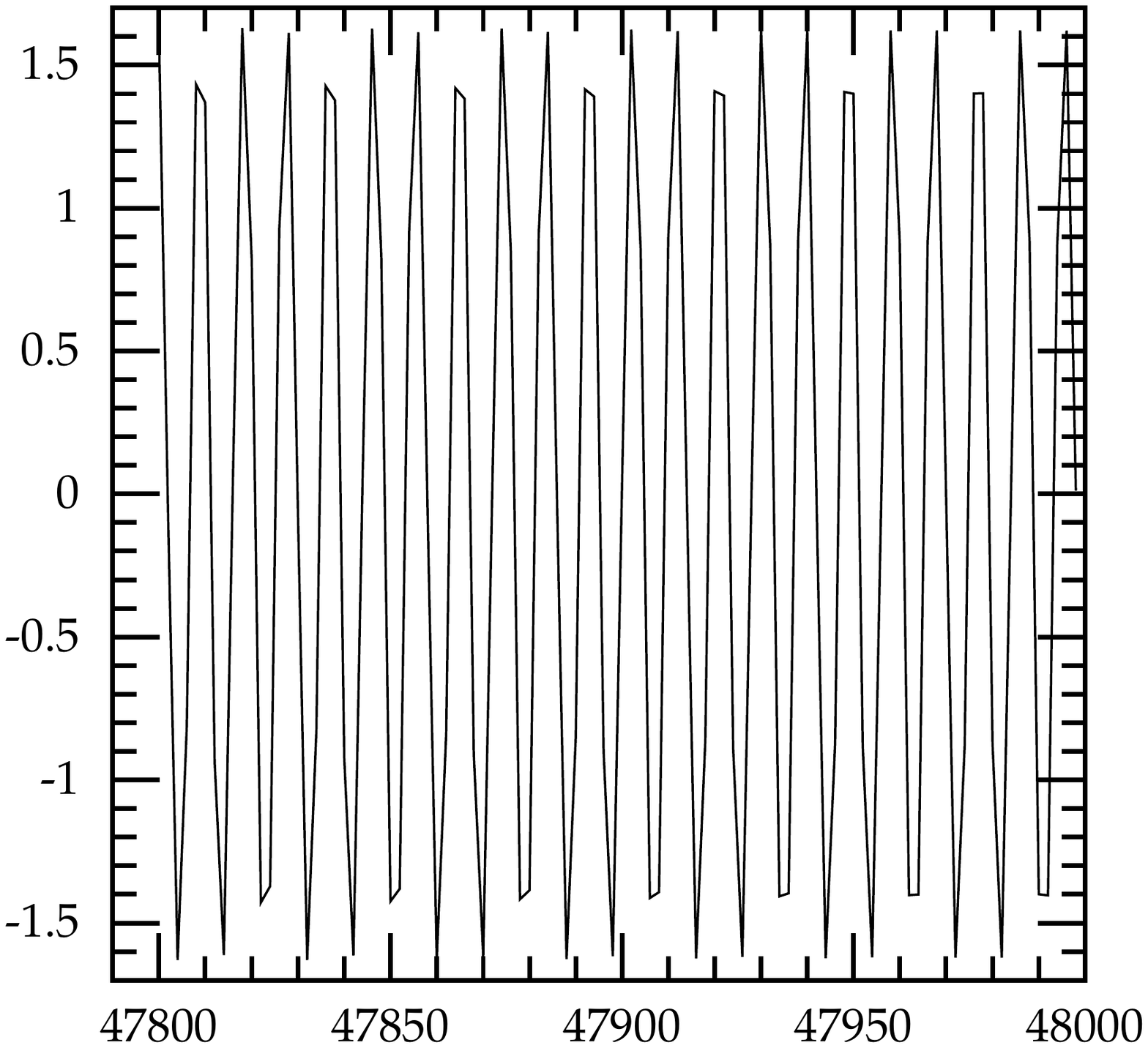}}
 \caption{Breather's simulation in the theory \rf{model} with initial configuration \rf{initialphi},  and  with parameters given by $\varepsilon=0.01$, $\nu=0.5$  and $\gamma=0.0$. The plots show the time dependence of (adimensional units): 
   (a) and (c) the energy \rf{energy};  (b) and (d) the field $\phi$ at position $x=0$ in the grid. The figures (a) and (c) correspond to the moment just before when the quasi-breather has started moving. Figures (e) and (f) show a blow-up of the amplitude of the field $\phi$ at the beginning of simulation (e) and at the time just before the soliton has started moving (f). }
  \label{fig:0010500}
\end{figure}

\begin{figure}
  \centering
  \subfigure[]{\includegraphics[trim = 0cm 0cm 1.8cm 1.8cm, width=0.45\textwidth]{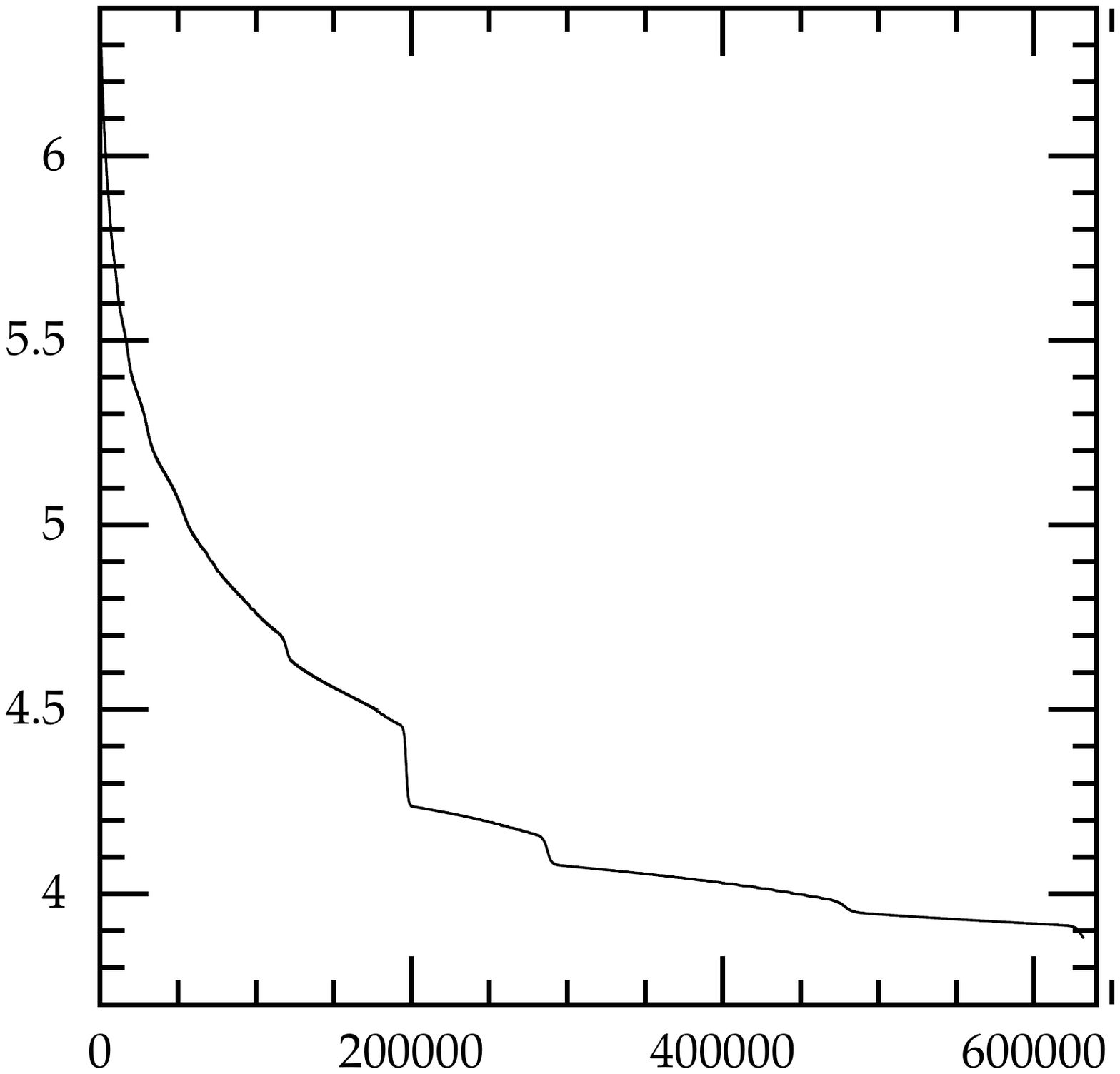}}                
  \subfigure[]{\includegraphics[trim = 0cm 0cm 1.8cm 1.8cm,  width=0.45\textwidth]{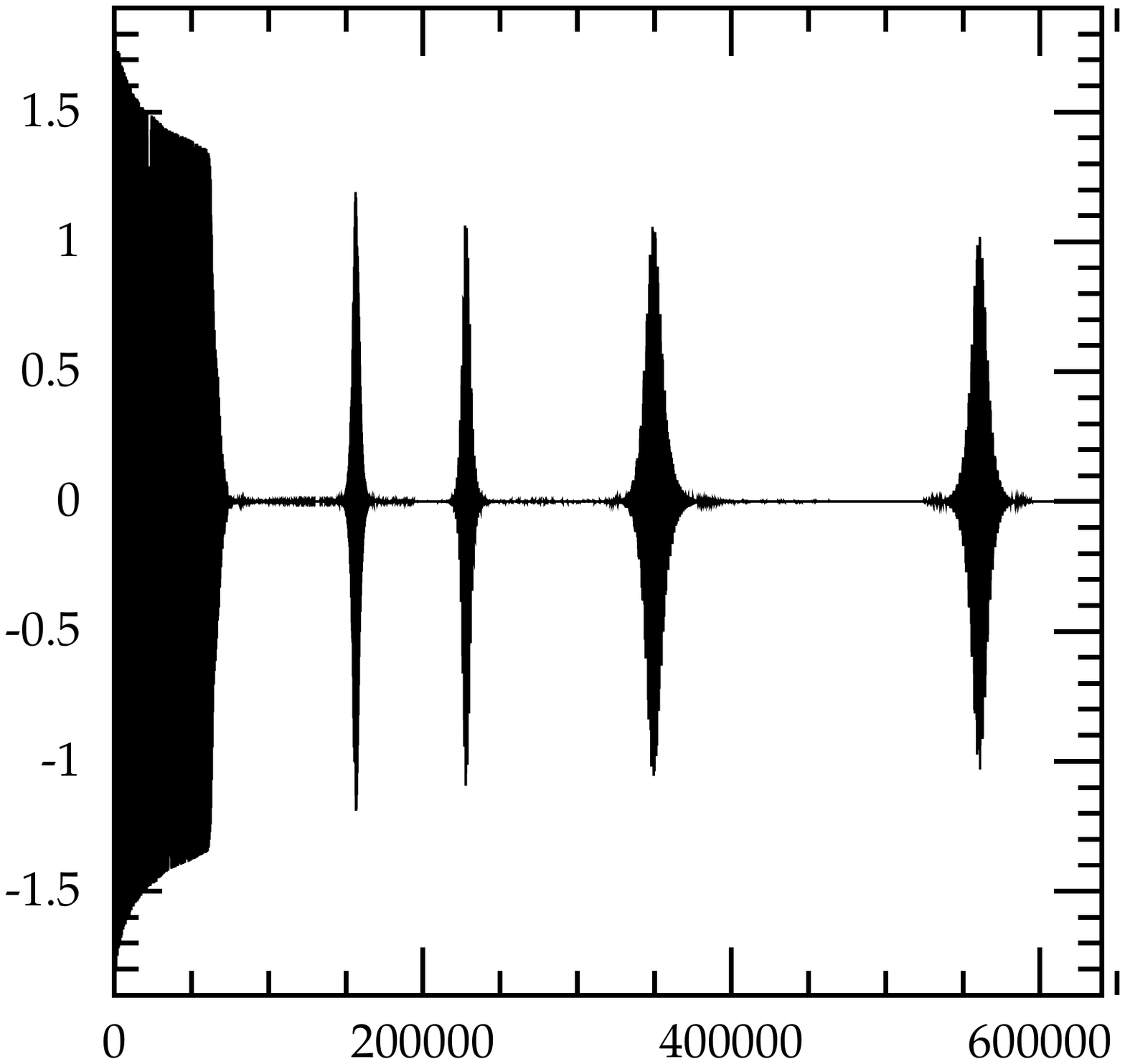}}
  \caption{Breather's simulation in the theory \rf{model} with initial configuration \rf{initialphi},  and with parameters given by $\varepsilon=0.01$, $\nu=0.5$ and $\gamma=0.2$. The plots show the time dependence of (adimensional units): 
   (a) the energy \rf{energy};  (b) the field $\phi$ at position $x=0$ in the grid.}
  \label{fig:0010502}
\end{figure}

\begin{figure}
  \centering
  \subfigure[]{\includegraphics[trim = 0cm 0cm 1.8cm 1.8cm, width=0.45\textwidth]{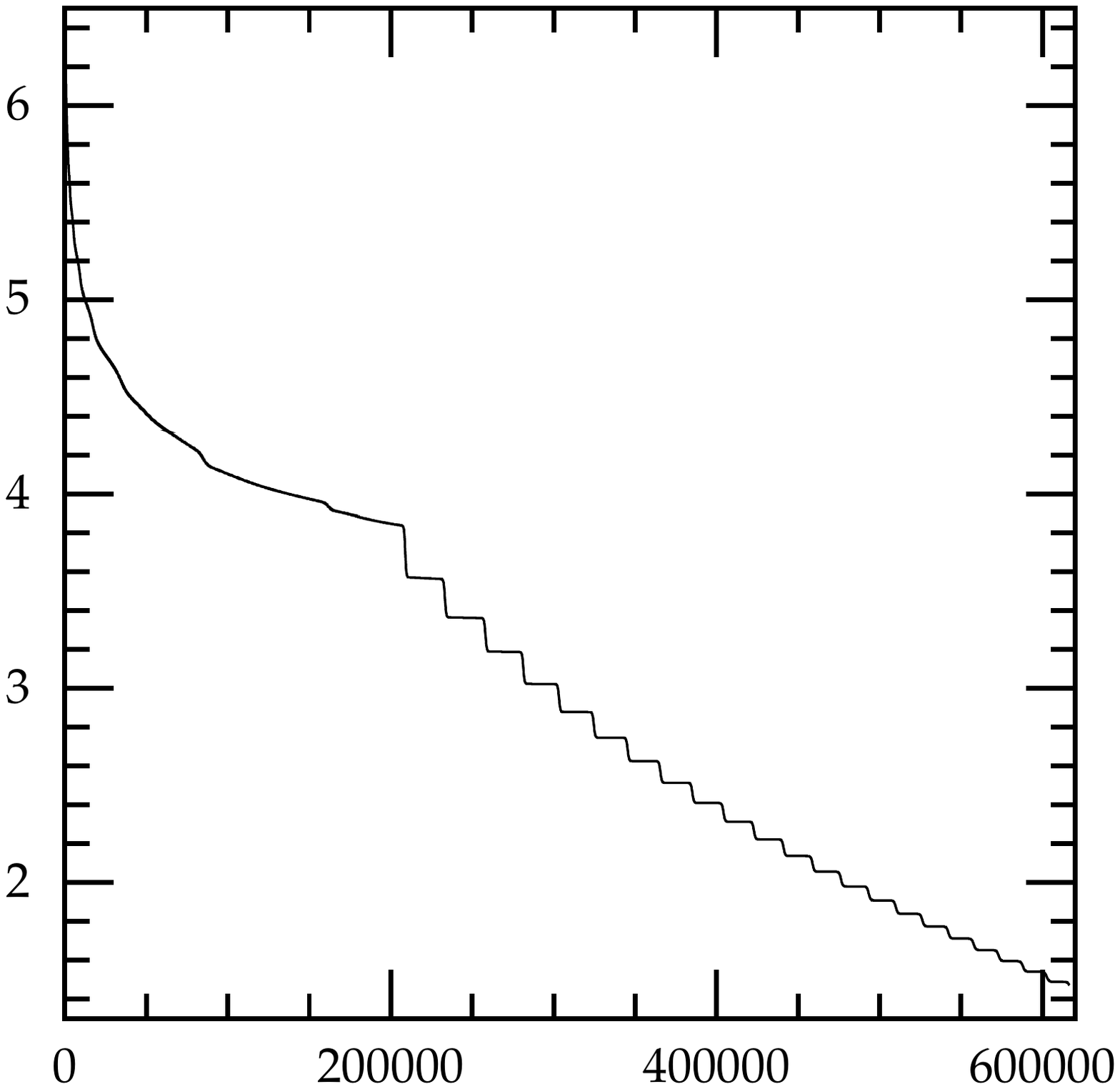}}                
  \subfigure[]{\includegraphics[trim = 0cm 0cm 1.8cm 1.8cm,  width=0.45\textwidth]{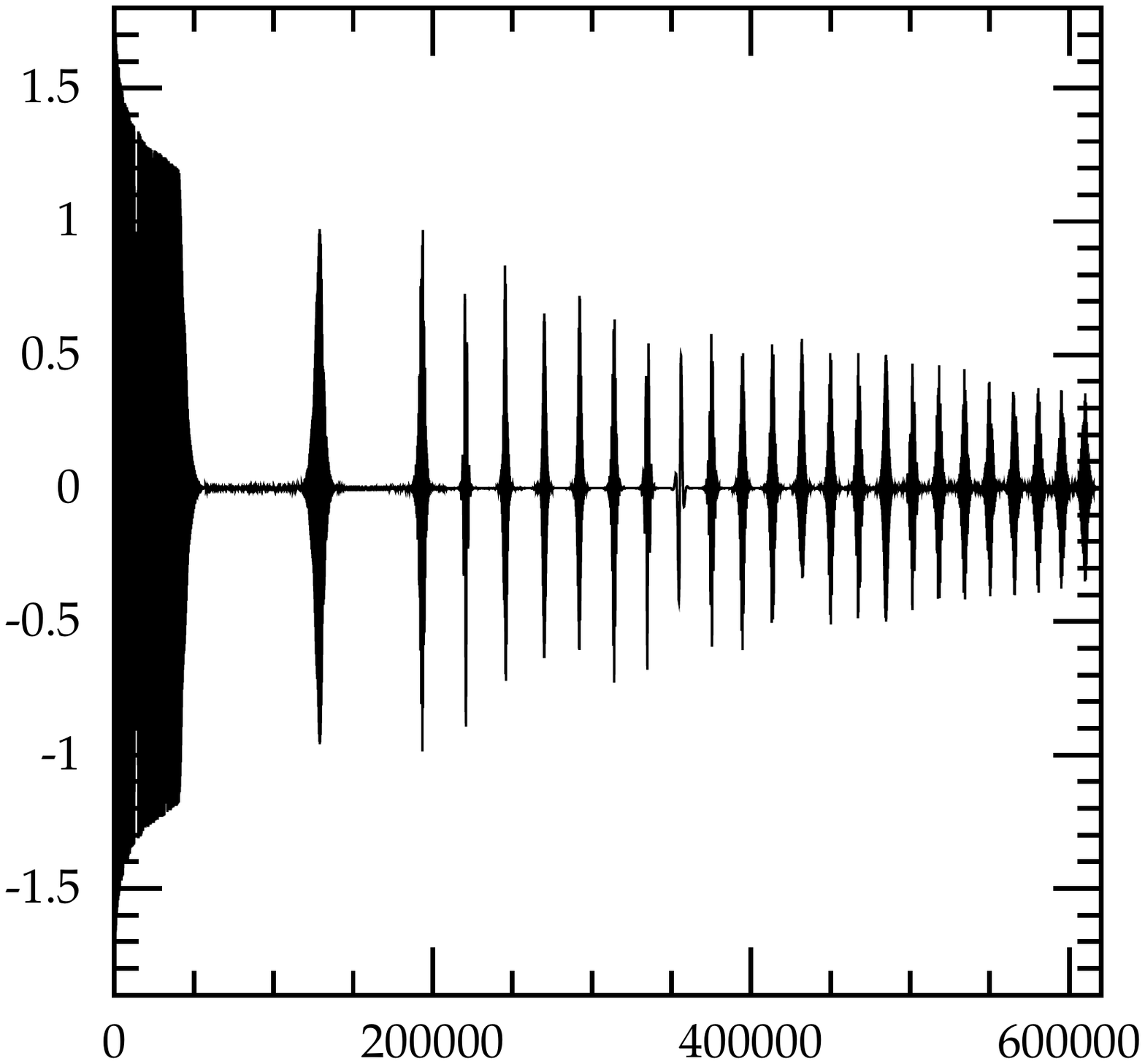}}
  \caption{Breather's simulation in the theory \rf{model} with initial configuration \rf{initialphi},  and with parameters given by $\varepsilon=0.01$, $\nu=0.5$ and $\gamma=0.5$. The plots show the time dependence of (adimensional units): 
   (a) the energy \rf{energy};  (b) the field $\phi$ at position $x=0$ in the grid.}
  \label{fig:0010505}
\end{figure}

\begin{figure}
  \centering
  \subfigure[]{\label{fig:energy600a}\includegraphics[trim = 0cm 0cm 1.8cm 1.8cm, width=0.45\textwidth]{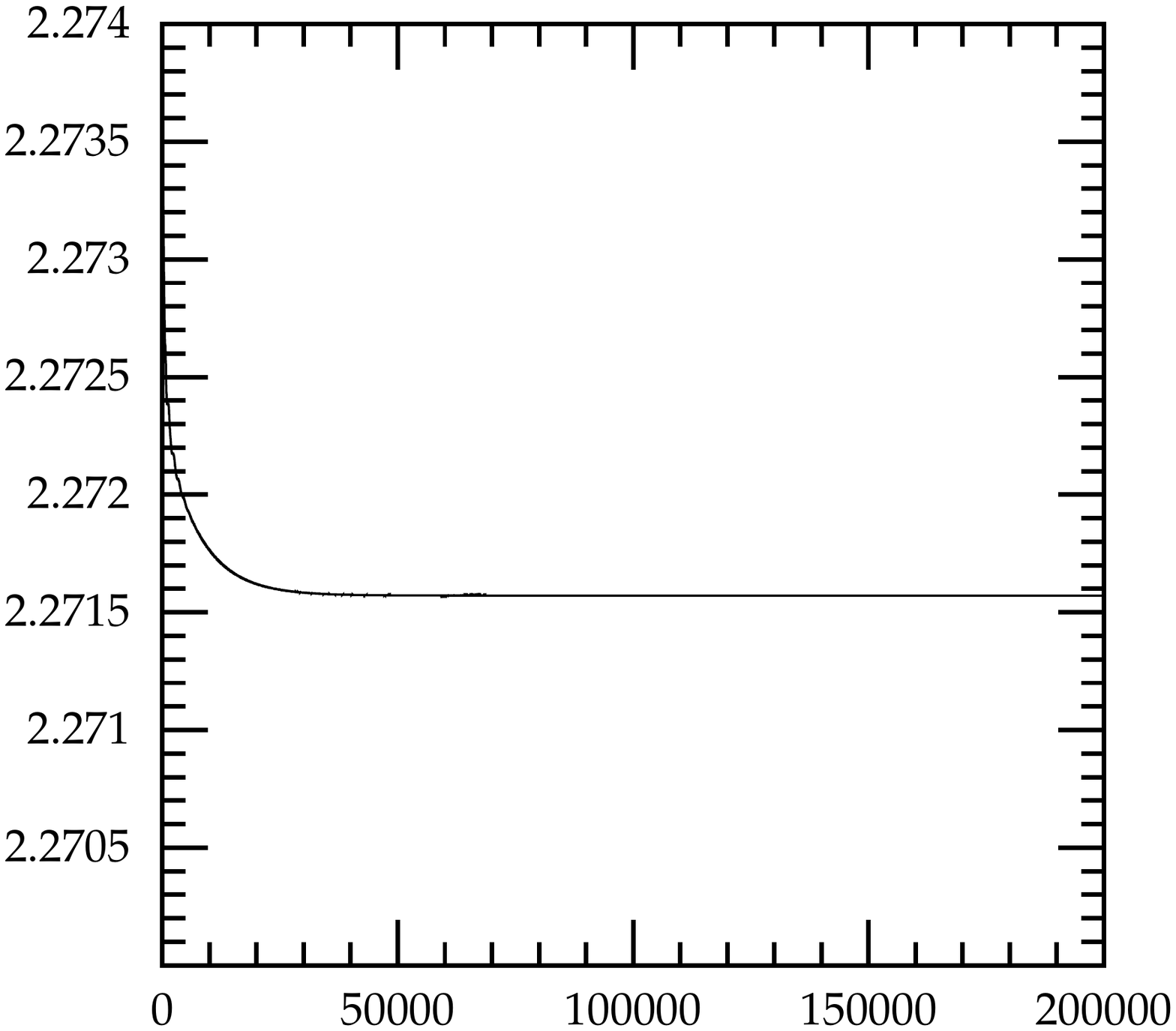}}  
  \subfigure[]{\label{fig:energy600b}\includegraphics[trim = 0cm 0cm 1.8cm 1.8cm, width=0.45\textwidth]{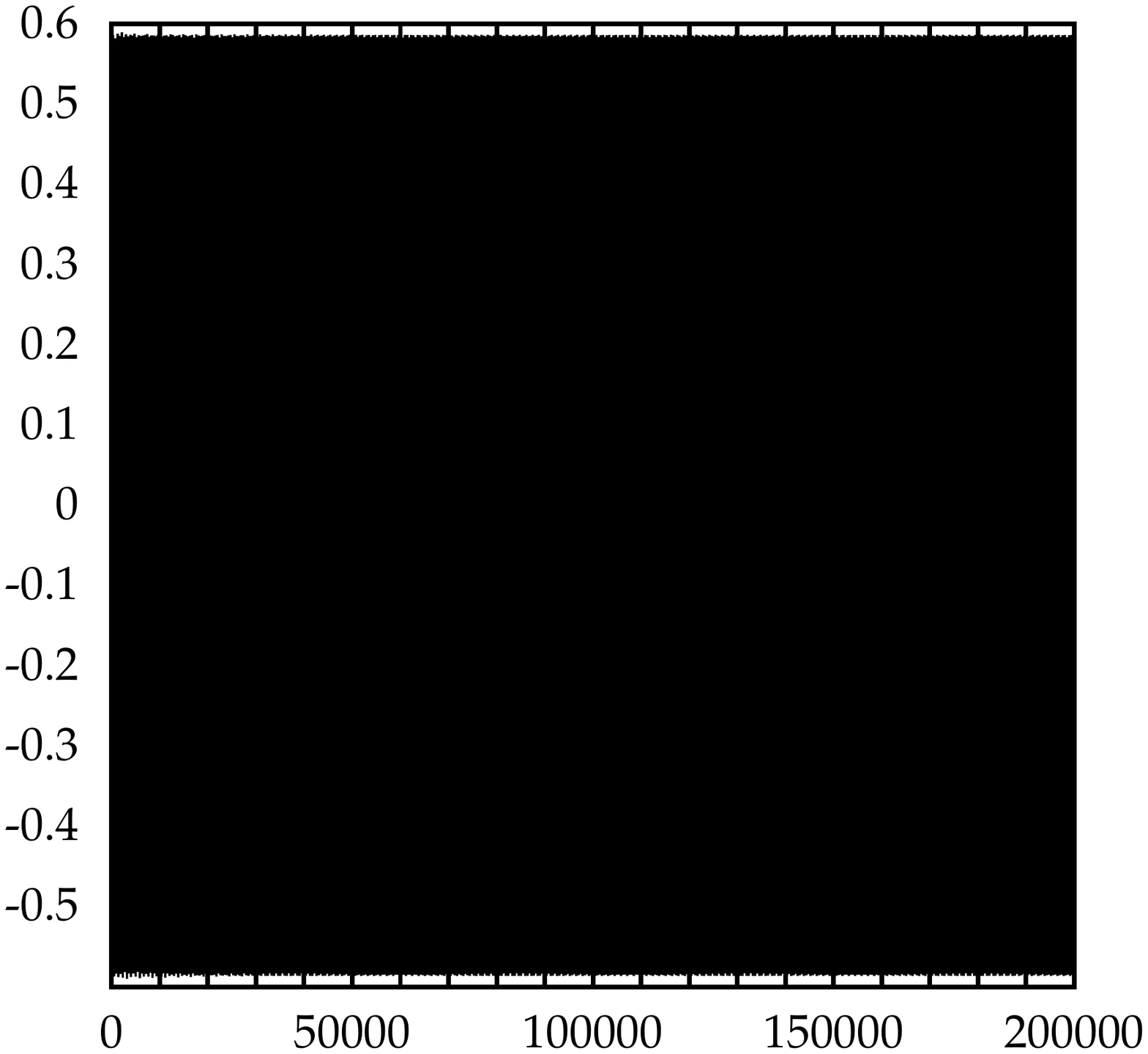}}                              
  \subfigure[]{\label{fig:anomaly600a}\includegraphics[trim = 0cm 0cm 1.8cm 1.8cm,  width=0.45\textwidth]{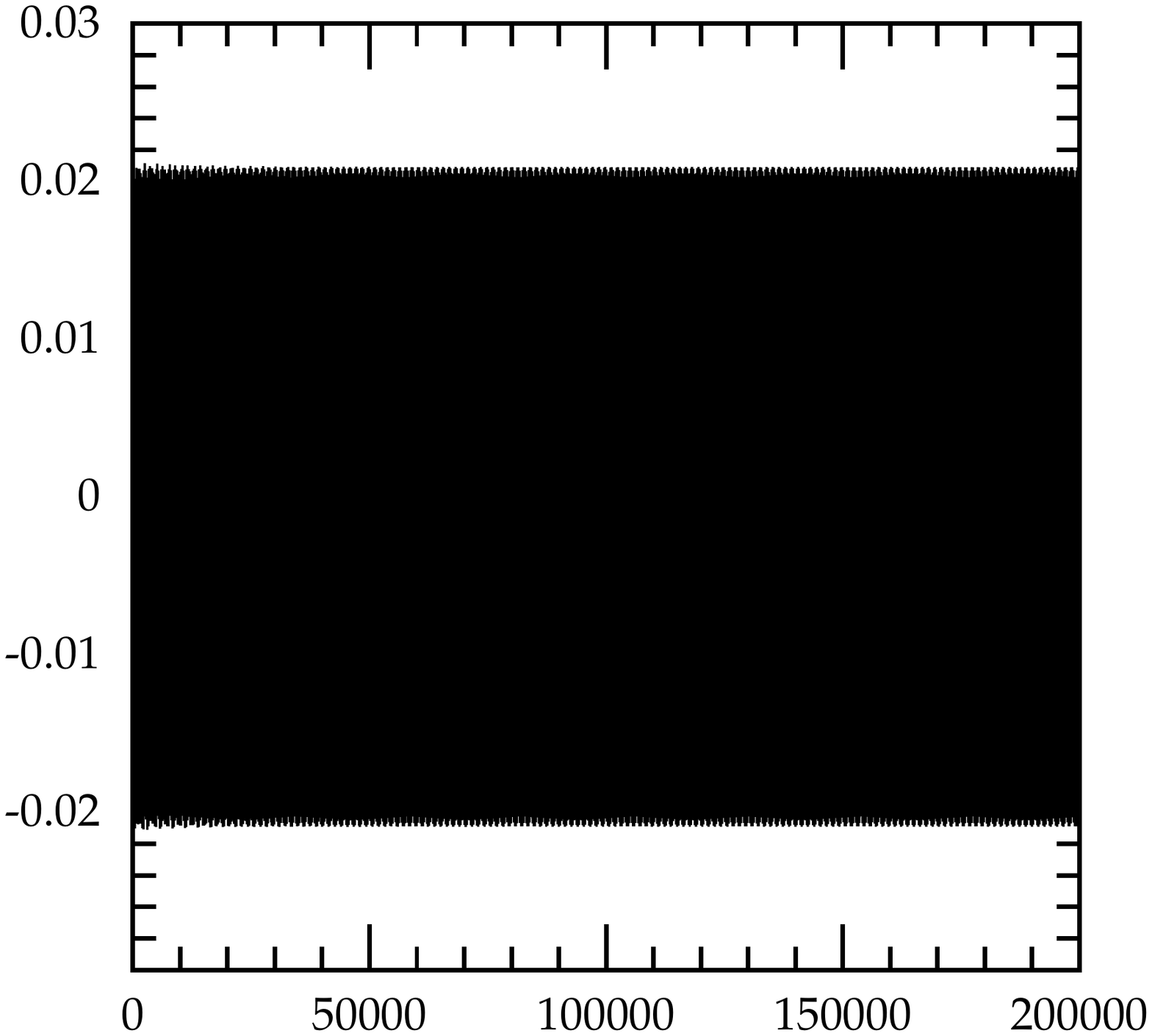}}
  \subfigure[]{\label{fig:anomaly600b}\includegraphics[trim = 0cm 0cm 1.8cm 1.8cm,  width=0.45\textwidth]{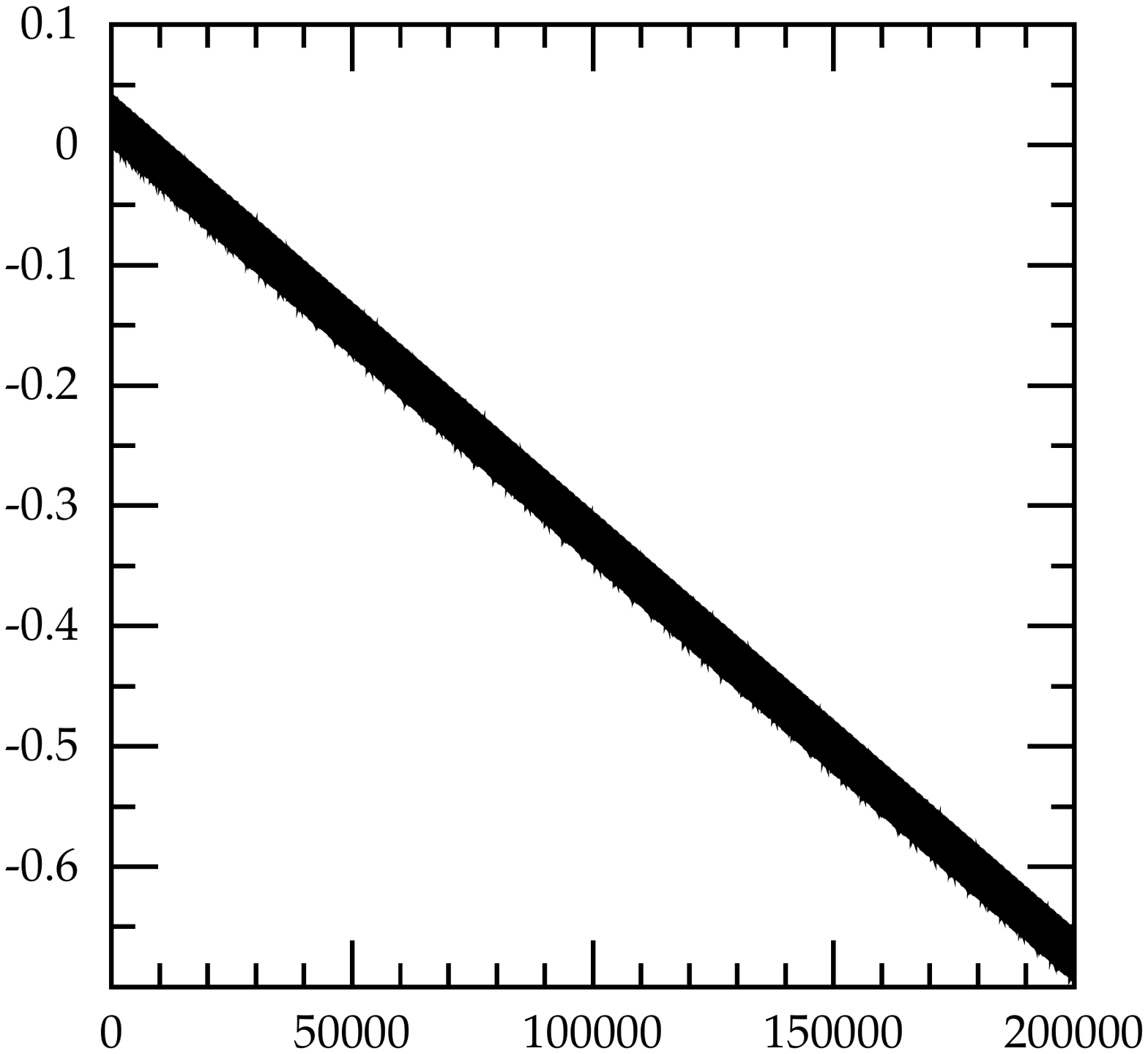}}
   \caption{Breather's simulation in the theory \rf{model} with initial configuration \rf{initialphi},  and with parameters given by $\varepsilon=0.01$, $\nu=0.95$ and $\gamma=0$. The plots show the time dependence of (adimensional units): 
   (a) the energy \rf{energy};  (b) the field $\phi$ at position $x=0$ in the grid; (c) the anomaly \rf{alpha3} and (d) the integrated anomaly \rf{beta3}. }
  \label{fig:600}
\end{figure}

\begin{figure}
  \centering
  \subfigure[]{\label{fig:energy900a}\includegraphics[trim = 0cm 0cm 1.8cm 1.8cm, width=0.45\textwidth]{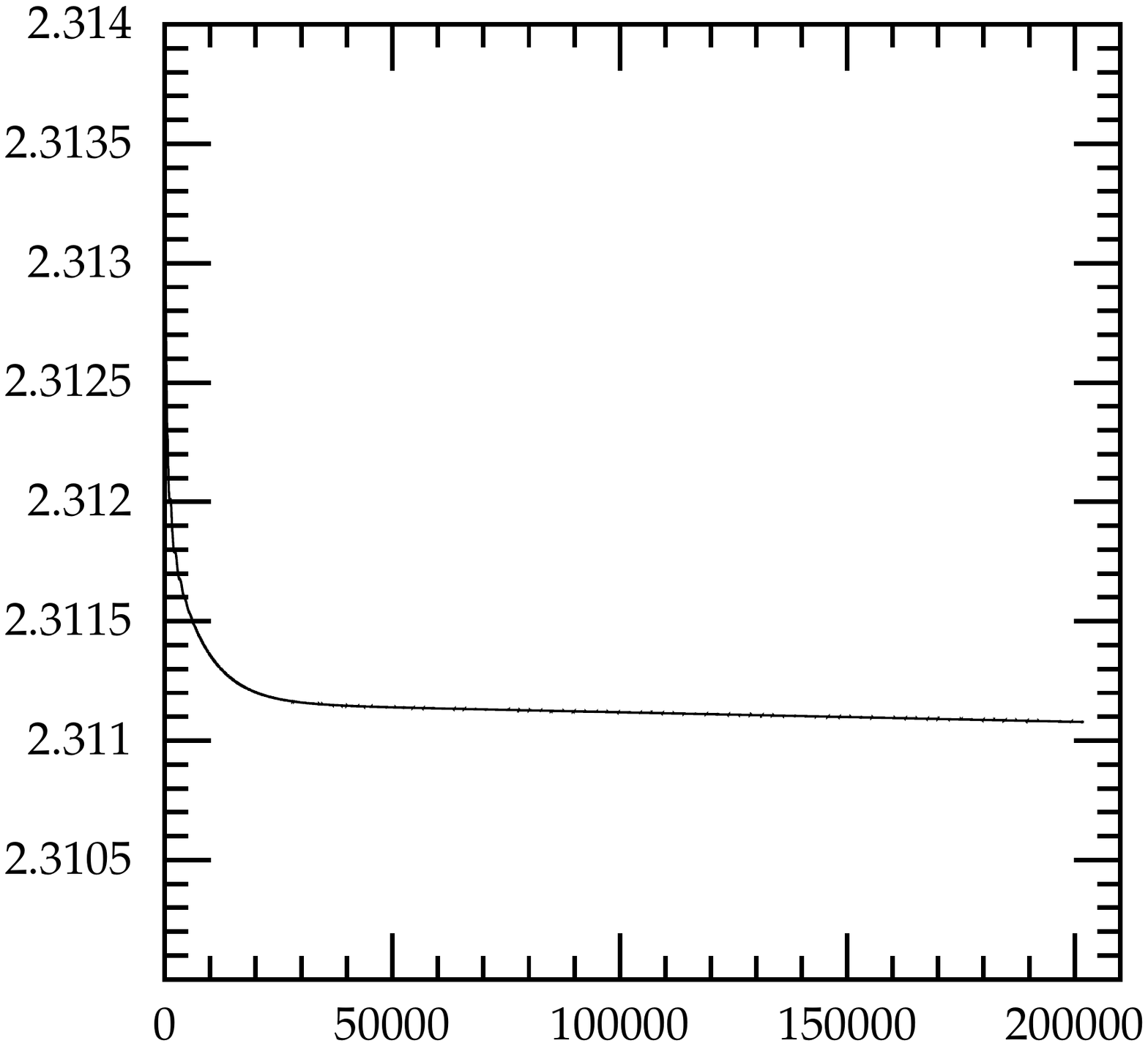}}
    \subfigure[]{\label{fig:energy900b}\includegraphics[trim = 0cm 0cm 1.8cm 1.8cm, width=0.45\textwidth]{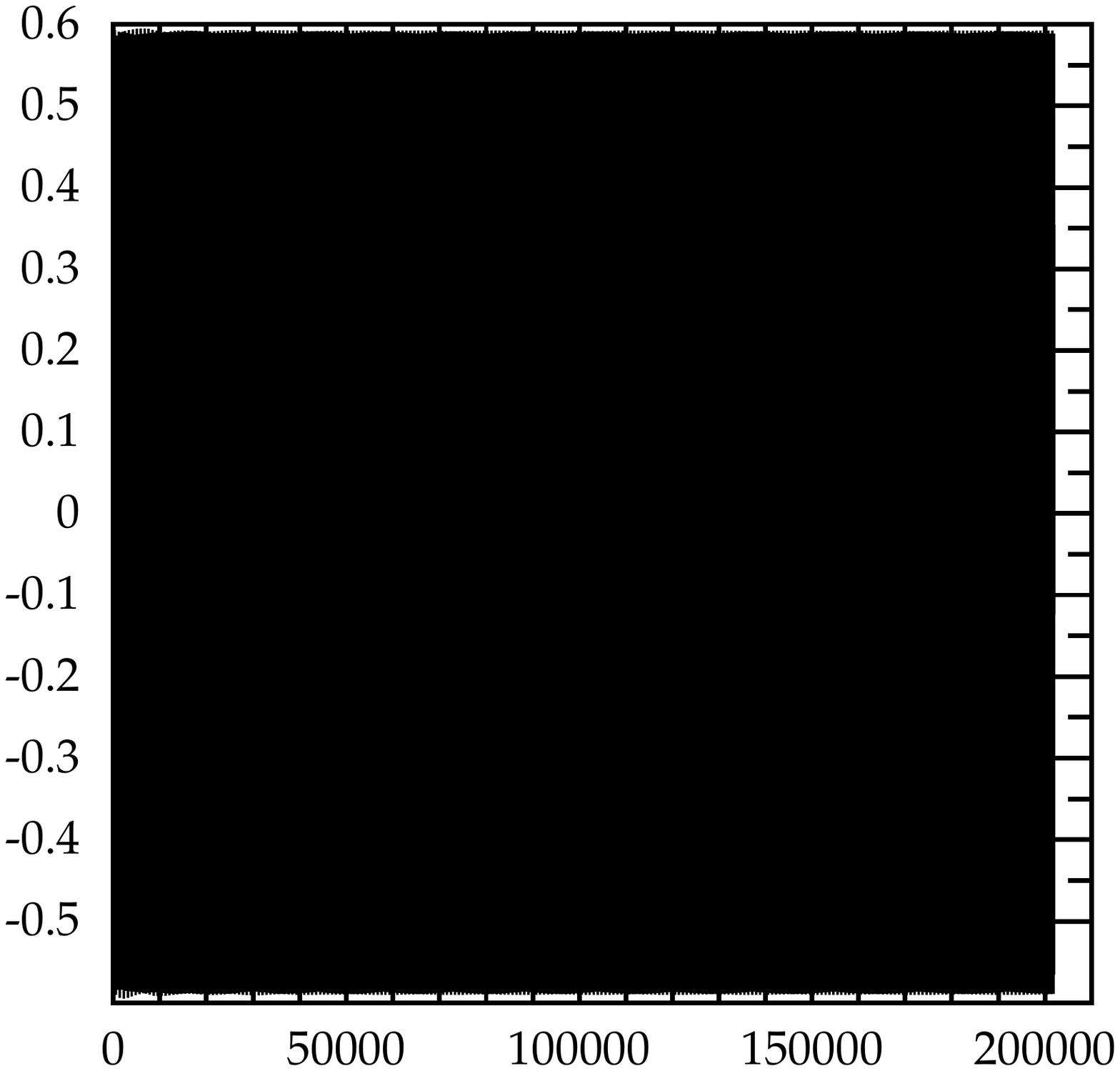}}                
  \subfigure[]{\label{fig:anomaly900a}\includegraphics[trim = 0cm 0cm 1.8cm 1.8cm,  width=0.45\textwidth]{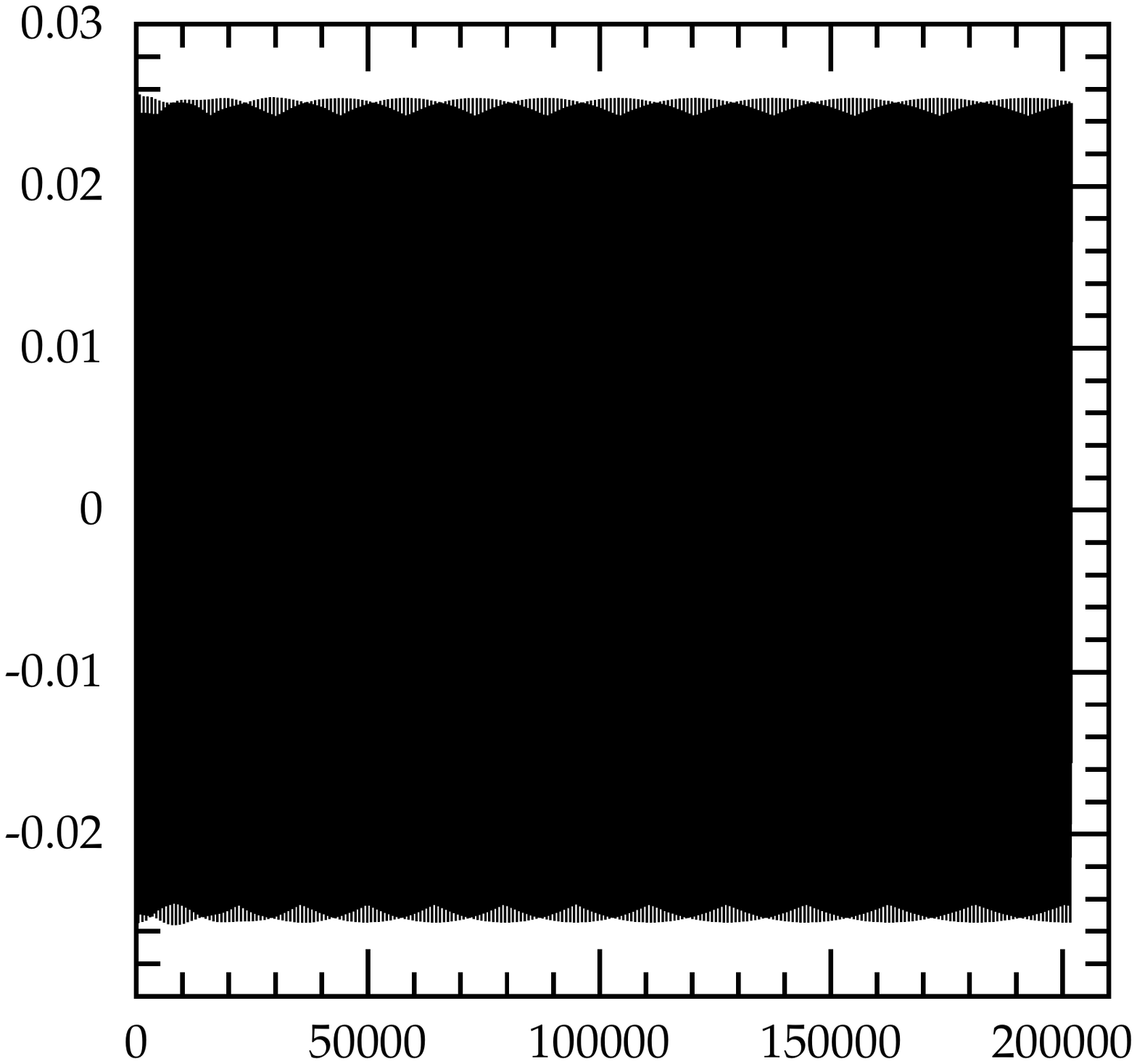}}
  \subfigure[]{\label{fig:anomaly900b}\includegraphics[trim = 0cm 0cm 1.8cm 1.8cm,  width=0.45\textwidth]{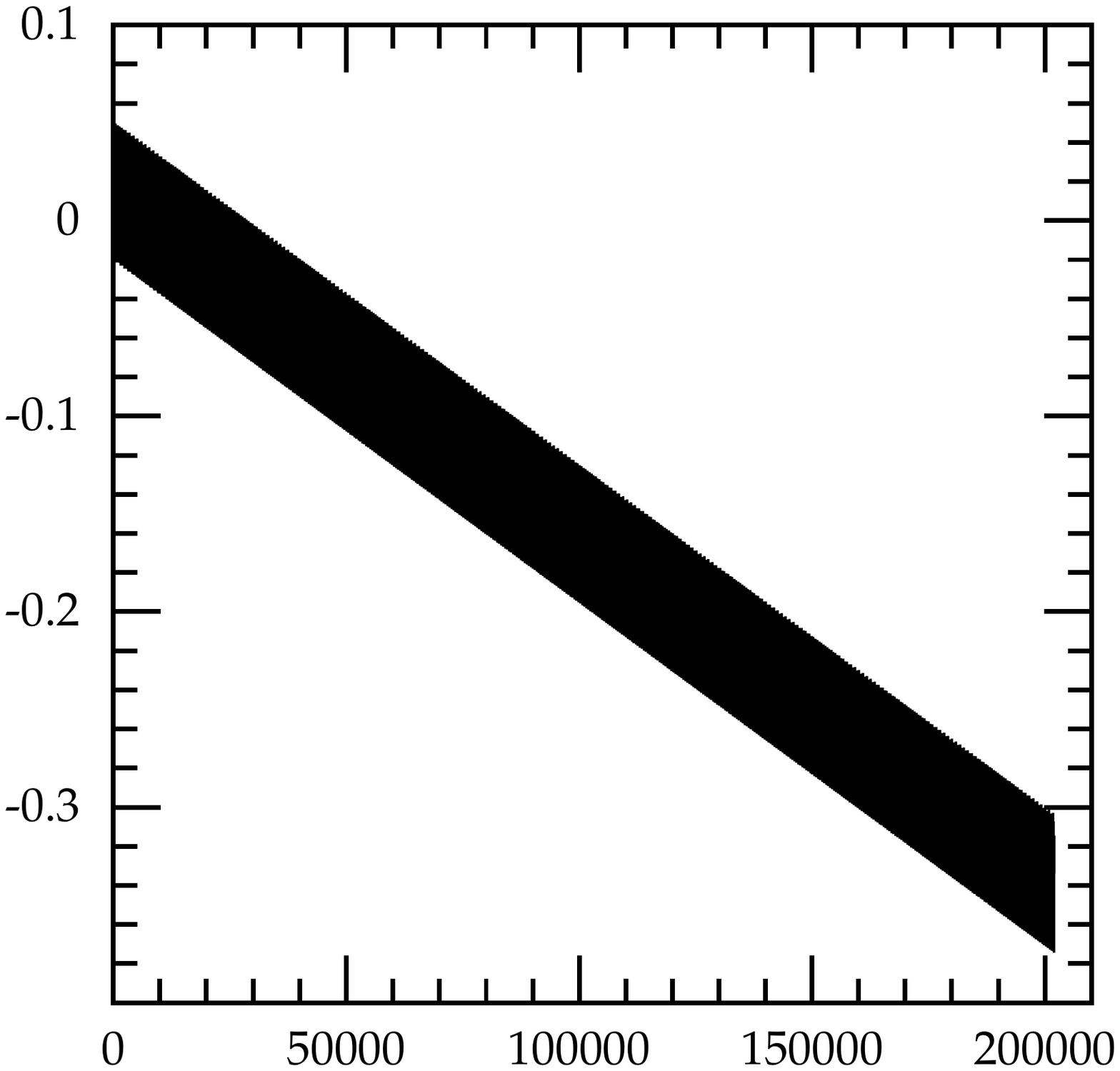}}
   \caption{Breather's simulation in the theory \rf{model} with initial configuration \rf{initialphi},  and with parameters given by $\varepsilon=0.01$, $\nu=0.95$ and $\gamma=0.3$. The plots show the time dependence of (adimensional units): 
   (a) the energy \rf{energy};  (b) the field $\phi$ at position $x=0$ in the grid; (c) the anomaly \rf{alpha3} and (d) the integrated anomaly \rf{beta3}. }
  \label{fig:900}
\end{figure}

\begin{figure}
  \centering
  \subfigure[]{\label{fig:energy150a}\includegraphics[trim = 0cm 0cm 1.8cm 1.8cm, width=0.45\textwidth]{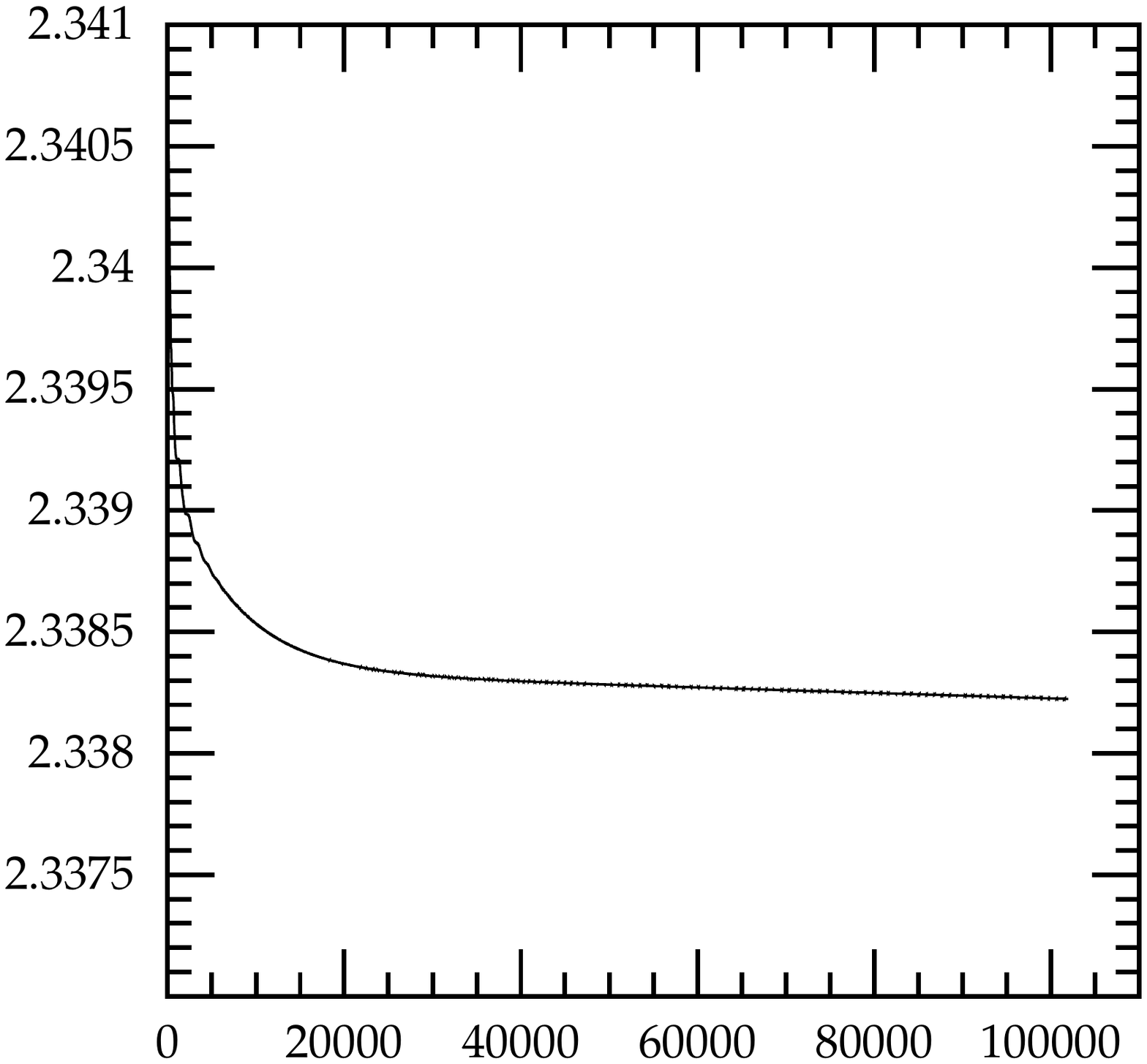}}
    \subfigure[]{\label{fig:energy150b}\includegraphics[trim = 0cm 0cm 1.8cm 1.8cm, width=0.45\textwidth]{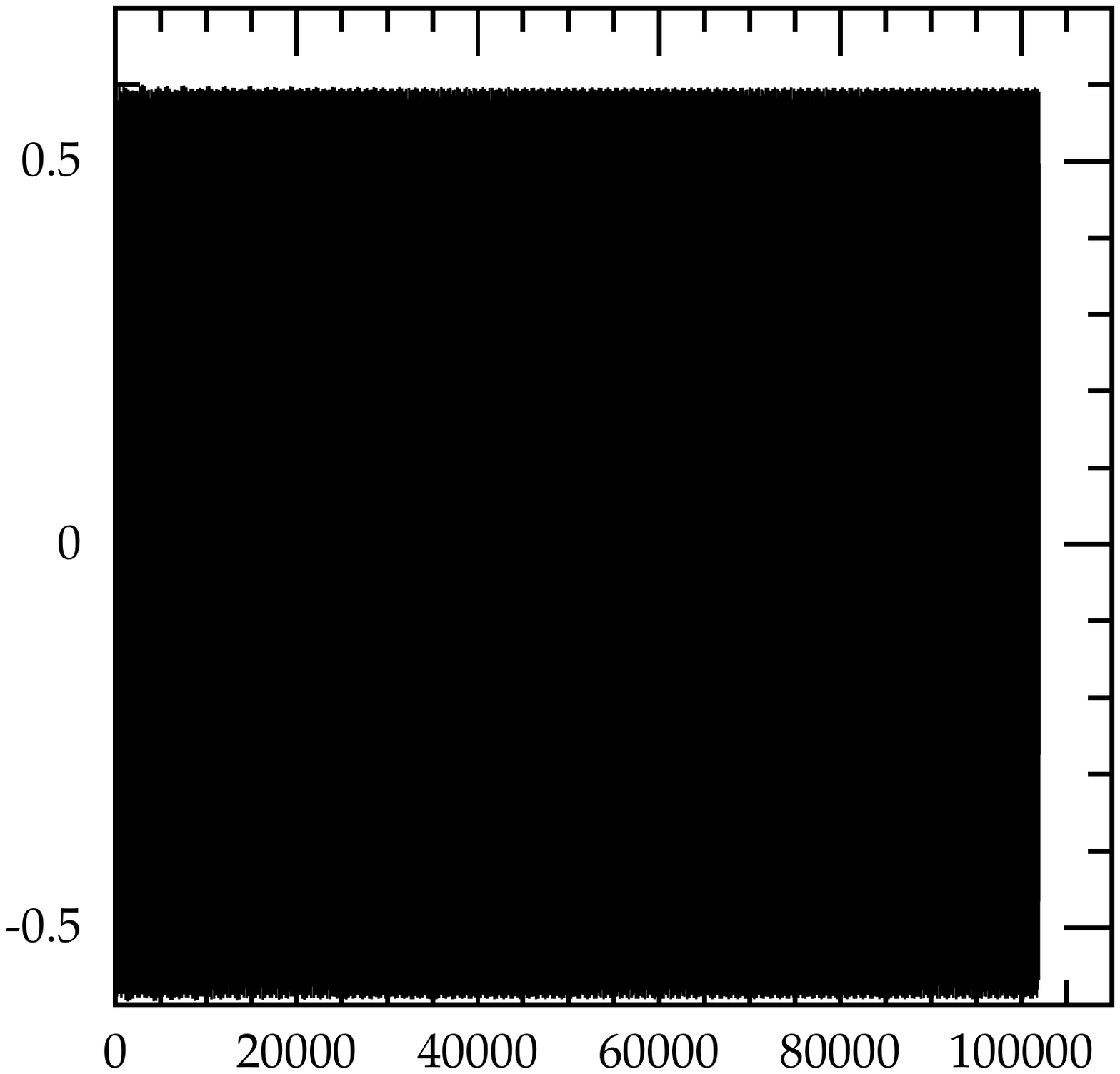}}                
  \subfigure[]{\label{fig:anomaly150a}\includegraphics[trim = 0cm 0cm 1.8cm 1.8cm,  width=0.45\textwidth]{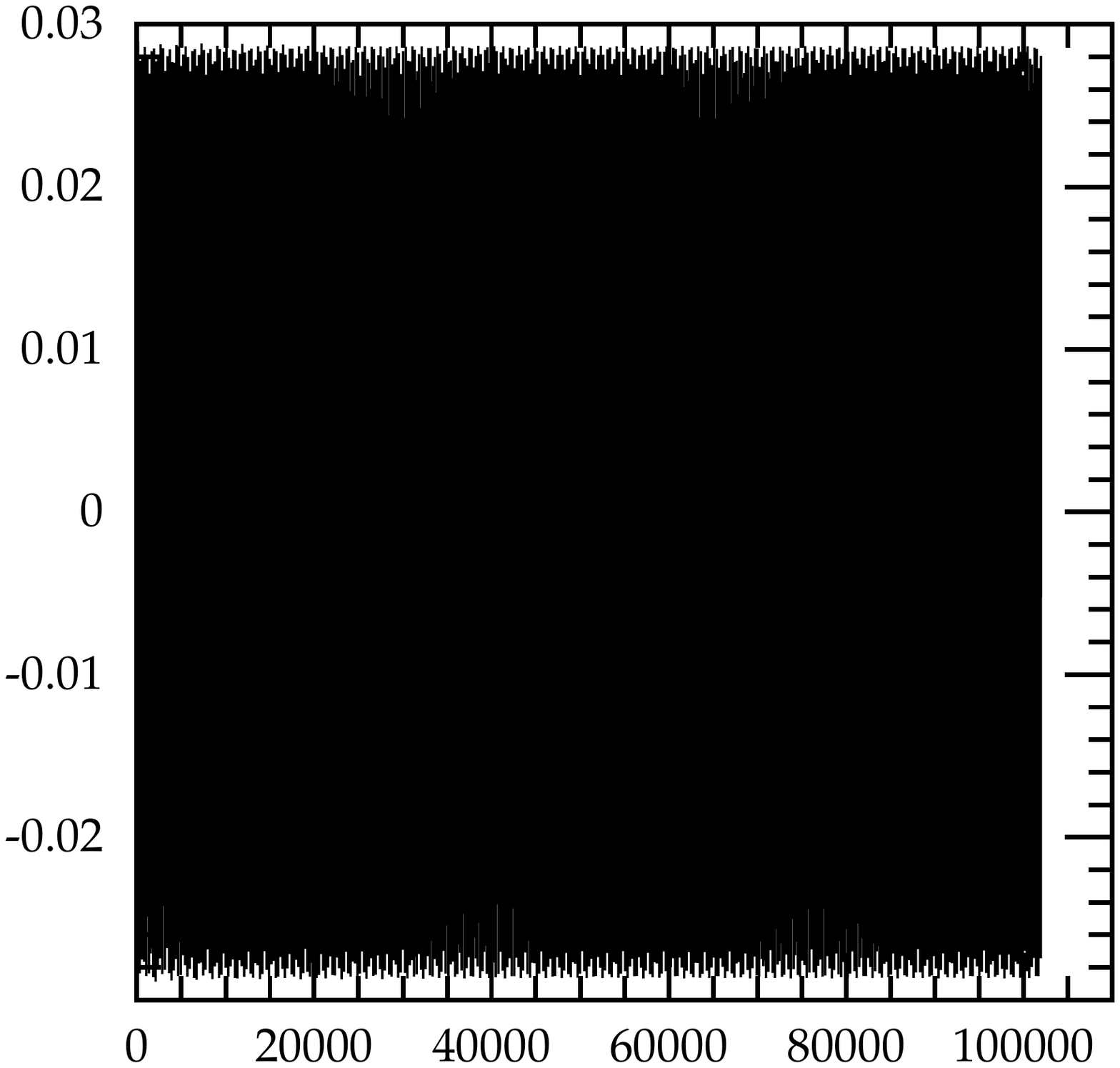}}
  \subfigure[]{\label{fig:anomaly150b}\includegraphics[trim = 0cm 0cm 1.8cm 1.8cm,  width=0.45\textwidth]{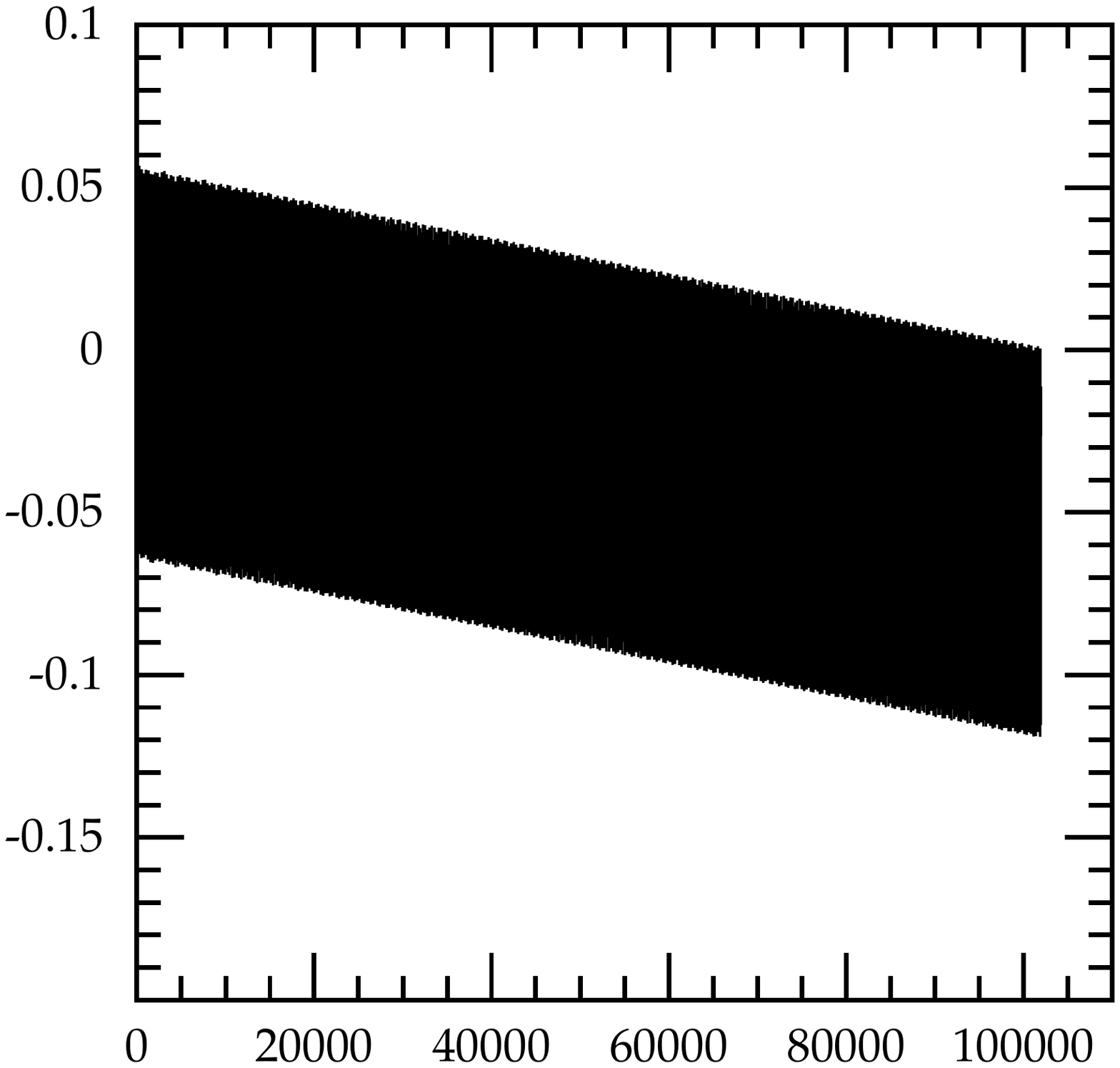}}
  \caption{Breather's simulation in the theory \rf{model} with initial configuration \rf{initialphi},  and with parameters given by $\varepsilon=0.01$, $\nu=0.95$ and $\gamma=0.5$. The plots show the time dependence of (adimensional units): 
   (a) the energy \rf{energy};  (b) the field $\phi$ at position $x=0$ in the grid; (c) the anomaly \rf{alpha3} and (d) the integrated anomaly \rf{beta3}. }
    \label{fig:150}
\end{figure}

\begin{figure}
  \centering
  \subfigure[]{\label{fig:energy350a}\includegraphics[trim = 0cm 0cm 1.8cm 1.8cm, width=0.45\textwidth]{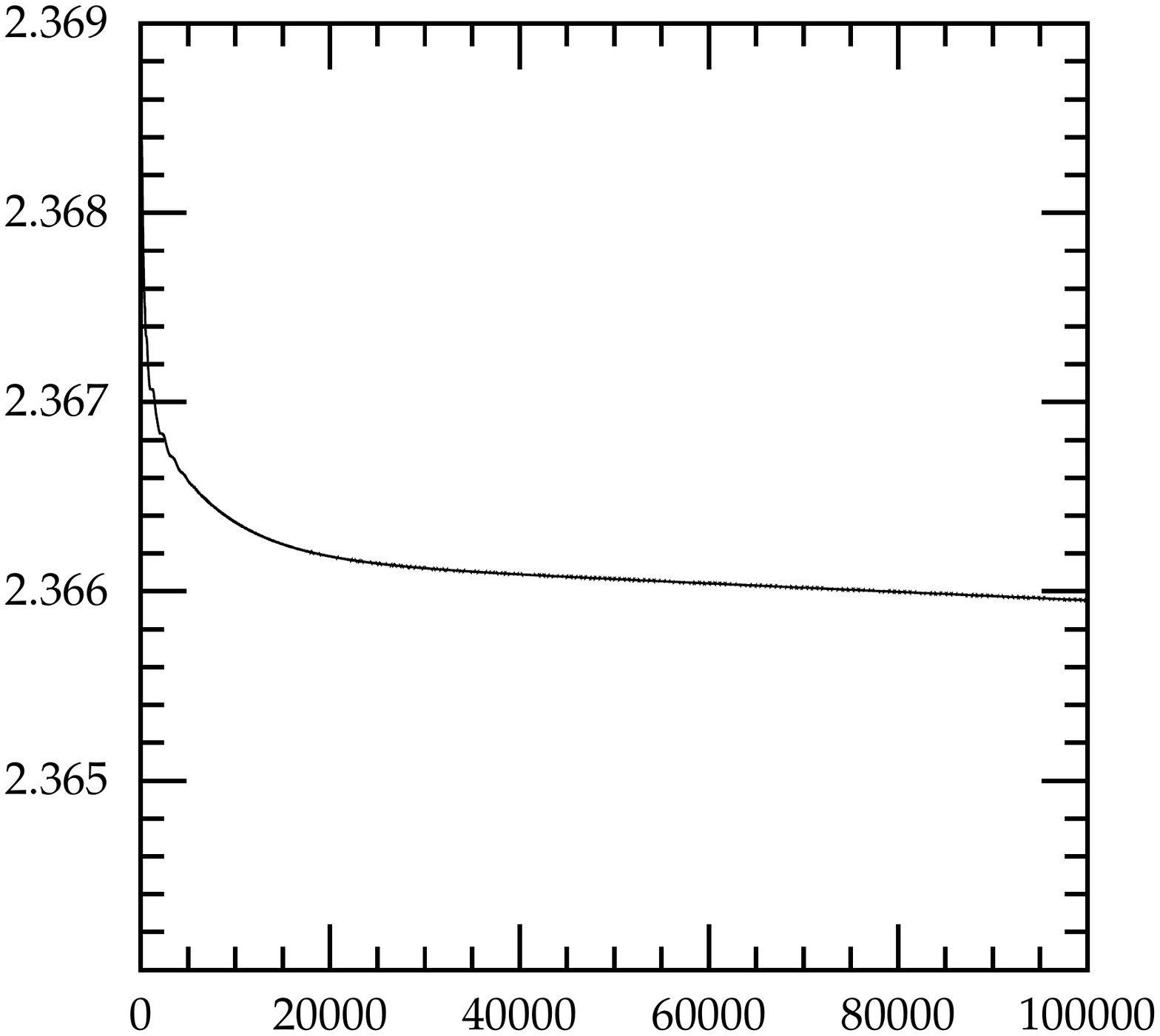}}
    \subfigure[]{\label{fig:energy350b}\includegraphics[trim = 0cm 0cm 1.8cm 1.8cm, width=0.45\textwidth]{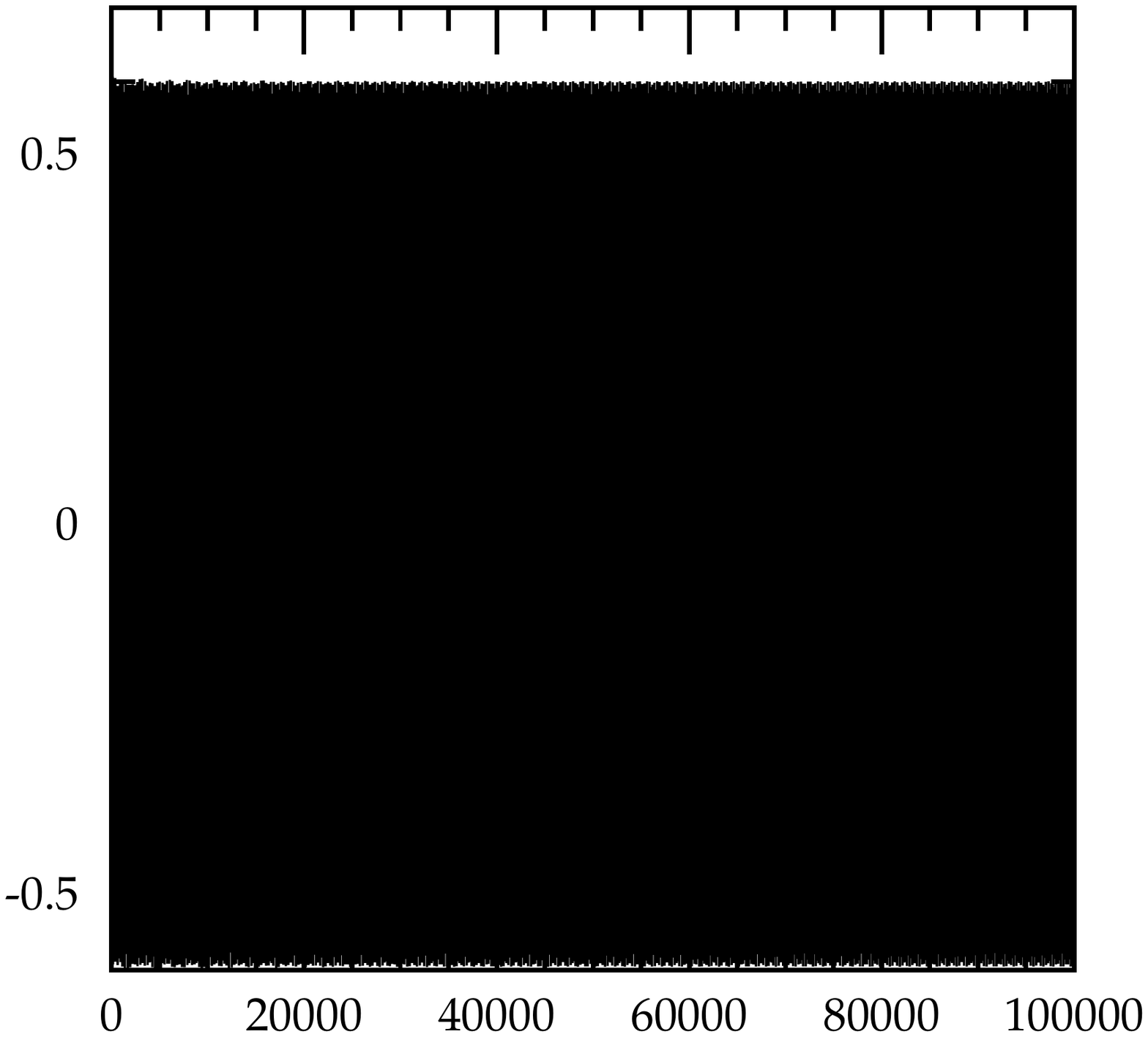}}                
  \subfigure[]{\label{fig:anomaly350a}\includegraphics[trim = 0cm 0cm 1.8cm 1.8cm,  width=0.45\textwidth]{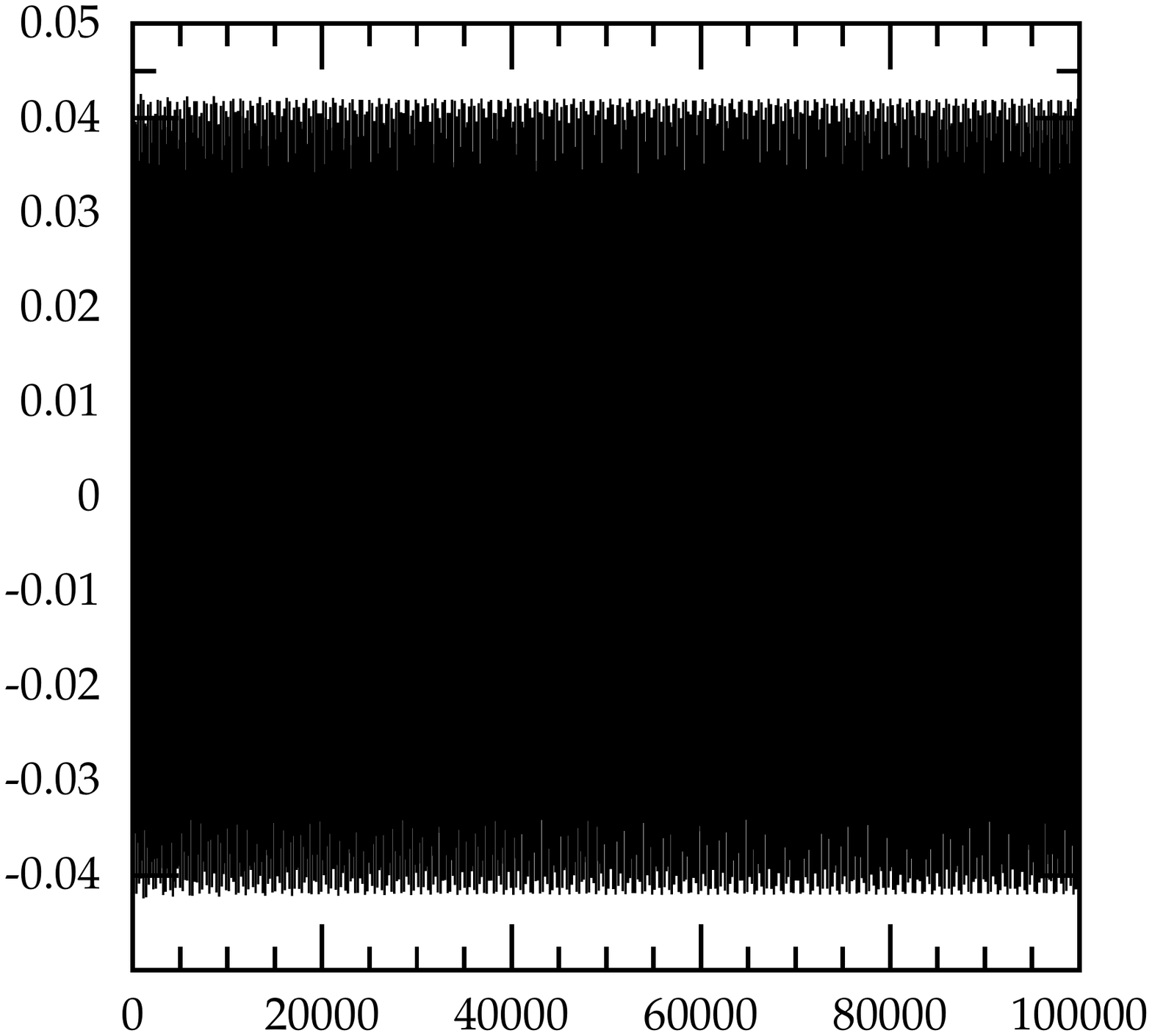}}
  \subfigure[]{\label{fig:anomaly350b}\includegraphics[trim = 0cm 0cm 1.8cm 1.8cm,  width=0.45\textwidth]{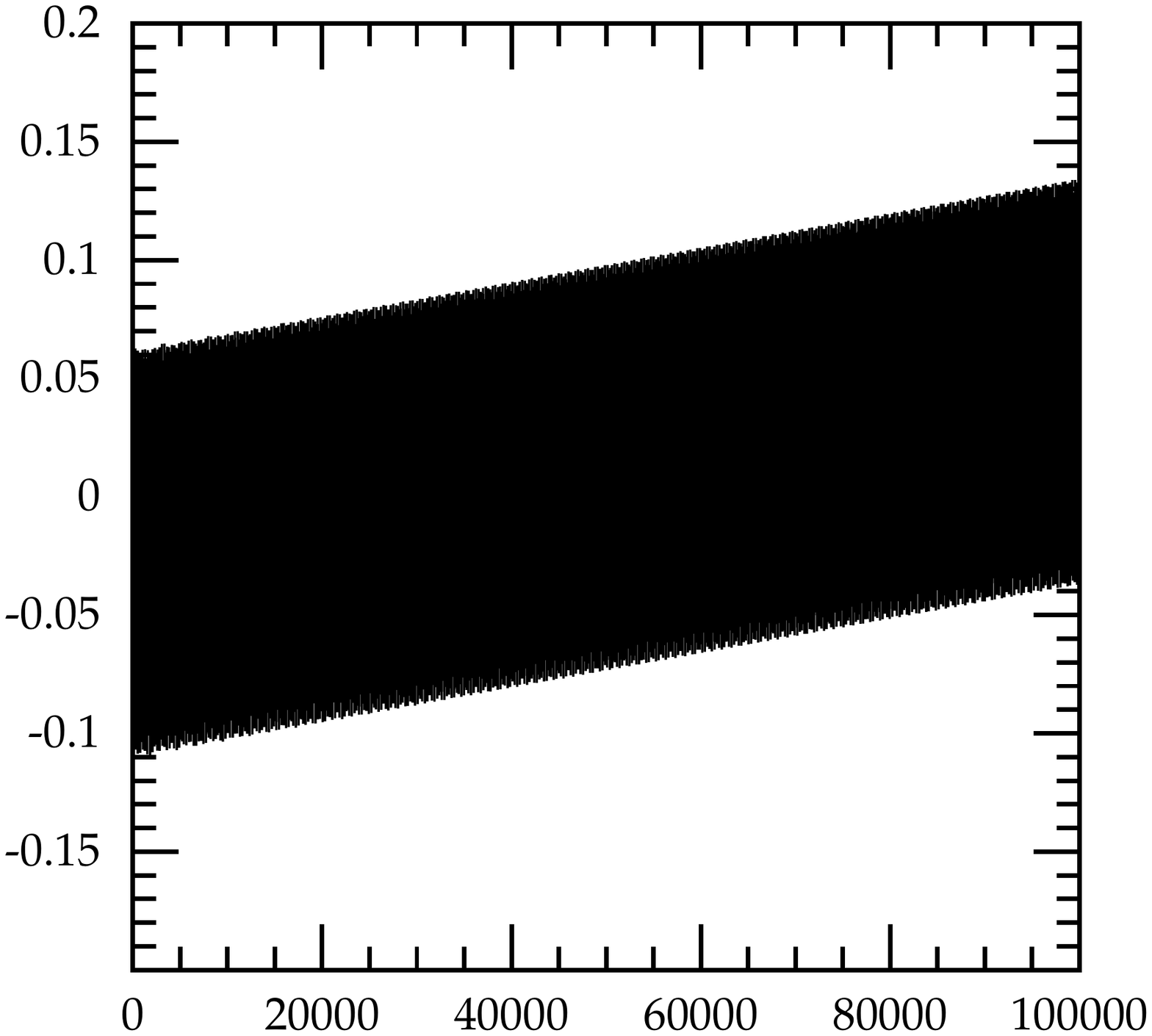}}
  \caption{Breather's simulation in the theory \rf{model} with initial configuration \rf{initialphi},  and with parameters given by $\varepsilon=0.01$, $\nu=0.95$ and $\gamma=0.7$. The plots show the time dependence of (adimensional units): 
   (a) the energy \rf{energy};  (b) the field $\phi$ at position $x=0$ in the grid; (c) the anomaly \rf{alpha3} and (d) the integrated anomaly \rf{beta3}. }
  \label{fig:350}
\end{figure}

We can clearly observe two types of effects playing a role in the outcomes of the simulations:

First, as predicted by our analytical calculations based on the parity argument, the breathers for the cases with $\gamma=0$ tend to live longer when compared to similar  breathers for the cases where $\gamma\neq 0$. In order to see this more clearly, let us look at the simulations shown in Figs. \ref{fig:fig2},  \ref{fig:4} and  \ref{fig:fig5}, all corresponding to $\ve =0.3$ and $\nu=0.5$. In Fig. \ref{fig:fig2} corresponding to the case $\gamma=0$ we see that the energy almost goes to a constant value after the initial stabilization of the solution and  the lifetime of the breather is extremely long. The simulation was stopped after  $1.1\times 10^{6}$ units of time and the energy was still almost constant (in fact it decreased all the time but extremely slowly). In 
Fig. \ref{fig:4}, corresponding to the case $\gamma =0.2$, we observe that the energy is still decreasing after $1.1\times 10^{6}$ units of time of simulation. So, despite the fact that the fall of the energy is slow we see the  $\gamma\neq 0$ effect playing an important role in increasing the speed of the decrease. Now, in Fig. \ref{fig:fig5}, corresponding to the case $\gamma =0.5$, we see that the energy drops much faster, and the breather starts moving after about $9\times 10^{4}$ units of time, and it has bounced off the edges of the grid twice during the simulation. So, by increasing $\gamma$ the phenomenon we have called {\em quasi-integrablility}, seems to have almost disappeared.  Notice also that for smaller values of $\ve$ the effect of $\gamma\neq 0$ is not so visible. However, we do notice in Figures \ref{fig:700} and \ref{fig:800}, corresponding to $\ve =0.01$ and $\nu=0.1$, that the breather for $\gamma=0.3$ is short-lived as compared to that for $\gamma=0$. The same effect is visible in Figures \ref{fig:0010500}, \ref{fig:0010502} and \ref{fig:0010505}, corresponding to $\ve =0.01$ and $\nu=0.5$. As the values of $\gamma$ are increased from $0.0$ to $0.2$ and then to $0.5$, the life-times  of the breathers decrease. The same effect is not seem however in figures  \ref{fig:600}, \ref{fig:900}, \ref{fig:150} and \ref{fig:350}, corresponding to the case $\ve =0.01$ and $\nu=0.95$.  We cannot see much difference in the behaviour of their energies as the value of $\gamma$ is increased. They seem to have stabilized quite well after $3\times 10^4$ units of time.  These  cases might be feeling the influence of the second effect which we will now describe. 

\begin{figure}
\centering
 \includegraphics[trim = 0cm 0cm 1.8cm 1.8cm, width=0.45\textwidth]{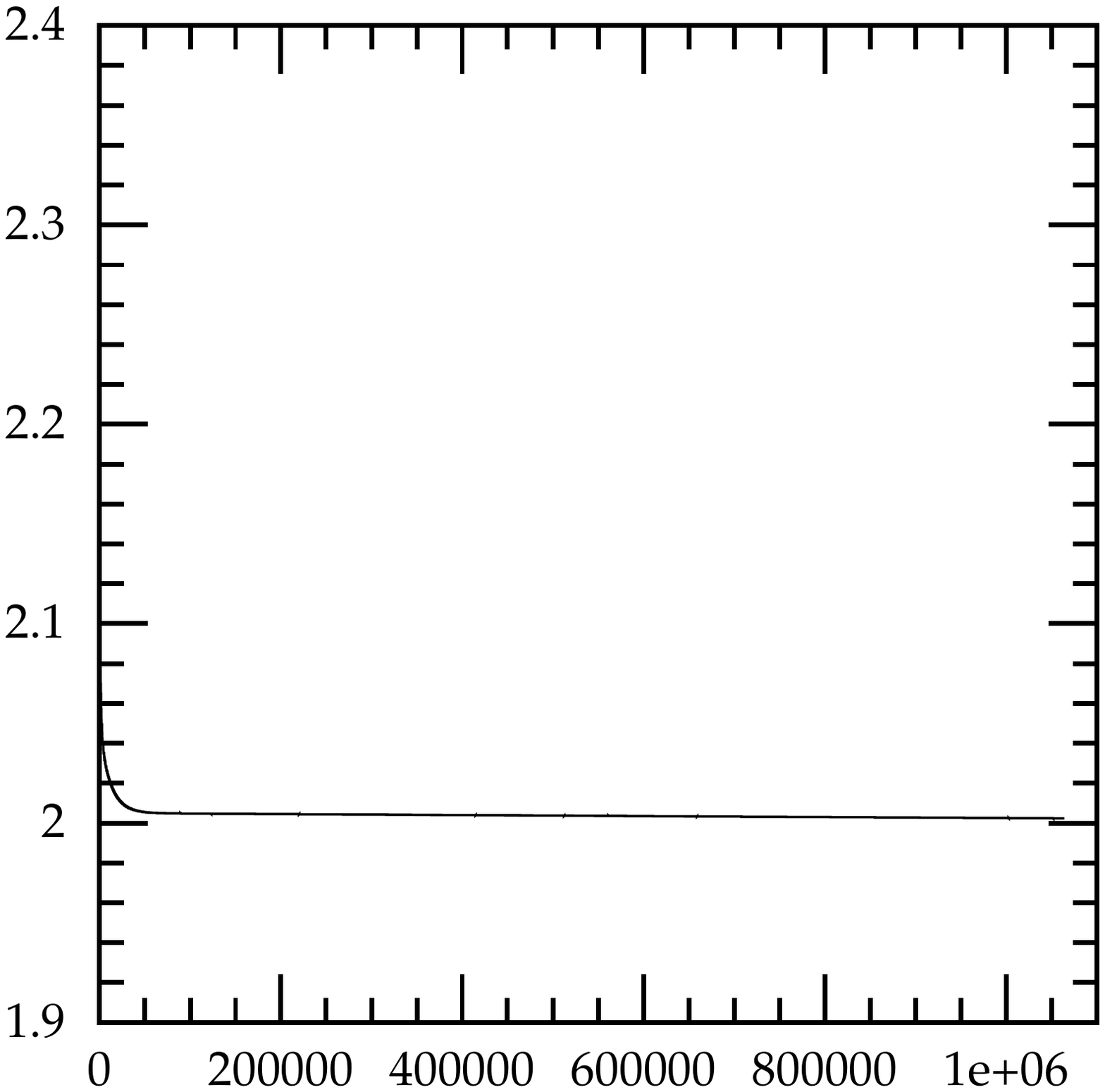}
\caption{Breather's simulation in the theory \rf{model} with initial configuration \rf{initialphi},  and with parameters given by $\varepsilon=0.2$, $\nu=0.5$ and $\gamma=0$. The plot shows the time dependence  (adimensional units) of the energy \rf{energy}.   }
\label{fig:02050}
\end{figure}

\begin{figure}
  \centering
  \subfigure[]{\label{fig:energy200a}\includegraphics[trim = 0cm 0cm 1.8cm 1.8cm, width=0.45\textwidth]{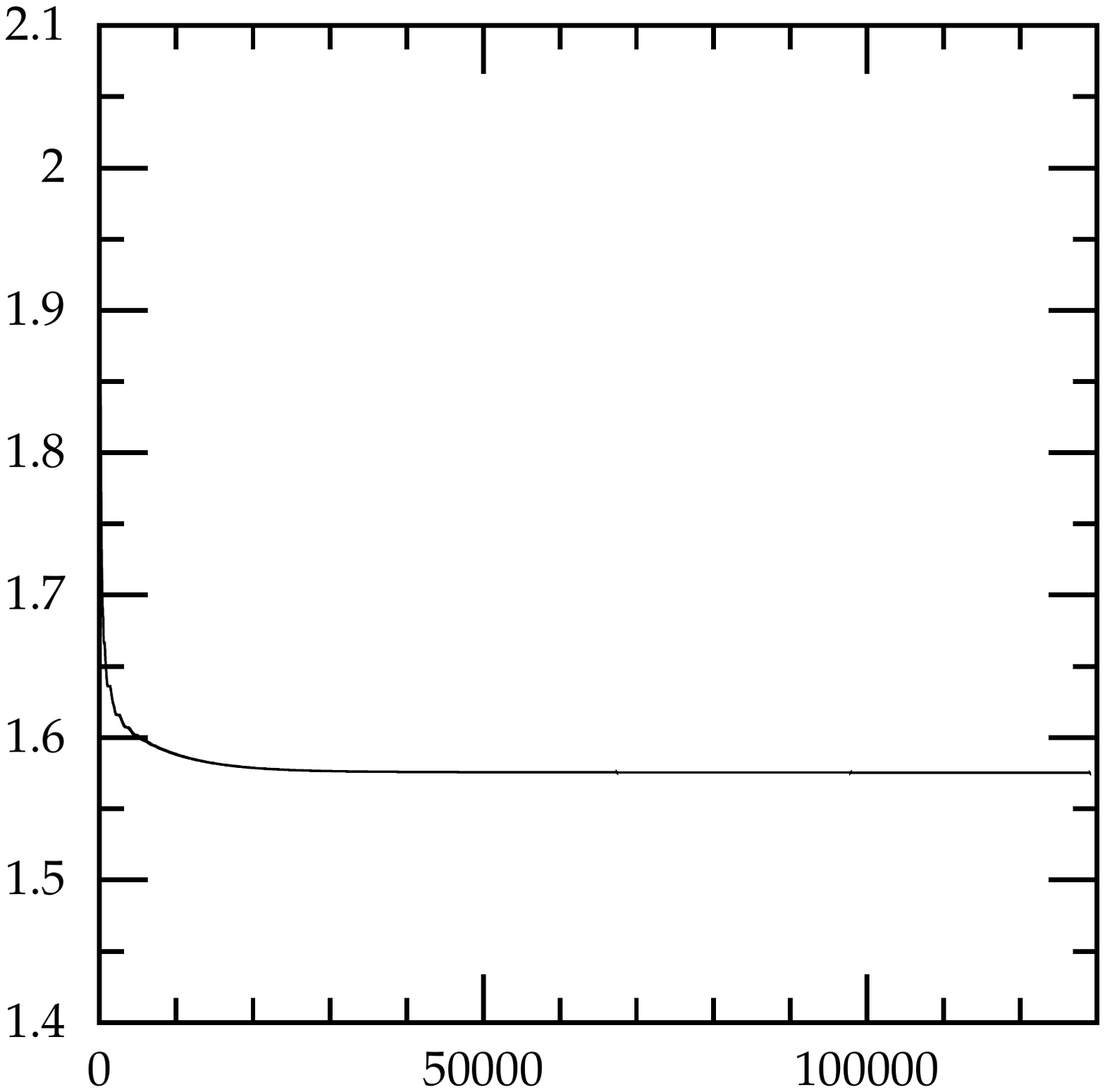}}   
  \subfigure[]{\label{fig:energy200b}\includegraphics[trim = 0cm 0cm 1.8cm 1.8cm, width=0.45\textwidth]{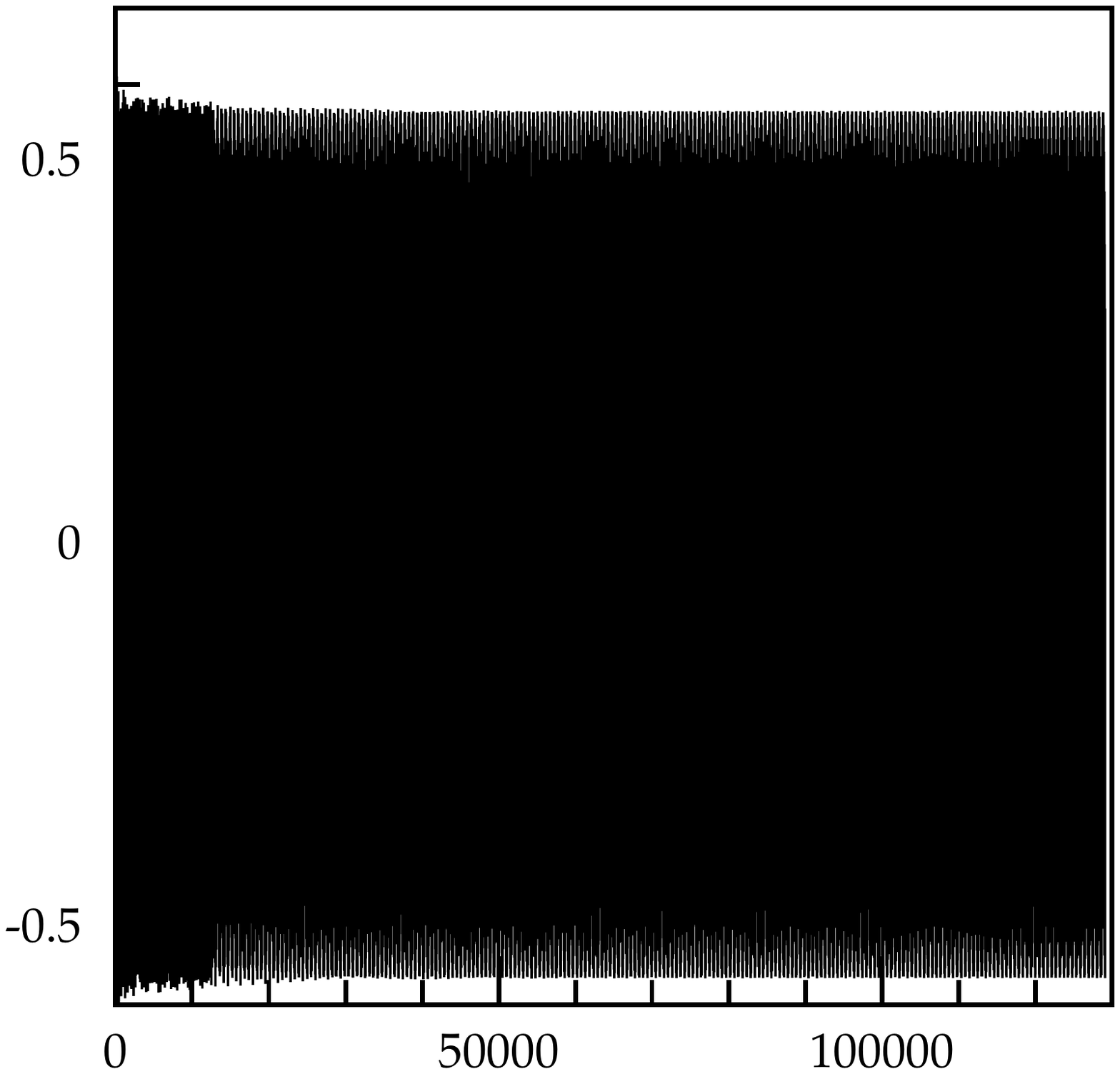}}                
  \subfigure[]{\label{fig:anomaly200a}\includegraphics[trim = 0cm 0cm 1.8cm 1.8cm,  width=0.45\textwidth]{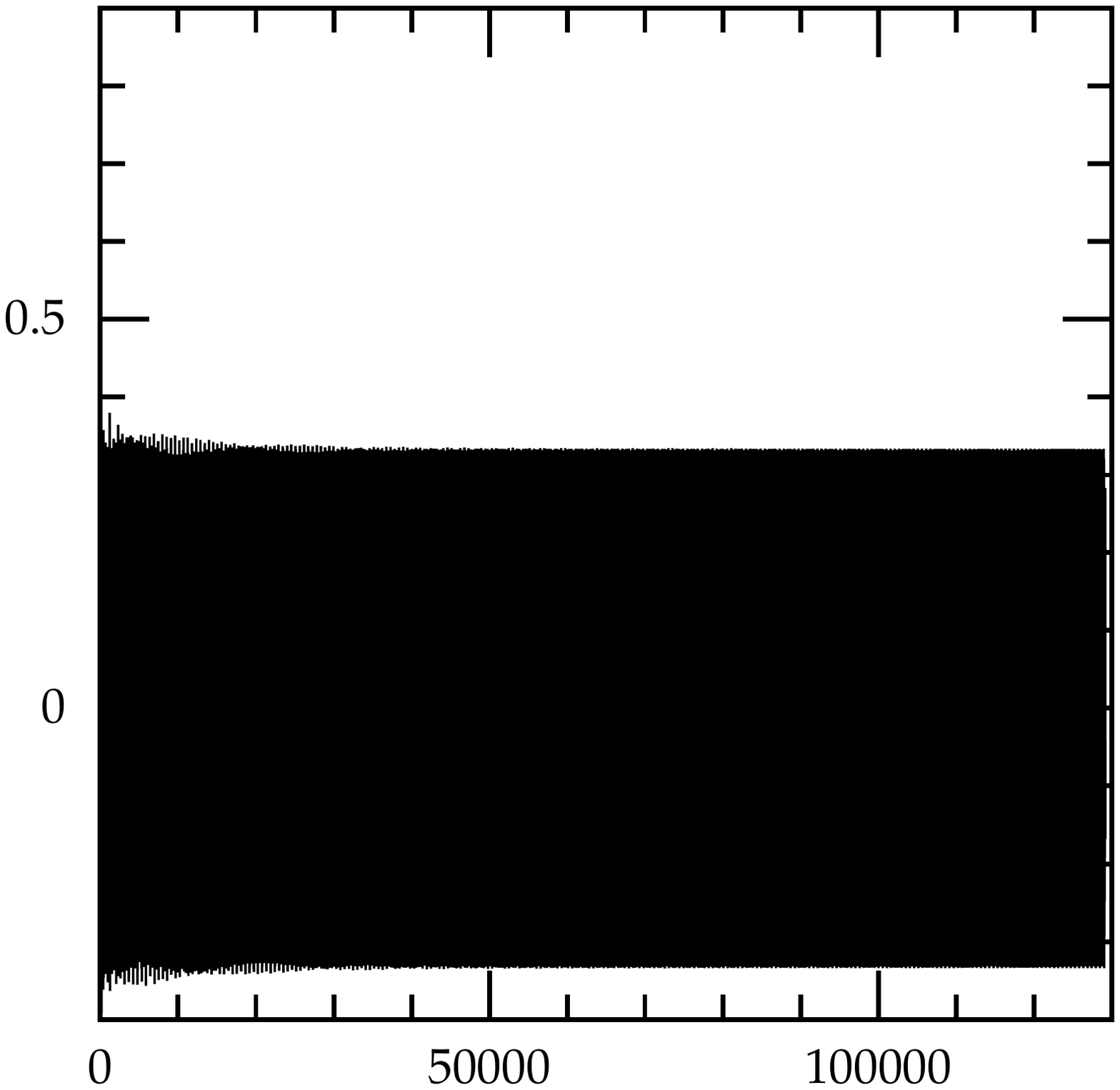}}
  \subfigure[]{\label{fig:anomaly200b}\includegraphics[trim = 0cm 0cm 1.8cm 1.8cm,  width=0.45\textwidth]{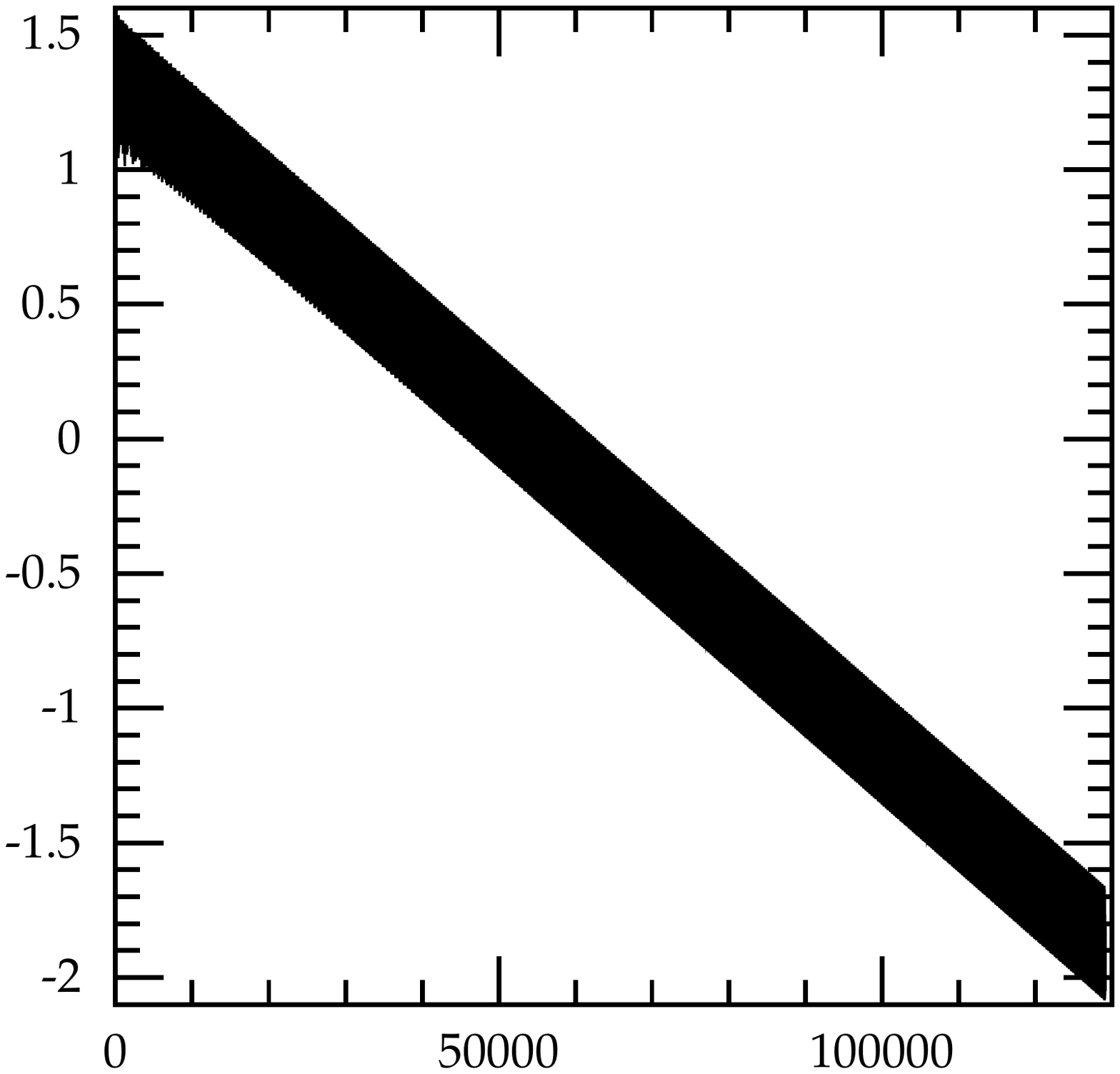}}
  \caption{Breather's simulation in the theory \rf{model} with initial configuration \rf{initialphi},  and with parameters given by of $\varepsilon=0.3$, $\nu=0.1$ and $\gamma=0$. The plots show the time dependence of (adimensional units): 
   (a) the energy \rf{energy};  (b) the field $\phi$ at position $x=0$ in the grid; (c) the anomaly \rf{alpha3} and (d) the integrated anomaly \rf{beta3}. }
    \label{fig:200}
\end{figure}

\begin{figure}
  \centering
  \subfigure[]{\label{fig:potential1}\includegraphics[trim = 0cm 0cm 1.8cm 1.8cm, width=0.45\textwidth]{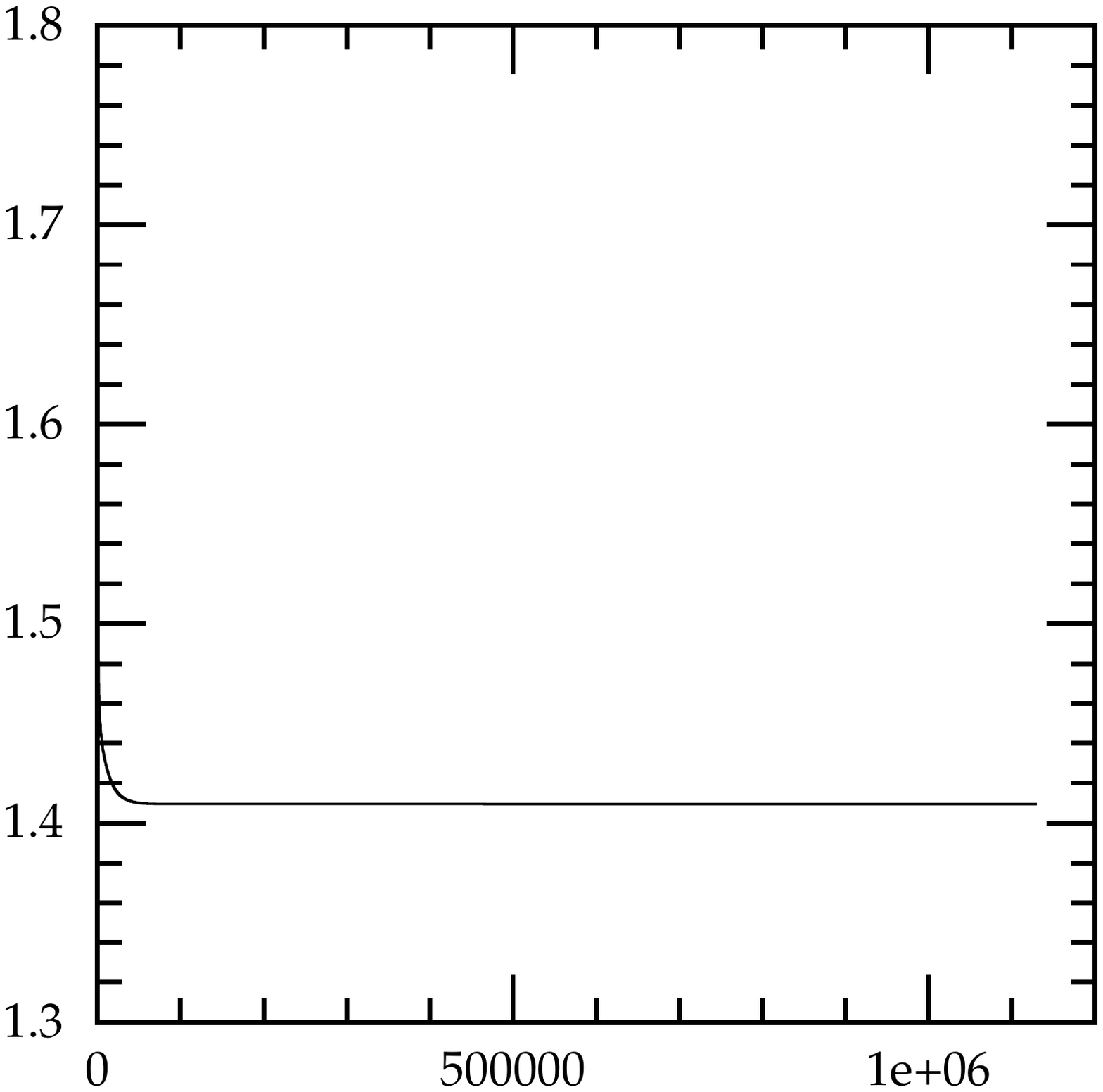}}                
  \subfigure[]{\label{fig:potential2}\includegraphics[trim = 0cm 0cm 1.8cm 1.8cm,  width=0.45\textwidth]{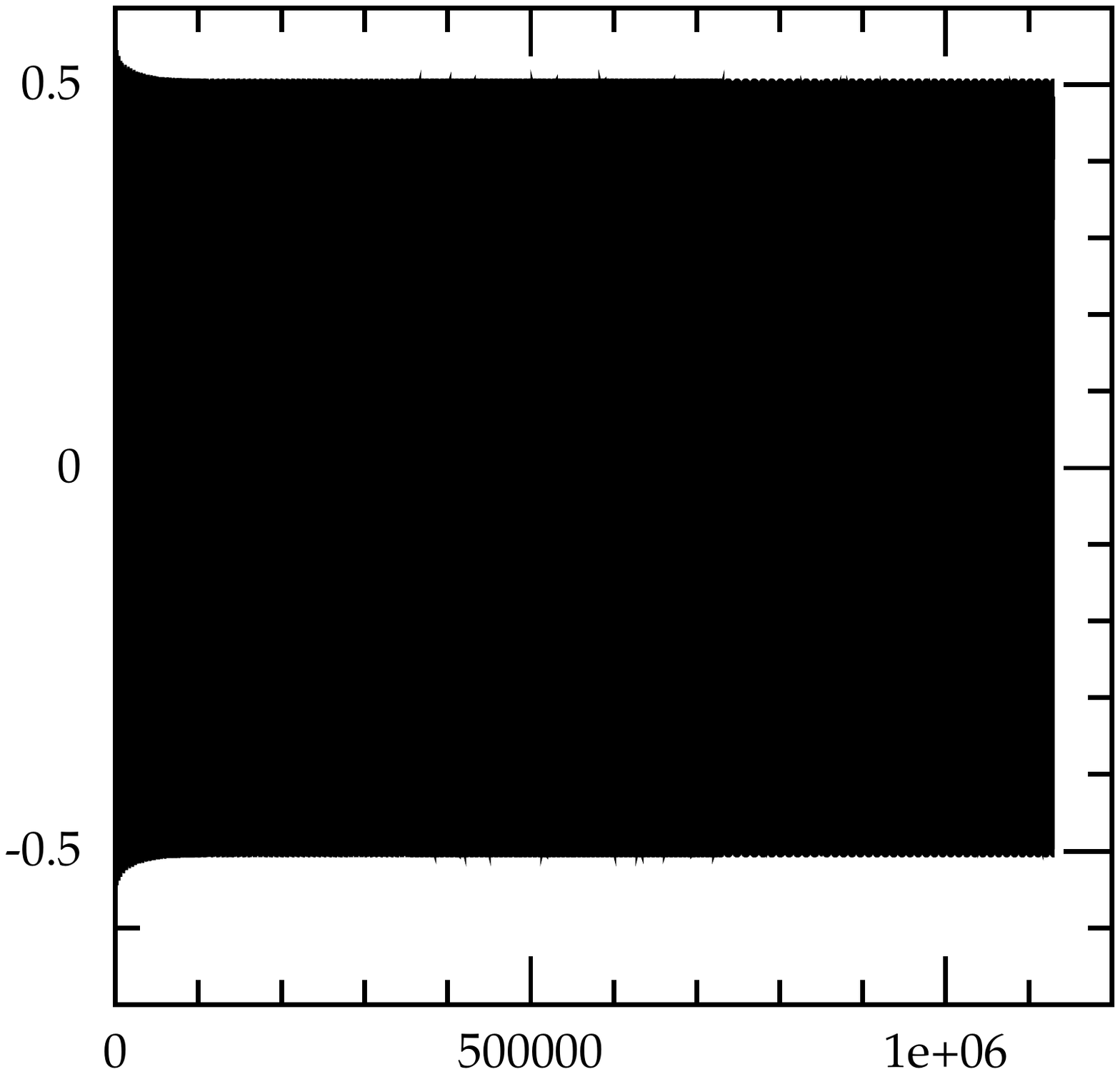}}
  \caption{Breather's simulation in the theory \rf{model} with initial configuration \rf{initialphi},  and with parameters given by of $\varepsilon=0.3$, $\nu=0.5$ and $\gamma=0$. The plots show the time dependence of (adimensional units): 
   (a) the energy \rf{energy}; and  (b) the field $\phi$ at position $x=0$ in the grid.  }
  \label{fig:fig2}
\end{figure}

\begin{figure}
\centering
 \includegraphics[trim = 0cm 0cm 1.8cm 1.8cm, width=0.45\textwidth]{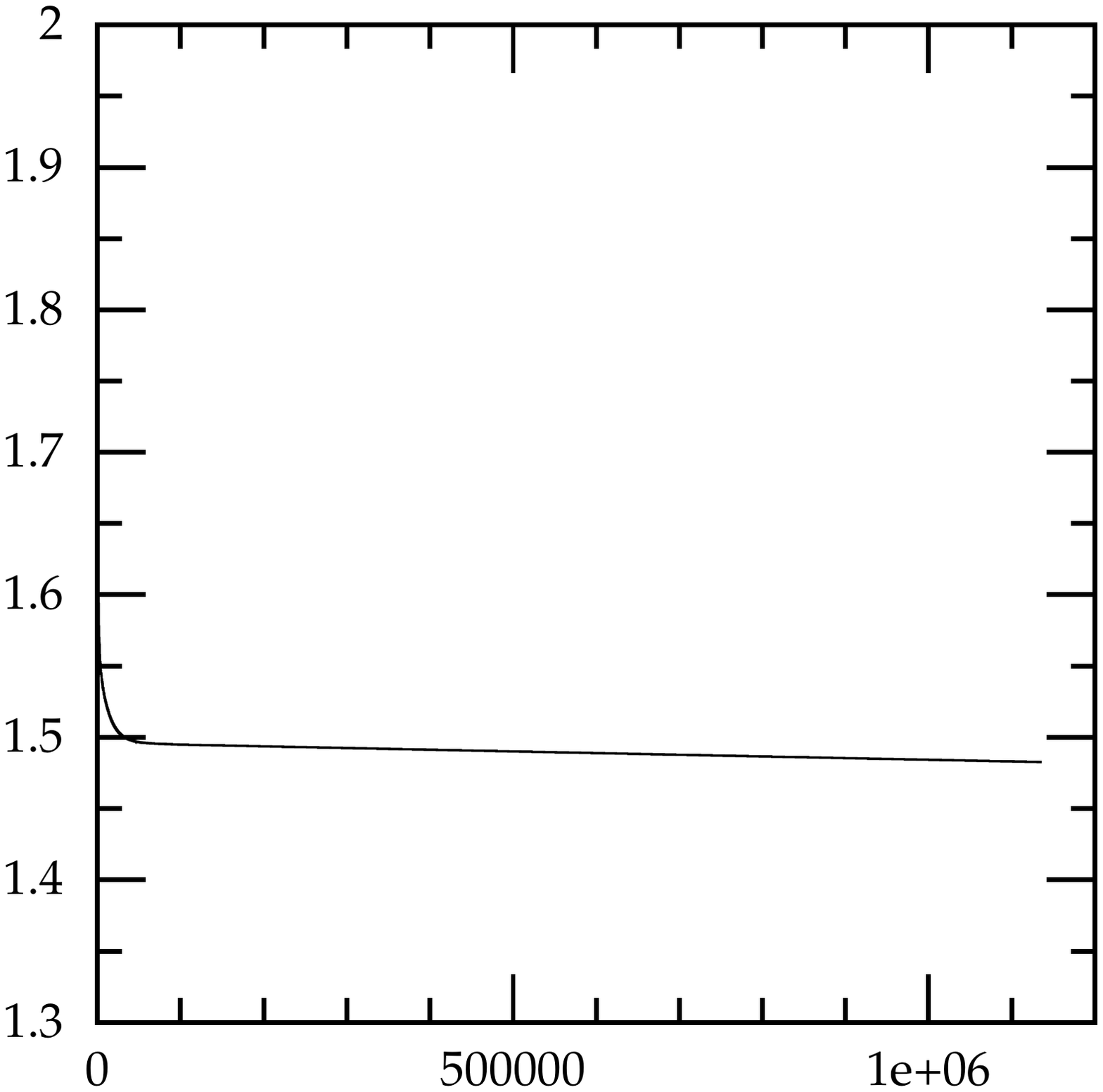}
 \caption{Breather's simulation in the theory \rf{model} with initial configuration \rf{initialphi},  and with parameters given by $\varepsilon=0.3$, $\nu=0.5$ and $\gamma=0.2$. The plot shows the time dependence  (adimensional units) of  the energy \rf{energy}.   }
\label{fig:4}
\end{figure}

\begin{figure}
  \centering
  \subfigure[]{\includegraphics[trim = 0cm 0cm 1.8cm 1.8cm, width=0.45\textwidth]{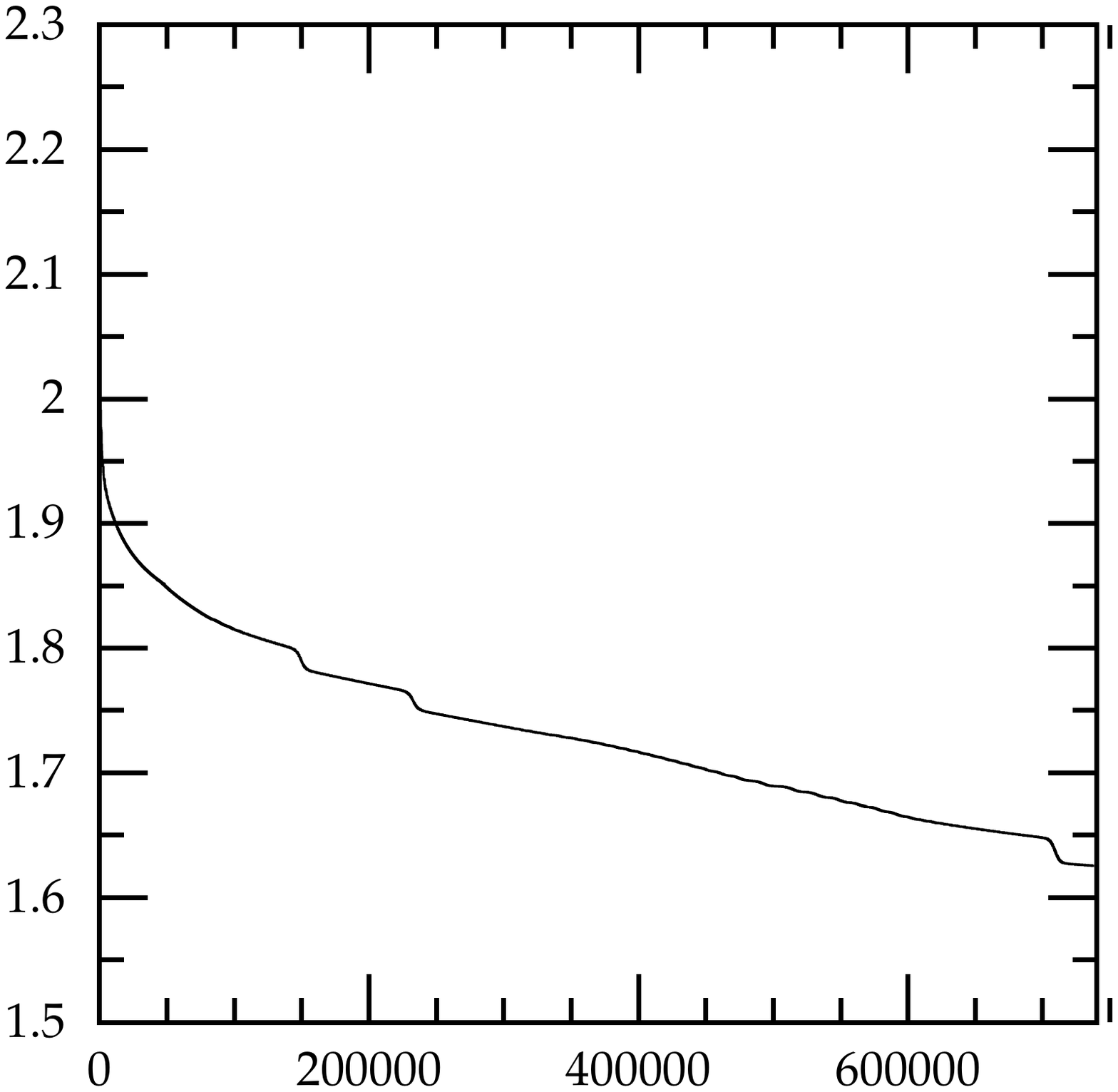}}                
  \subfigure[]{\includegraphics[trim = 0cm 0cm 1.8cm 1.8cm,  width=0.45\textwidth]{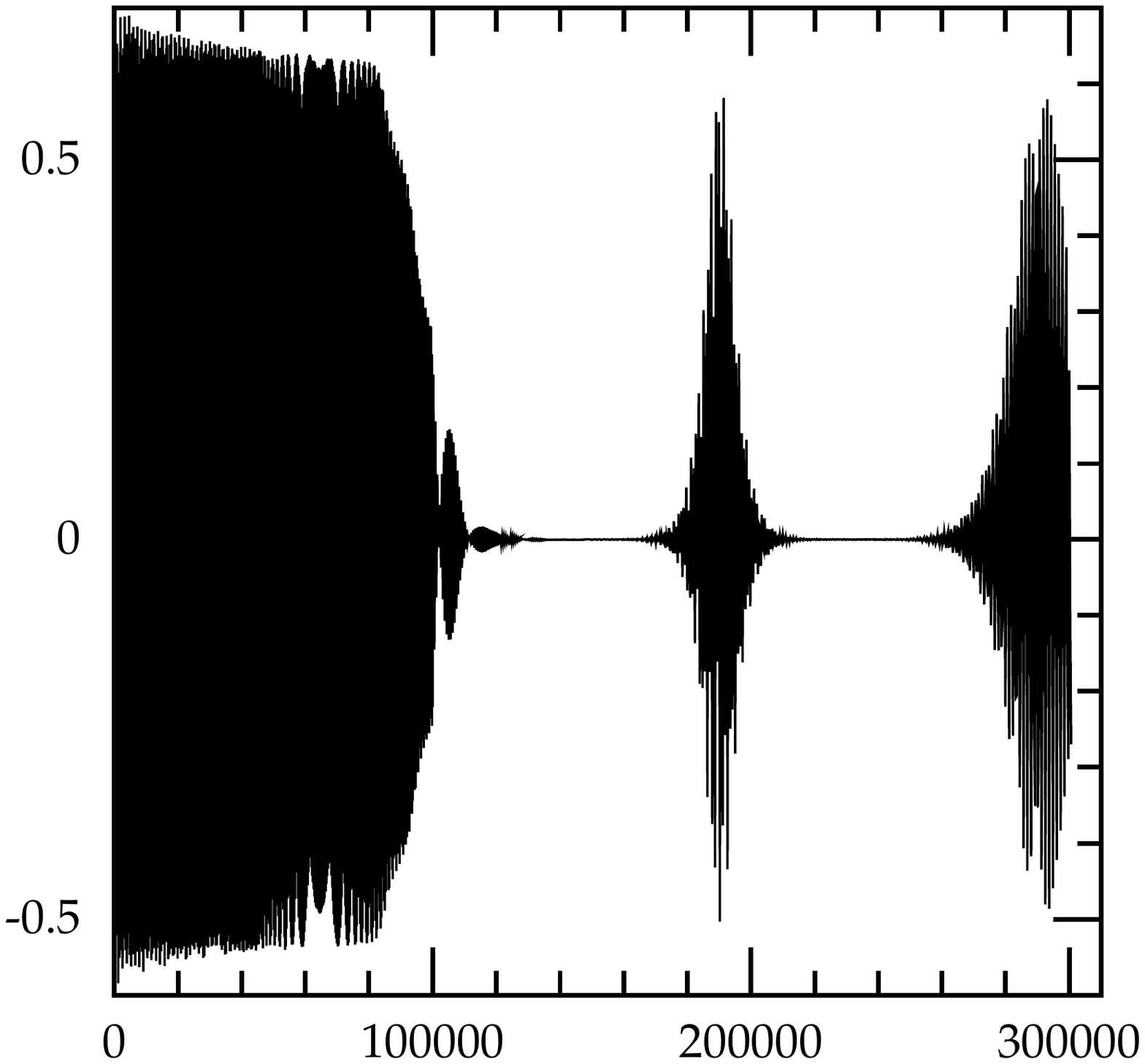}}
   \caption{Breather's simulation in the theory \rf{model} with initial configuration \rf{initialphi},  and with parameters given by of $\varepsilon=0.3$, $\nu=0.5$ and $\gamma=0.5$. The plots show the time dependence of (adimensional units): 
   (a) the energy \rf{energy}; and  (b) the field $\phi$ at position $x=0$ in the grid.  }
  \label{fig:fig5}
\end{figure}

\begin{figure}
  \centering
  \subfigure[]{\label{fig:energy300a}\includegraphics[trim = 0cm 0cm 1.8cm 1.8cm, width=0.45\textwidth]{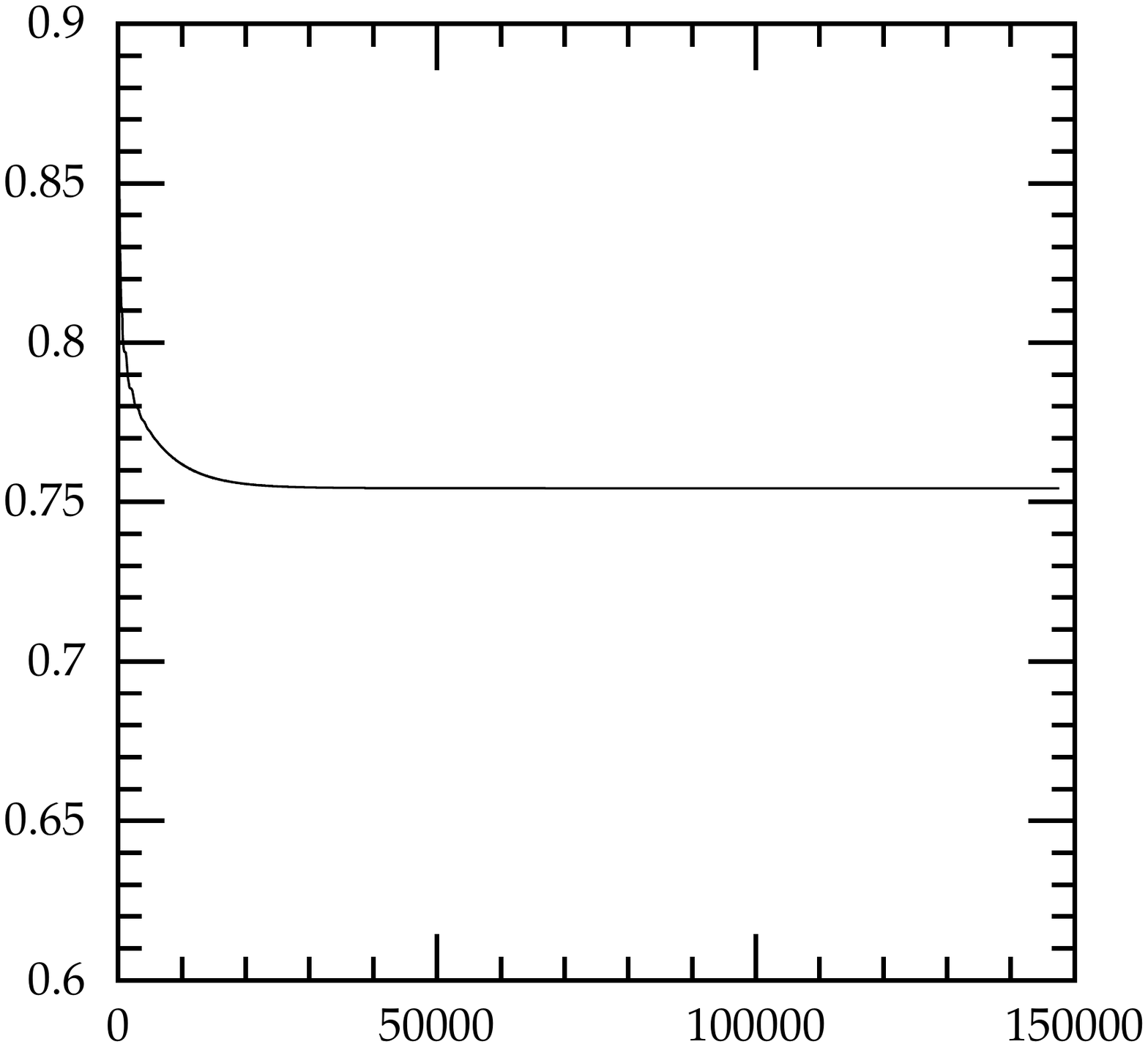}}   
  \subfigure[]{\label{fig:energy300b}\includegraphics[trim = 0cm 0cm 1.8cm 1.8cm, width=0.45\textwidth]{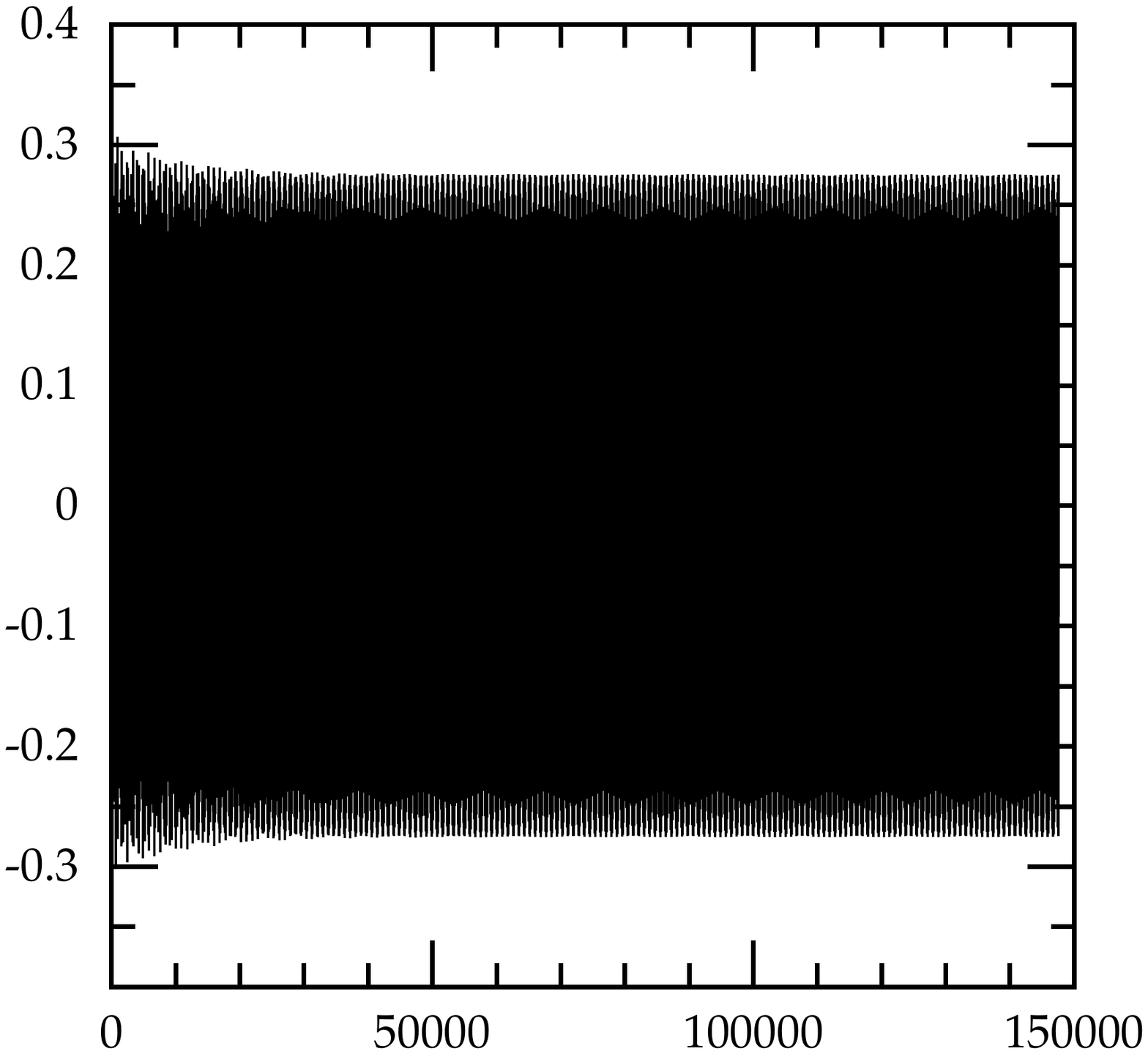}}                
  \subfigure[]{\label{fig:anomaly300a}\includegraphics[trim = 0cm 0cm 1.8cm 1.8cm,  width=0.45\textwidth]{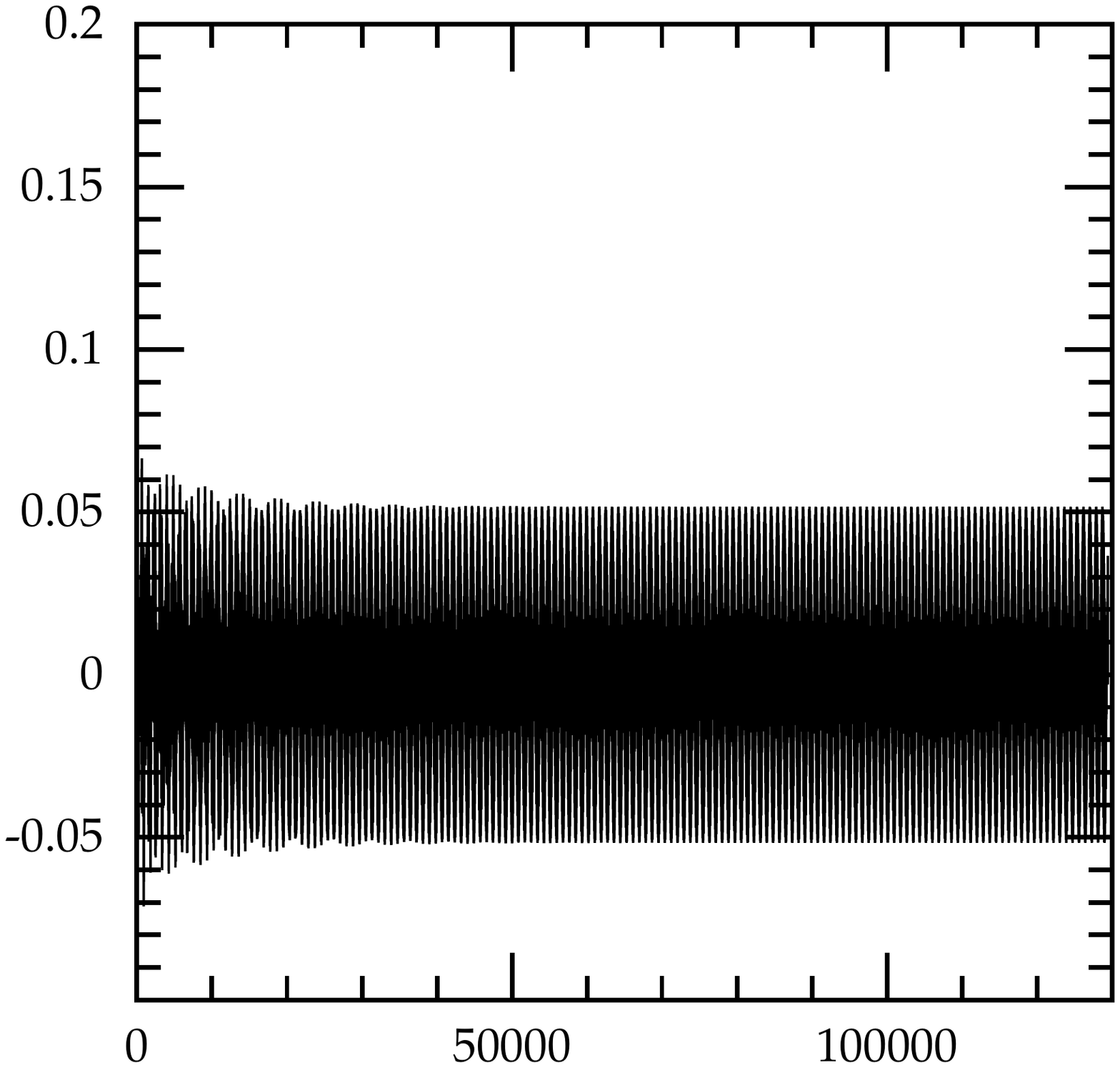}}
  \subfigure[]{\label{fig:anomaly300b}\includegraphics[trim = 0cm 0cm 1.8cm 1.8cm,  width=0.45\textwidth]{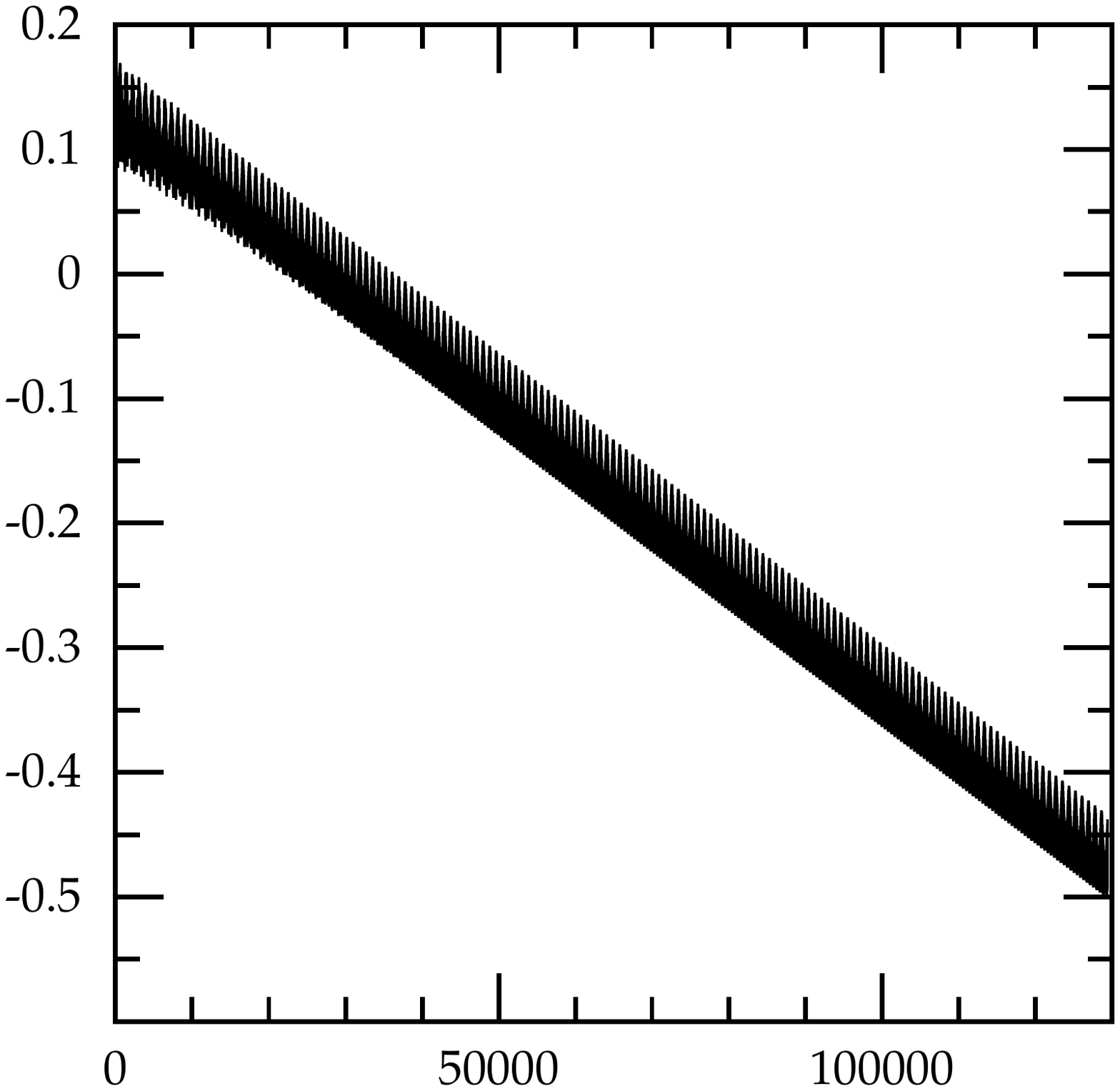}}
  \caption{Breather's simulation in the theory \rf{model} with initial configuration \rf{initialphi},  and with parameters given by of $\varepsilon=0.3$, $\nu=0.9$ and $\gamma=0$. The plots show the time dependence of (adimensional units): 
   (a) the energy \rf{energy};  (b) the field $\phi$ at position $x=0$ in the grid; (c) the anomaly \rf{alpha3} and (d) the integrated anomaly \rf{beta3}. }
    \label{fig:300}
\end{figure}

Looking at the table above one can easily spot a correlation between the values of the energy 
\rf{niceenergy} of the initial configuration \rf{initialphi} used in the simulations, and the lifetime of the breathers.  The higher the energy the shorter  the lifetime of the breathers. In fact, we have started all our simulations with $\phi=0$ on all sites of the grid. The energy \rf{niceenergy} corresponds to the total  kinetic energy given to the initial configuration. Note that the potential energy is zero because $V\(\phi=0\)=0$, and the elastic potential energy is also zero because $\frac{d\phi}{dx}=0$, on all sites of the grid at $t=0$. So, the smaller is the initial kinetic energy, the smaller is  the maximal value the $\phi$-field can reach during the oscillations of the breather. Indeed, by departing from $\phi=0$ the potential energy $V$ increases, as seen from the plot of the potential in Fig. \ref {fig:plotv}.  But if the value of $\phi$ remains small the breather oscillates only inside the part of the well of the potential around $\phi=0$ where $V$ varies very little with $\ve$. So, the breather can stay close to the breather solution of the sine-Gordon model which is integrable. Therefore, one would expect such a breather to live longer.  Indeed, looking at Figures \ref{fig:600} - \ref{fig:300} one observes that for energies smaller than $\sim 2.3$, the amplitude of oscillations of the field $\phi$ is never larger than $0.6$. For these values of amplitudes one sees from Fig. \ref {fig:plotv} that the field $\phi$ does not reach regions where the potential departs significantly  from the sine-Gordon potential. In all these cases the breathers live quite long, except for the case of Figure \ref{fig:fig5} where $\gamma=0.5$, and as discussed above the lack of good parity properties of the solution makes it  short-lived. The dependence on $\gamma$ only affects the initial drop of energy but is not very significant from the point of view of whether the breather is long-lived or not. This is clearly seen from the cases involving initially high $\nu$ as shown in figures \ref{fig:600}, \ref{fig:900}, \ref{fig:150} 
and \ref{fig:350}. All these cases correspond to $\gamma=$ 0, 0.3, 0.5 and 0.7, respectively.
We note that all these breathers are long-lived. Their initial energies increase (but very little) with $\gamma$ and, in the initial evolution decrease more for larger $\gamma$ but then the decrease
slows down and the breathers appear to be very long-lived. It would be interesting to check
whether, at some stage, their energies 'cross' but the decrease is so slow and the gaps are still large enough that one would have to wait extremely long to observe this, if it ever happens, so this is not practical.

The only exception to our observation above is the case shown in Fig. \ref{fig:fig5} where $\gamma=0.5$ (and $\nu$ is smaller) and the effect of the lack of parity properties in this case,  as discussed above, plays an important role and makes the breather to die faster. The fact that the energy \rf{niceenergy} decreases with the increase of the value of $\ve$, makes it possible for us to  find very long lived breathers for large values of $\ve$. Note that the increase of the frequency $\nu$ of the initial configuration also plays a role in favouring the long lifetime of breathers since it decreases the value of the initial energy \rf{niceenergy}.

We can also observe a correlation among the anomaly $\alpha^{(3)}$, given in \rf{alpha3}, and the integrated anomaly   $\beta^{(3)}$ given in \rf{beta3}, with the life-time of the breather solutions. In Figures \ref{fig:600}, \ref{fig:900}, \ref{fig:200} and \ref{fig:300}, where the breathers are long lived, we see that the anomaly $\alpha^{(3)}$ oscillates steadily within a fixed interval. The integrated anomaly  $\beta^{(3)}$ on the other hand presents a very slow drift of the order of one part in $10$ for $10^4$-$10^5$ units of time. This is  quite a long range  of time integration, and it could well be inside the numerical errors which are difficult to estimate in these cases. In Figures \ref{fig:700} and \ref{fig:800}, where the breathers are short lived we see that the anomaly $\alpha^{(3)}$ does not oscillate within a fixed interval and the integrated anomaly $\beta^{(3)}$ does vary a lot. Thus, there is indeed a correlation between long lived/short lived breathers and well/badly behaved anomalies. However, the effect of the $\gamma$ parameter is not very visible in the behaviour of the anomalies. It seems that the other effect discussed above in connection with the low initial kinetic energy seems to be predominant in these cases. 

\begin{figure}
  \centering
  \subfigure[]{\label{fig:period1}\includegraphics[trim = 0cm 0cm 1.8cm 1.8cm, width=0.45\textwidth]{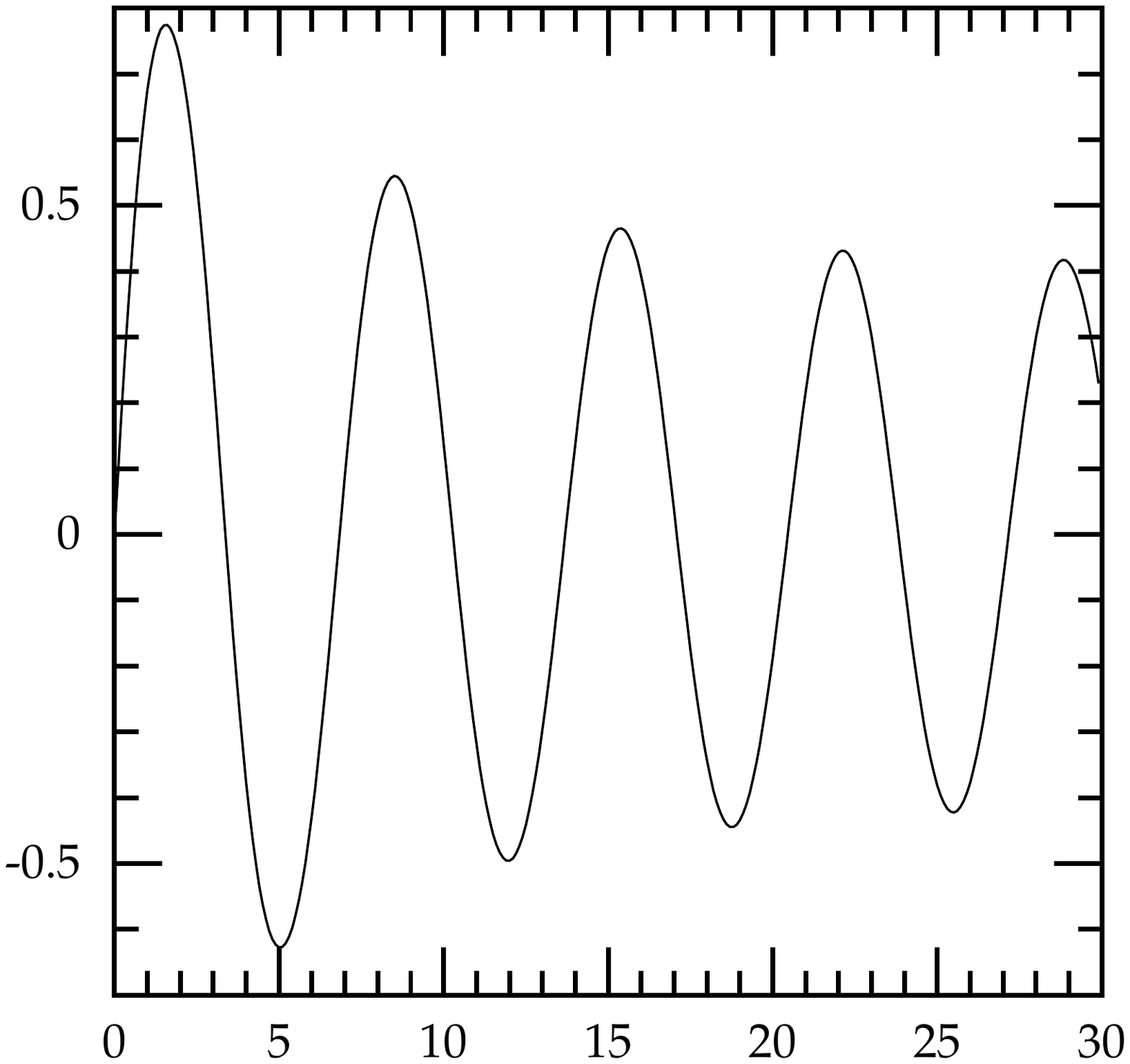}}                
  \subfigure[]{\label{fig:period2}\includegraphics[trim = 0cm 0cm 1.8cm 1.8cm,  width=0.45\textwidth]{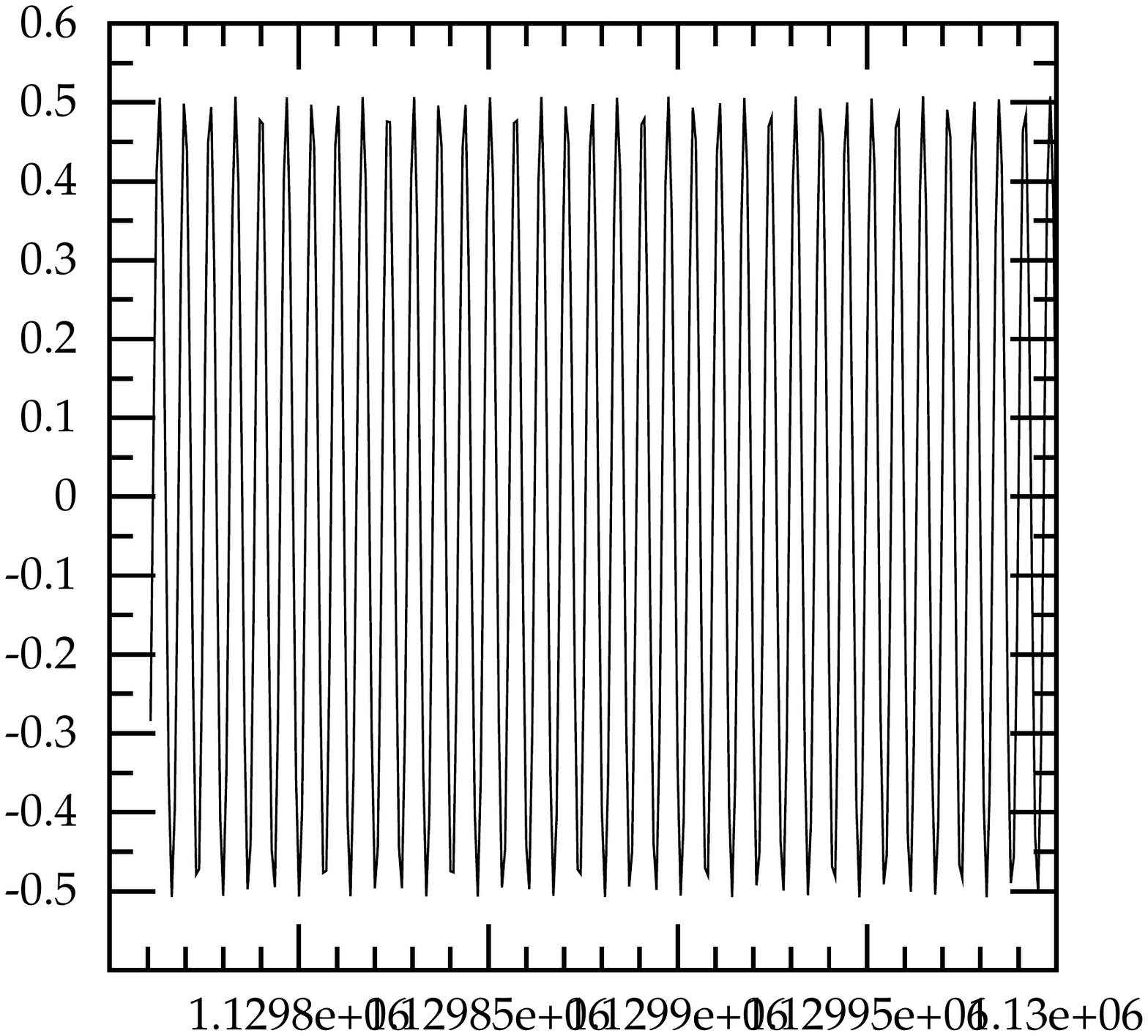}}
  \caption{Breather's simulation in the theory \rf{model} with initial configuration \rf{initialphi},  and with parameters given by of $\varepsilon=0.3$, $\nu=0.5$  and $\gamma=0.0$. The plots show the time dependence of (adimensional units): 
  (a) time dependence of the field $\phi$ at position $x=0$ in the grid, at the beginning of the simulation and (b) the same time dependence of the filed at a much later time.  }
  \label{fig:period}
\end{figure}

We have also observed that the energy of the breather-like solutions, after they have stabilized, seems to depend on the frequency in a way very similar to the exact sine-Gordon breather, {\em i.e.} $E\sim \sqrt{1-\nu^2}$. In Figure \ref{fig:period} we show, for the case of $\varepsilon=0.3$, $\nu=0.5$  and $\gamma=0.0$, the time dependence of the field $\phi$ at $x=0$,  in the left figure at the beginning of the simulation and in the right figure the same time dependence of the field at a much later time. 
From these two plots it is very clear that at first the breather oscillates with a period of $\sim T=6.8$ and much much later this period has decreased to $\sim T=6.628571$. Thus the frequency of this quasi-breather has increased from $\nu \sim 0.923997839264706$ to $\nu \sim 0.947894335107758$. Initially the energy was $E_{in}=1.7491569855$ and at the end $E_{fin}=1.409446715$. 
Thus the reduction of energy was roughly $1.7491569855/1.409446715=1.2410238476989$
which is quite small.

Note that for genuine breathers of the sine Gordon model the energy is proportional to
$E\sim \sqrt{1-\nu^2}$, so let us check what would have happened had we used this fact to estimate
the energies in this case too (as our field configuration resembles the sine-Gordon model's breather so well). We would have had
 \be
\left(\frac{E_{in}}{E_{fin}}\right)^2\,=\,\frac{1-\nu_{in}^2}{1-\nu_{fin}^2}.
\lab{ratioofenergies}
\ee
In our case the left and right hand sides of this formula are given by
\br
\lab{expratio}
(1.2410238476989)^2=1.54014019055743,
\\
 \frac{1-(0.923997839264706)^2}{1-(0.947894335107758)^2}\,=\,1.44072198272432
 \nonumber
\er
thus showing that this approximation is good within $\sim 7\%$.

We have also looked at the energy drop in simulations involving small $\varepsilon$. In such cases
we had the initial drop of the energy (like for larger values of $\varepsilon$) followed by a motion 
of the breather towards the boundaries with a reflection of it from the boundaries (producing a further
sharp drop of the energy at each reflection). This is clear from the simulation shown in figure 
\ref{fig:0010500} (see figures (c) and (d))

 Given the irregularity of the energy drop and the motion of the breathers it makes little sense to perform a comparison of the energy loss to the increase of the frequency of the oscillation that we have performed for larger values of $\varepsilon$.  
So if we want to perform a similar calculation we have to restrict our attention to the initial 
(non-moving) times of the breather, ({\it i.e.} consider it only for $t$ up to $\sim 48000$).
Hence, in the first two figures (a) and (b)  of Fig. \ref{fig:0010500} we present the blow up of the previous figures for the range of $t$ up to this value. Of course the plot of the variation of the field at $x=0$ is still not very clear 
so in the last two figures of Fig. \ref{fig:0010500}, namely (e) and (f) we present the variation of the field at the beginning of this interval and at the end of it. The plots cover the range of $t$ of 200 units and they show that at the beginning of the simulation the system performed $\sim 18.5$ oscillations in 200 units of time, while 
at $t\sim 47900$ this has increased to about 21.2

This shows that the frequency of the breather has increased from $0.581194640897501$ to about 
$0.6628760498885$. During the same time the energy changed from $6.3058416093$ to $5.7858560502$.

The square of the ratio of energies is given by $(\frac{6.3058416093}{5.7858560502})^2=1.1878206388$ and this is in an excellent agreement with the ratio
\be
\frac{1-(0.581194640897501)^2}{1-(0.6628760498885)^2}\,=\,1.18126701955371.
\ee
Hence again we have results in good agreement with our expectations.


In  figures \ref{fig:0010502} and \ref{fig:0010505} we present plots of the time dependence of the energy
and the field at $x=0$ for $\varepsilon=0.01$ but this time for larger values of $\gamma$; namely $\gamma=0.2$ (\ref{fig:0010502}) and  $\gamma=0.5$ (\ref{fig:0010505}).
The results are not that different from what we saw for $\gamma=0$ except that the decrease 
of energy is progressively greater.
In fact, in each case, the energy start decreasing quite fast but then the decrease slows down.
Again, the breathers  start moving and so, again, we could calculate the increase of frequencies
of oscillations and, like before, compare our expressions with the decrease of energies.

In the case of $\gamma=0.2$ the energies are 6.3787993275 and 5.0901576543, respectively.
Hence 
\be
\left(\frac{6.3787993275}{5.0901576543}\right)^2\sim 1.57041853413227.
\ee

The frequencies are $\nu_i=0.584336233551001$ and $\nu_f=0.75398223684$, respectively
and so we get 
\be
\frac{1-\nu_i^2}{1-\nu_f^2}\,=\,1.52615226946135
\ee
showing that, again, both sets of numbers are very close together.

For $\gamma=0.5$ (Fig:\ref{fig:0010505})  the corresponding energies are $6.4914572256$ and $4.5034440606$.
The frequencies are $\nu_i=0.596902604183049$ and $\nu_f=0.8168140899347$.
Thus, as before, we get 
\be
\left(\frac{6.4914572256}{4.5034440606}\right)^2=2.07775747962054
\ee
and
\be
\frac{1-\nu_i^2}{1-\nu_f^2}\,=\,1.9341309111334
\ee
in a very good agreement with the correspodning ratio from the energies.

\section{Further Comments and Some Conclusions}
\label{sec:conclusions}
\setcounter{equation}{0}

In this paper we have performed further studies of quasi-integrability  based on the observation in \cite{us}, \cite{us2} about the behaviour of the anomaly $X$ of the curvature \rf{zc} of the Lax potentials,  which distinguishes integrable models
from non-integrable ones. We have seem that the  anomaly integrated in time (see for instance \rf{beta3})  also vanishes in some non-integrable models for field configurations which possess the parity  symmetries discussed in Section \ref{sec:model}.

This observation was originally made in some very specific models and here we have tried to assess its general validity.  
So, in \cite{recent} we  constructed three  classes of models (one with symmetry, one without it and one (dependent on two parameters)) which would allow us to interpolate between the two. Our results have confirmed the validity of our assumption 
(and so extended the class of models in which our observation holds) and have also allowed us to study the way the anomaly varies as we move away from the models with this extra symmetry.
These results were first tested in great detail for the scattering of kinks of these models.
Of course, in such scatterings the kinks were interacting with each other only over very short
periods of time (when they were close to each other). So we decided to look also at the systems
involving breather-like structures on which the kinks and anti-kinks, being bound into breathers, 
interact with other all the times. In our previous work \cite{recent}
we have only glanced at such configurations. As the breather-like configurations depend on many parameters  in this paper we have concentrated our attention at looking at them in detail, in particular when the symmetry is present and when it is not. 
As expected, we have found that the symmetry helps a lot in the validity of ideas of quasi-integrability. When the symmetry is present the energy decrease is much reduced and the configurations
resemble, in their behaviour, the sine-Gordon breathers. When the symmetry is broken the breathers
decay quite rapidly and the range of validity of quasi-integrability is much reduced.

However, the symmetry is only one of many topics to investigate for the breather-like
configurations. We have also looked at the way the decay takes place and the dependence of this behaviour
on various parameters of the model. One of the most interesting outcomes of our simulations is the understanding of the way the decay takes place. The breathers increase the frequency of their 
oscillations. This is quite clear from the energy of our initial configuration \rf{niceenergy}
but it is amazing to see as the breather-like field looses its energy this formula still holds true, as only 
$\nu$ in it increases.
In our simulations some breather-like configurations started moving and lost their energies also by reflecting from the edges of the grid.
For them the frequency increased even more so that their total energy (consisting of the energy of the oscillation and the energy
of the motion was comparable to the final energy of the field, and much lower than the original energy).

Finally our work has lead to the discovery of the existence of many long lived breather-like fields a long way away ({\it i.e.} for large 
perturbations) from the sine-Gordon model.
This was particularly true for the cases with symmetry and so it does extend the range of validity of quasi-integrability.

\vspace{2cm}

{\bf Acknowledgements:} 
The work reported in this paper was started when the authors
were ``researching in pairs'' at the Matematisches Forschunginstitute in Oberwolfach. They would like to thank the Oberwolfach Institute for its hospitality. The work was finished when WJZ was visiting the S\~ao Carlos Institute of Physics of the  University of S\~ao Paulo (IFSC/USP).
His visit was supported by a grant from FAPESP. He would like to thank FAPESP for this grant
and the IFSC/USP for its hospitality.
LAF's work was partially supported by CNPq (Brazil).

\newpage

\vspace{1cm}

\end{document}